\pdfoutput=1

\documentclass[11pt]{article}
\usepackage{jheppub}

\usepackage{amsmath}
\usepackage{amssymb}
\usepackage{graphicx,color,slashed,mathdots}
\usepackage{ifpdf}
\usepackage[english]{babel}
\usepackage{url}
\usepackage{dsfont}
\usepackage{float}
\usepackage{color}

\allowdisplaybreaks[1]

\font\tenshuffle=shuffle10 \font\sevenshuffle=shuffle7 \font\fiveshuffle=shuffle7 at 5pt
\def\shuffle{{%
\def\Dshuffle{\mathbin{\hbox{\tenshuffle\char'001}}}%
\def\Sshuffle{\mathbin{\hbox{\sevenshuffle\char'001}}}%
\def\SSshuffle{\mathbin{\hbox{\fiveshuffle\char'001}}}%
\mathchoice{\Dshuffle}{\Dshuffle}{\Sshuffle}{\SSshuffle}}}

\restylefloat{figure}
\definecolor{dgreen}{rgb}{0,0.70,0.30}
\definecolor{gold}{rgb}{0.85,.66,0}
\definecolor{purple}{rgb}{1.0,0.3,0.6}

\usepackage{tikz}
\usetikzlibrary{calc} \usetikzlibrary{patterns} \usetikzlibrary{decorations.pathreplacing} \usetikzlibrary{decorations.markings} \usetikzlibrary{decorations.pathmorphing} \usetikzlibrary{positioning}

\def\be{\begin{equation}}
\def\ee{\end{equation}}
\def\ba{\begin{array}}
\def\ea{\end{array}}

\newcommand{\bea}{\begin{eqnarray}}
\newcommand{\eea}{\end{eqnarray}}

\def\beq{\begin{equation}}
\def\eeq{\end{equation}}

\newcommand{\dd}{\mathrm{d}}
\newcommand{\te}{\textrm}
\newcommand{\ap}{{\alpha'}}
\newcommand{\la}{\lambda}

\newcommand{\AAA}{\mathbb A}
\newcommand{\FF}{\mathbb F}

\newcommand{\Mfrak}{\mathfrak M}
\newcommand{\Nfrak}{\mathfrak N}
\newcommand{\YMF}{${\rm YM}{+} F^3{+}  F^4$}

\title{Berends--Giele currents in Bern--Carrasco--Johansson \\ gauge for 
$F^3$- and $F^4$-deformed Yang--Mills amplitudes}
\author[a,b]{Lucia M. Garozzo,}
\author[a]{Leonel Queimada,}
\author[a,c]{Oliver Schlotterer}

\affiliation[a]{Perimeter Institute for Theoretical Physics,
Waterloo, ON N2L 2Y5, Canada}
\affiliation[b]{Department of Physics and Astronomy, Uppsala University, 75108 Uppsala, Sweden}
\affiliation[c]{Max--Planck--Institut f\"ur Gravitationsphysik,
Albert--Einstein--Institut,
14476 Potsdam, Germany}

\emailAdd{lucia.garozzo@physics.uu.se}
\emailAdd{lquintaqueimada@perimeterinstitute.ca}
\emailAdd{olivers@aei.mpg.de}

\date{\today}

\abstract{We construct new representations of tree-level amplitudes in $D$-dimensional gauge theories with
deformations via higher-mass-dimension operators $\alpha' F^3$ and $\alpha'^{2} F^4$. Based on Berends--Giele 
recursions, the tensor structure of these amplitudes is compactly organized via off-shell currents. 
On the one hand, we present manifestly cyclic representations, where the complexity 
of the currents is systematically reduced. On the other hand, the duality between color and kinematics
due to Bern, Carrasco and Johansson is manifested by means of non-linear gauge transformations of the
currents. We exploit the resulting notion of Bern--Carrasco--Johansson gauge to provide explicit and
manifestly local double-copy representations for gravitational amplitudes involving $\alpha' R^2$ and $\alpha'^2 R^3$ operators.
}

\preprint{UUITP-40/18 \\
}

\begin{document}

\linespread{0.96}

\maketitle{}

\setcounter{tocdepth}{2}
%\tableofcontents

\numberwithin{equation}{section}

\newpage

%%%%%%%%%%%%%%%%%%%%%%%%%%%%%%%%%%%%%%%%%%%%%%%%
%%%%%%%%%%%%%%%%%%%%%%%%%%%%%%%%%%%%%%%%%%%%%%%%
%%%%%%%%%%%%%%%%%%%%%%%%%%%%%%%%%%%%%%%%%%%%%%%%
%%%%%%%%%%%%%%%%%%%%%%%%%%%%%%%%%%%%%%%%%%%%%%%%
%%%%%%%%%%%%%%%%%%%%%%%%%%%%%%%%%%%%%%%%%%%%%%%%

\linespread{1.00}

\section{Introduction}
\label{sec:1}

Recent investigations of scattering amplitudes in gauge theories and gravity revealed
a wealth of mathematical structures and surprising connections between different theories.
For gravitational theories in $D$ spacetime dimensions,
traditional methods for tree amplitudes and loop integrands
naively give rise to an exasperating proliferation of terms. 
Still, the final answers for these quantities across various loop- and leg orders
take a strikingly simple form: The dependence on the 
spin-two polarizations can often be reduced to squares of suitably chosen 
gauge-theory quantities.

The double-copy structure of perturbative gravity originates from string theory where Kawai
Lewellen and Tye (KLT) identified universal relations between open- and closed-string tree-level
amplitudes \cite{Kawai:1985xq}. The KLT relations have been later on reformulated in a field-theory framework 
by Bern, Carrasco and Johansson (BCJ) \cite{BCJ, loopBCJ, Bern:2017yxu} such as to flexibly
address multiloop integrands. In this way, numerous long-standing questions on the ultraviolet 
properties of supergravity theories  have been resolved \cite{Bern:2012uf, Bern:2012cd, Bern:2013uka, 
Bern:2014sna, Bern:2017ucb, Bern:2018jmv}, bypassing the spurious explosion of terms in intermediate steps.

This double-copy approach to gravitational amplitudes takes a particularly elegant form once a hidden 
symmetry of gauge-theory amplitudes is manifested -- the duality between color and kinematics due
to BCJ \cite{BCJ}. At tree level, the BCJ duality in gauge theories has 
not only been explained and manifested in string theories \cite{BjerrumBohr:2009rd, Stieberger:2009hq, 
Tye:2010dd, Mafra:2011kj, Mafra:2014oia, Fu:2018hpu} but also extends to various constituents of string-theory
amplitudes \cite{Mafra:2011nv, Broedel:2012rc, Zfunctions, Stieberger:2014hba, 
Huang:2016tag, Azevedo:2018dgo}. In particular, the following terms in the gauge-field
effective action of the open bosonic string\footnote{The low-energy effective
action of the open bosonic string involves another operator $\sim \zeta_2 \alpha'^2 F^4$ at the mass 
dimensions in (\ref{pert0.1}) which will not be discussed in this article. Said $\zeta_2 \alpha'^2 F^4$-operator 
is also known from the superstring and cannot be reconciled with the BCJ duality \cite{Broedel:2012rc}.} 
in $D$ spacetime dimensions preserve the BCJ duality to the order of $\alpha'^2$ \cite{Broedel:2012rc},
\begin{align}
{\cal S}_{{\rm YM}+F^3+F^4} &= \int \dd^D x \ \te{Tr} \Big\{ \, \frac{1}{4}\, F_{\mu \nu} \, F^{\mu \nu} + \frac{ 2\ap}{3} \,F_\mu{}^\nu \, F_\nu{}^\la \,F_{\la}{}^\mu + \frac{ \ap^2}{4}  \,  [F_{\mu \nu}, F_{\la \rho}]  [F^{\mu \nu}, F^{\la \rho}] \, \Big\} \, ,
\label{pert0.1}
\end{align}
where $F^{\mu \nu}$ and $\alpha'$ denote the non-abelian field strength and the inverse string tension, respectively. 
In presence of the effective action (\ref{pert0.1}), KLT formulae and
BCJ double-copy representations known from Einstein gravity  
extend\footnote{See \cite{BjerrumBohr:2003vy, BjerrumBohr:2003af} 
for earlier work on
the interplay of the KLT relations at the three- and four-point level with
gravitational matrix elements of $R^2,R^3$ operators and $F^3,F^4$-deformed gauge-theory amplitudes.} 
to gravitational tree amplitudes\footnote{In slight abuse of terminology, we will usually refer to the matrix
elements from higher-mass-dimension operators as ``amplitudes''. In the case at hand, we will be interested
in contributions from single- or double-insertions of $\alpha' R^2$ operators and single-insertions of $\alpha'^2 R^3$} from $\alpha' R^2+\alpha'^2 R^3$ operators \cite{Broedel:2012rc} involving higher powers in the Riemann curvature $R$. The schematic notation $R^2$ and $R^3$ for operators in the gravitational effective 
action is understood to comprise additional couplings of a $B$-field and a dilaton $\varphi$ (such as $e^{-2\varphi }R^2$)
known from the low-energy regime of the closed bosonic string \cite{Metsaev:1986yb}.

The interplay of higher-mass-dimension operators $D^{2m} F^n$ and $D^{2m}R^n$ in string theories
with the BCJ duality and double copy is well understood from the worldsheet description of tree-level amplitudes
\cite{Broedel:2012rc, Stieberger:2014hba, Huang:2016tag, Azevedo:2018dgo}. Also, $D$-dimensional amplitudes of the 
$F^3$ operator and their double copy have been studied in the CHY formalism \cite{He:2016iqi}.
The purpose of this work is to
explore a complementary approach and to manifest the BCJ duality of the $\alpha' F^3+\alpha'^2 F^4$
operators directly from the Feynman rules of the action (\ref{pert0.1}).
We will follow some of the ideas in earlier work on ten-dimensional super-Yang--Mills (SYM) 
\cite{Mafra:2011kj, Mafra:2014oia, Lee:2015upy, Mafra:2015vca} and realize the 
BCJ duality at the level of Berends--Giele currents \cite{Berends:1987me} -- up to the order of $\ap^2$.

We will reorganize the Feynman-diagrammatics of $(\alpha' F^3+\alpha'^2 F^4)$-deformed
Yang--Mills (YM) theory such as to find an explicit off-shell realization of the BCJ duality. 
The key idea is to remove the deviations from the BCJ duality by applying a concrete 
non-linear gauge transformation to the generating series of Berends--Giele currents.
Our starting point for the currents is Lorenz gauge, and their transformed versions
which obey the color-kinematics duality are said to implement {\it BCJ gauge} in 
$(\alpha' F^3+\alpha'^2 F^4)$-deformed YM theory\footnote{See 
\cite{Lee:2015upy, Mafra:2015vca} for generating series of Berends--Giele currents, their non-linear 
gauge transformations and BCJ gauge in ten-dimensional SYM.}.

Particular emphasis will be put on the locality properties of our construction, i.e.\ the 
absence of spurious kinematic poles in the gauge-theory constituents.
Like this, the gravitational amplitudes from $\alpha' R^2+\alpha'^2 R^3$ operators obtained
via double copy reflect the propagator structure of cubic-vertex diagrams and 
facilitate loop-level applications based on the unitarity method 
\cite{Bern:1994zx, Bern:1994cg, Bern:1997sc, Britto:2004nc, Bern:2007ct}.
Moreover, locality of the gauge-theory building blocks will be crucial for one of our main results: a kinematic derivation
of the BCJ relations \cite{BCJ} among color-ordered amplitudes of 
(\YMF) \cite{Broedel:2012rc}, a manifestly gauge invariant formulation of the BCJ duality.

Finally, the complexity of the Berends--Giele currents of (\YMF)
will be systematically shortened by adapting techniques \cite{Mafra:2010ir, Mafra:2010jq, Mafra:2011kj, Mafra:2014oia}
from ten-dimensional SYM. Our manipulations resemble BRST integration by parts of the 
pure-spinor superstring \cite{Berkovits:2000fe} and allow for manifestly cyclic amplitude representations
as well as streamlined expressions for the gauge parameter towards BCJ gauge.

The results of this work on the currents and amplitudes of (\YMF) are valid 
up to and including the order of $\alpha'^2$. At higher orders in $\alpha'$, effective operators 
including $\alpha'^3 D^2F^4$ as provided by the bosonic string are required to maintain the 
BCJ duality \cite{Broedel:2012rc, Huang:2016tag}.
Moreover, our results hold in any number $D$ of spacetime dimensions: Apart from the critical dimension
$D=26$ of the bosonic string and the phenomenologically interesting situation with $D=4$, this allows for a
flexible unitarity-based investigation of loop integrands in various dimensions and dimensional 
regularization, see e.g.\ \cite{Ellis:2008ir, Davies:2011vt}.

By its close contact with Lagrangians, the construction in this work
resonates with recent developments in scalar theories with color-kinematics duality and double-copy structures
\cite{Chen:2013fya, Cachazo:2014xea}:
For the color-kinematics duality of the non-linear sigma model (NLSM) of Goldstone bosons \cite{Chen:2013fya},
a Lagrangian origin along with the structure constants of a kinematic algebra has been 
identified in \cite{Cheung:2016prv}. 
This new formulation of the NLSM can be derived from higher dimensional YM theory \cite{Cheung:2017yef},
and a string-inspired higher-derivative extension of the NLSM\footnote{Said higher-derivative extension
of the NLSM is defined by the $\zeta_2\ap^2$-order of abelian Z-theory \cite{Carrasco:2016ldy}.} \cite{Carrasco:2016ldy} has been recently obtained
from the analogous dimensional reduction of $\alpha' F^3$ in a companion paper \cite{NLSMperturbiners}.
In view of these connections, we hope that the notion of BCJ gauge inspires a reformulation of the (\YMF)-Lagrangian
(\ref{pert0.1}) where -- similar to \cite{Cheung:2016prv} -- the $D$-dimensional kinematic algebra is manifest\footnote{See \cite{Bern:2010yg, Tolotti:2013caa} for earlier Lagrangian-based approaches to the BCJ duality and \cite{Fu:2016plh} for a connection with the Drinfeld double of the Lie algebra of vector fields. Also see \cite{Monteiro:2011pc} for the kinematic algebra
in the self-dual sectors of $D=4$ YM theory and gravity.}.

Another source of motivation for this work stems from the renewed interest in the 
gravitational $\alpha'R^2+\alpha'^2R^3$ interactions in $D\neq 4$ dimensions. While $R^3$ is well-known to be the 
 first (non-evanescent) two-loop counterterm for pure gravity \cite{Goroff:1985sz, Goroff:1985th}, the evanescent 
 one-loop counterterm 
$R^2$ was recently found to contaminate dimensional regularization at two loops \cite{Bern:2015xsa, Bern:2017puu}.
Moreover, evanescent matrix elements of $R^2$ are closely related to certain anomalous amplitudes of ${\cal N}=4$ supergravity
\cite{Carrasco:2013ypa} through double copy \cite{Bern:2017tuc}. Finally, when viewed as ambiguities in defining 
quantum theories, matrix elements of higher dimensional operators can be crucial to restore symmetries when
using a non-ideal regulator for loop amplitudes \cite{Bern:2017rjw}.
We hope that our $D$-dimensional double-copy representations for tree-level amplitudes of 
$(\alpha'R^2+\alpha'^2R^3)$-deformed gravity shed further light into these loop-level 
topics: either by unitarity or by using the BCJ-gauge currents as building blocks for loop amplitudes that universally represent
tree-level subdiagrams\footnote{See for instance \cite{Mafra:2014gja, Mafra:2015mja, Berg:2016fui, He:2017spx} 
for the use of tree-level Berends--Giele currents in $D>4$-dimensional loop amplitudes of gauge theories with maximal and 
half-maximal supersymmetry.}.

%%%%%%%%%%%%%%%%%%%%%%%%%%%%%%%%%%%%%%%%%%%%%%%%
%%%%%%%%%%%%%%%%%%%%%%%%%%%%%%%%%%%%%%%%%%%%%%%%

\subsection{Outline}
\label{sec:1.1}

This work is organized as follows: In section \ref{sec:2}, we review the basics 
of Berends--Giele recursions, the BCJ duality as well as the double copy and establish
the associated elements of notation. Section \ref{sec:3} is dedicated to amplitudes
of (\YMF) in different types of Berends--Giele representations including a systematic
reduction of the rank of the currents. In section \ref{sec:4}, an explicit off-shell realization of the 
BCJ duality is obtained from the Berends--Giele setup. Finally, section
\ref{sec:5} relates this realization of the BCJ duality to non-linear gauge freedom and
combines the off-shell ingredients from the previous section to manifestly local amplitude representations of
(\YMF) and gravity with $\alpha'R^2+\alpha'^2R^3$ operators. A derivation of the BCJ relations to the order
of $\ap^2$ from purely kinematic arguments is given in section \ref{sec:5.2}.

%%%%%%%%%%%%%%%%%%%%%%%%%%%%%%%%%%%%%%%%%%%%%%%%
%%%%%%%%%%%%%%%%%%%%%%%%%%%%%%%%%%%%%%%%%%%%%%%%
%%%%%%%%%%%%%%%%%%%%%%%%%%%%%%%%%%%%%%%%%%%%%%%%
%%%%%%%%%%%%%%%%%%%%%%%%%%%%%%%%%%%%%%%%%%%%%%%%
%%%%%%%%%%%%%%%%%%%%%%%%%%%%%%%%%%%%%%%%%%%%%%%%

\section{Review and notation}
\label{sec:2}

In this section, we set up notation and review the key ideas and applications of 
Berends--Giele recursions for tree-level amplitudes in YM theory, in particular
\begin{itemize}
\item the resummation of Berends--Giele currents to obtain perturbiner solutions
to the non-linear field equations
\item manifestly cyclic Berends--Giele representations of YM amplitudes involving
currents of smaller rank than naively expected.
\end{itemize}
We will also review the BCJ duality and the double copy from a perspective
which later on facilitates the implementation of these features in tree
amplitudes and Berends--Giele currents of (\YMF) as well as gravity with $\alpha'R^2+\alpha'^2R^3$ operators.

%%%%%%%%%%%%%%%%%%%%%%%%%%%%%%%%%%%%%%%%%%%%%%%%
%%%%%%%%%%%%%%%%%%%%%%%%%%%%%%%%%%%%%%%%%%%%%%%%

\subsection{Berends--Giele recursions}
\label{sec:2.1}

An efficient approach to determine the tensor structure of $D$-dimensional tree amplitudes
in pure YM theory has been introduced by Berends and Giele in 1987 \cite{Berends:1987me}.
The key idea of the reference is to recursively combine all color-ordered Feynman diagrams 
involving multiple external on-shell legs and a single off-shell leg. This recursion is implemented via
currents $J^\mu_{12\ldots p}$ that depend on the polarization vectors $e^\mu_i$ and
lightlike momenta $k_i^\mu$ of the external particles $i=1,2,\ldots,p$ subject to the following on-shell constraints
\beq
e_i\cdot k_i = k_i \cdot k_i = 0 \ \forall \ i=1,2,\ldots \, .
\label{pert1.minus1}
\eeq
While Latin letters $i,j,\ldots$ refer to external-state labels, Lorentz-indices $\mu,\nu,\ldots = 0,1,\ldots,$ $D{-}1$ 
are taken from the Greek alphabet. 

Currents of arbitrary multiplicity can be efficiently computed from the Berends-Giele recursion \cite{Berends:1987me}
\beq
J_i^\mu = e^\mu_i \ , \ \ \ \ \ \ 
s_P J^\mu_P = \sum_{XY=P} [J_X , J_Y]^\mu + \sum_{XYZ=P} \{J_X,J_Y,J_Z\}^\mu\, ,
\label{pert1.0}
\eeq
where
\begin{align}
[J_X , J_Y]^\mu &= (k_Y \cdot J_X) J_Y^\mu -  (k_X \cdot J_Y) J_X^\mu  + \frac{1}{2}(k_X^\mu - k_Y^\mu)(J_X\cdot J_Y) \label{pert1.0A} \\
 \{J_X,J_Y,J_Z\}^\mu &= (J_X\cdot J_Z) J_Y^\mu-\frac{1}{2}(J_X\cdot J_Y) J_Z^\mu-\frac{1}{2} (J_Y\cdot J_Z) J_X^\mu  \, .\label{pert1.0B}
\end{align}
The external states have been grouped into multiparticle labels or words $P=12\ldots p$.
We will represent multiparticle labels by capital letters $P,Q,X,Y,\ldots$ and denote
their length, i.e.\ the number of labels in $P=12\ldots p$, by $|P|=p$. The summation
over $XY=P$ on the right-hand side of (\ref{pert1.0}) instructs to deconcatenate $P$ into
non-empty words $X=12\ldots j$ and $Y=j{+}1\ldots p$ with $j=1,2,\ldots,p{-}1$ and therefore generates
$|P|{-}1$ terms\footnote{For instance, the summation over $XY=P$ with $P=1234$ of length four incorporates the pairs
$(X,Y) = (123,4),\, (12,34)$ and $(1,234)$.}. Similarly, 
$XYZ=P$ encodes $\frac{1}{2}(|P|{-}1)(|P|{-}2)$ deconcatenations into non-empty words 
$X=12\ldots j$, $Y=j{+}1\ldots l$ and $Z=l{+}1\ldots p$ with $1\leq j<l \leq p{-}1$.

\begin{figure}[h]
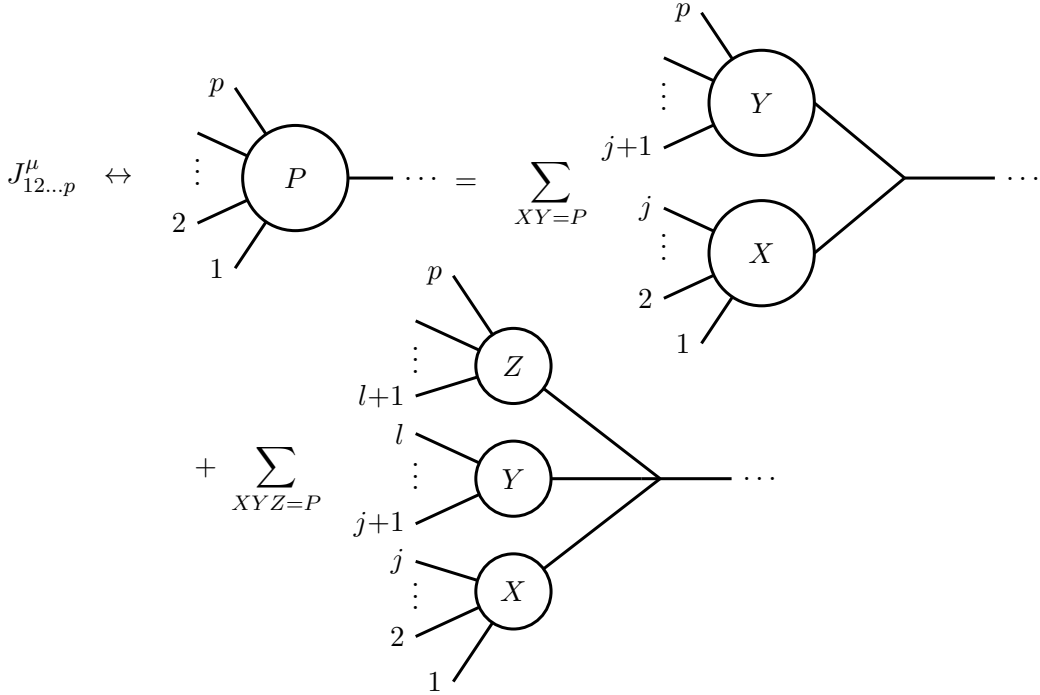

\begin{center}
\tikzpicture[line width=0.40mm]
\draw (0,0) -- (-0.8,-1.2) node[left]{$1$};
\draw (0,0) -- (-1.3,-0.6) node[left]{$2$};
\draw (0,0) -- (-0.8,1.2) node[left]{$p$};
\draw (0,0) -- (-1.3,0.6);
\draw (-1.3,0.2)node{$\vdots$};
\draw(0,0) -- (1.3,0)node[right]{$\ldots$};
\draw[fill=white] (0,0) circle(0.7cm);
\draw(0,0)node{$P$};
\draw(-3,0)node{$J^\mu_{12\ldots p} \ \ \leftrightarrow $};
%%%%
\draw (3,-0.2)node{$\displaystyle= \ \ \ \sum_{XY=P}$};
\scope[xshift=6.2cm,yshift=1cm]
\draw (0,0) -- (-0.8,1.2) node[left]{$p$};
\draw (0,0) -- (-1.3,-0.6) node[left]{$j{+}1$};
\draw (0,0) -- (-1.3,0.6) ;
\draw (-1.3,0.2)node{$\vdots$};
%
%\draw(0,0) -- (1.3,0);
\draw[fill=white] (0,0) circle(0.7cm);
\draw(0,0)node{$Y$};
\endscope
\scope[xshift=6.2cm,yshift=-1cm]
\draw (0,0) -- (-1.3,0.6) node[left]{$j$};
\draw (0,0) -- (-1.3,-0.6) node[left]{$2$} ;
\draw (0,0) -- (-0.8,-1.2) node[left]{$1$};
\draw (-1.3,0.2)node{$\vdots$};
%
%\draw (0,0) -- (1.3,0);
\draw[fill=white] (0,0) circle(0.7cm);
\draw(0,0)node{$X$};
\endscope
%11.5
\draw (8.1,0) -- (6.9,1);
\draw (8.1,0) -- (6.9,-1);
\draw(8.1,0) -- (9.3,0)node[right]{$\ldots$};
%%%%%%%%%%%%%%%%%%%%%%%%%%%%%%%%%%%%%%%%%%%%%%%%
%%%%%%%%%%%%%%%%%%%%%%%%%%%%%%%%%%%%%%%%%%%%%%%%
\scope[yshift=-4cm,xshift= - 3.5cm]
\scope[xshift=6.2cm,yshift=1cm]
\draw (0.2,0.5) -- (-0.6,1.7) node[left]{$p$};
\draw (0.2,0.5) -- (-1.1,0.1) node[left]{$l{+}1$};
\draw (0.2,0.5) -- (-1.1,1.1) ;
\draw (-1.1,0.7)node{$\vdots$};
%
%\draw(0,0) -- (1.3,0);
\draw[fill=white] (0.2,0.5) circle(0.5cm);
\draw(0.2,0.5)node{$Z$};
\endscope
\scope[xshift=6.2cm,yshift=-1cm]
\draw (0.2,-0.5) -- (-1.1,-0.1) node[left]{$j$};
\draw (0.2,-0.5) -- (-1.1,-1.1) node[left]{$2$} ;
\draw (0.2,-0.5) -- (-0.6,-1.7) node[left]{$1$};
\draw (-1.1,-0.45)node{$\vdots$};
%
%\draw (0,0) -- (1.3,0);
\draw[fill=white] (0.2,-0.5) circle(0.5cm);
\draw(0.2,-0.5)node{$X$};
\endscope
%11.5
\draw (8.35,0) -- (6.8,1.2);
\draw (8.35,0) -- (6.8,-1.2);
\draw(8.1,0) -- (9.3,0)node[right]{$\ldots$};
\draw(8.1,0) -- (6.5,0);
\draw (6.4,0) -- (5.1,0.6) node[left]{$l$};
\draw (5.1,0.15)node{$\vdots$};
\draw (6.4,0) -- (5.1,-0.6) node[left]{$j{+}1$};
\draw[fill=white] (6.4,0.0) circle(0.5cm);
\draw(6.4,0)node{$Y$};
\draw(3,0)node{$\displaystyle + \ \sum_{XYZ=P}$};
\endscope
\endtikzpicture
\caption{Berends--Giele currents $J^\mu_{12\ldots p}$ of rank $p$
combine the diagrams and propagators of a color-ordered 
$(p{+}1)$-point YM tree amplitude with an off-shell leg $\cdots$.
The sums in (\ref{pert1.0}) gather all combinations of cubic and 
quartic Feynman vertices that preserve the color order. Like this, $J^\mu_{12\ldots p}$ can be computed from 
quadratic contributions $\sim J^\nu_{12\ldots j}J^\la_{j+1\ldots p}$ with $j=1,2,\ldots,p{-}1$ and 
trilinear ones $\sim J^\nu_{12\ldots j} J^\la_{j+1\ldots l}J^\rho_{l+1\ldots p}$
with $1\leq j<l\leq p{-}1$.}
\label{f:numone}
\end{center}
\end{figure}

Moreover, the right-hand side of (\ref{pert1.0}) involves multiparticle momenta $k_P$ through 
Mandelstam invariants or inverse propagators $s_P$
\beq 
k_{P=12\ldots p}^\mu = k_1^\mu + k_2^\mu + \ldots + k_p^\mu
\ , \ \ \ \ \ \ 
s_P = \frac{1}{2} k_P^2  \, .
\label{pert1.mand}
\eeq
Finally, the brackets in (\ref{pert1.0A}) and (\ref{pert1.0B}) capture the cubic and quartic
Feynman vertices of pure YM theory in Lorenz gauge.
As depicted in figure \ref{f:numone}, the role of the deconcatenations $XY=P$ and $XYZ=P$ in
(\ref{pert1.0}) is to connect lower-rank currents $J_X^\mu, J_Y^\nu$ and $J^\lambda_Z$ via
Feynman vertices in all possible ways that preserve the color order of the on-shell legs
in the word $P=12 \ldots p$. 

Accordingly, color-ordered on-shell amplitudes at $n=p{+}1$ points are recovered by
taking the off-shell leg in the rank-$p$ current $J^\mu_{P}$ on shell: This on-shell limit
is implemented by contraction with the polarization vector $J^\mu_n=e^\mu_n$ of the last leg
and removing the propagator $s^{-1}_{12\ldots p}$ in the $p$-particle channel of $J^\mu_P$ which would
diverge by $n$-particle momentum conservation $k_{12\ldots p}^2\rightarrow (-k_n)^2 = 0$ 
\cite{Berends:1987me}\footnote{Here and in later equations of this work, we keep both instances 
of a contracted Lorentz index in
the uppercase position to avoid interference with the multiparticle labels of the currents. The signature
of the metric is still taken to be Minkowskian, regardless of the position of the indices.},
\beq
{\cal A}_{\rm YM}(1,2,\ldots,n{-}1,n) = s_{12\ldots n-1} J^\mu_{12\ldots n-1} J^\mu_n \, .\label{pert1.1}
\eeq
For instance, the rank-two current due to (\ref{pert1.0}) with $X=1$ and $Y=2$ yields the
following representation of the three-point amplitude
\begin{align}
s_{12} J_{12}^\mu &= (k_2\cdot e_1) e_2^\mu -  (k_1\cdot e_2) e_1^\mu + \frac{1}{2}(k_1^\mu - k_2^\mu)(e_1\cdot e_2)  \label{pert1.2}
\\
{\cal A}_{\rm YM}(1,2,3) &= s_{12} J^\mu_{12} J^\mu_3 =  (k_2\cdot e_1) (e_2\cdot e_3) - (k_1\cdot e_2) (e_1\cdot e_3)+ {1\over 2}(e_1\cdot e_2) e_3\cdot(k_1{-}k_2) \, ,
\notag
\end{align}
where cyclicity may be manifested via $e_3\cdot k_2 = - e_3\cdot k_1$ by 
means of on-shell constraints and momentum conservation. Note that Berends--Giele formulae
similar to (\ref{pert1.1}) have been given for tree amplitudes in ten-dimensional SYM \cite{Mafra:2010jq},
doubly-ordered amplitudes of bi-adjoint scalars \cite{Mafra:2016ltu} and worldsheet integrals for 
tree-level scattering of open strings \cite{Mafra:2016mcc}.

The symmetry properties $[J_X,J_Y]= - [J_Y,J_X]$ and
$\{J_X,J_Y,J_Z\} + {\rm cyc}(X,Y,Z)=0$ of the brackets in (\ref{pert1.0A}) and (\ref{pert1.0B})
imply that the currents in (\ref{pert1.0}) obey shuffle symmetry \cite{Berends:1988zn, Lee:2015upy}\footnote{The shuffle product 
$P \shuffle Q$ of words $P=p_1p_2\ldots p_{|P|}$ and $Q= q_1q_2\ldots q_{|Q|}$ is recursively defined by
\[
P\shuffle \emptyset= \emptyset \shuffle P = P \ , \ \ \ \ 
P \shuffle Q = p_1(p_2\ldots p_{|P|} \shuffle Q) + q_1(q_2\ldots q_{|Q|} \shuffle P) \, .
\]
All currents or amplitudes in this work are understood to obey a linearity property $J_{X+Y}^\mu = J_X^\mu + J_Y^\mu$ when 
formal sums of words appear in a subscript, e.g.\ $J^{\mu}_{1\shuffle 2}= J^{\mu}_{1 2+21}= J^{\mu}_{12}+J^{\mu}_{21}$ 
from $1\shuffle 2=12+21$.}
\beq
J^\mu_{P\shuffle Q} = 0 \ \forall \ P,Q\neq \emptyset \, .
\label{pert1.3cur}
\eeq
As pointed out in \cite{Mafra:2015vca}, the amplitude formula (\ref{pert1.1}) propagates the shuffle 
symmetry of the currents to the Kleiss--Kuijf (KK) relations \cite{Kleiss:1988ne, DelDuca:1999rs} 
\beq
{\cal A}_{\rm YM}((P\shuffle Q),n) = 0 \ \forall \ P,Q\neq \emptyset\, ,
\label{pert1.3}
\eeq
where the words $P$ and $Q$ involve external-state labels $1,2,\ldots,n{-}1$.
In the same way as shuffle symmetry (\ref{pert1.3cur}) leaves $(p{-}1)!$ independent
permutations of rank-$p$ currents $J^\mu_{12\ldots p}$, KK relations (\ref{pert1.3})
allow to expand color-ordered amplitudes in an $(n{-}2)!$-element set \cite{Kleiss:1988ne, DelDuca:1999rs},
\beq
J^\mu_{P1Q} = (-1)^{|P|} J^\mu_{1(\tilde P \shuffle Q)}
\, , \ \ \ \ \ \ 
{\cal A}_{\rm YM}(P,1,Q,n) = (-1)^{|P|} {\cal A}_{\rm YM}(1,(\tilde P \shuffle Q),n) \, ,
\label{pert1.3KK}
\eeq
where $\tilde P = p_{|P|}\ldots p_2 p_1$ denotes the reversal of the word $P=p_1p_2\ldots p_{|P|}$.

%%%%%%%%%%%%%%%%%%%%%%%%%%%%%%%%%%%%%%%%%%%%%%%%
%%%%%%%%%%%%%%%%%%%%%%%%%%%%%%%%%%%%%%%%%%%%%%%%

\subsection{Perturbiners as generating series of Berends--Giele currents}
\label{sec:2.2}

The Berends--Giele construction of the previous section can be related to solutions of the 
non-linear field equations: Generating series of Berends--Giele currents turn out
to solve the equations of motion from the action ${\cal S}_{{\rm YM}}$ of pure YM theory
\beq
{\cal S}_{{\rm YM}} = \frac{1}{4} \int \dd^D x \ \te{Tr}  (\FF_{\mu \nu} \FF^{\mu \nu} ) \ , \ \ \ \ \ \
\frac{ \delta {\cal S}_{{\rm YM}}  }{\delta \AAA_\la} = [\nabla_\mu , \FF^{\la \mu} ] \, .
\label{pert1.5}
\eeq
We use the following conventions in deriving the Lie-algebra valued gluon field $\AAA^\mu$ and its non-linear field strength $\FF^{\mu \nu}$ from a connection $\nabla_\mu$,
\beq
\nabla_\mu = \partial_\mu - \AAA_\mu \ , \ \ \ \ \ \ \FF_{\mu \nu} = - [\nabla_\mu,\nabla_\nu] 
= \partial_\mu \AAA_\nu - \partial_\nu \AAA_\mu - [\AAA_\mu,\AAA_\nu]  \, .
\label{pert1.4}
\eeq
The relation of tree-level amplitudes with solutions of the field equations via generating series
goes back to the ``perturbiner'' formalism \cite{Rosly:1996vr, Rosly:1997ap, Selivanov:1997aq, Selivanov:1999as, Bardeen:1995gk}.
In these references, generating series of MHV amplitudes are derived from self-dual
YM theory, see \cite{Selivanov:1998hn} for supersymmetric extensions. The connection 
between perturbiner solutions and the dimension-agnostic Berends--Giele currents of \cite{Berends:1987me} was 
established in \cite{Lee:2015upy, Mafra:2015vca} and will now be reviewed.

Lorenz gauge $\partial_\mu \AAA^\mu=0$ simplifies the
equations of motion $[\nabla_\mu , \FF^{\la \mu} ]=0$ to the wave equation
with the notation $\Box = \partial_\mu \partial^\mu$ for the d'Alembertian,
\begin{align}
\Box \AAA^\la &= [ \AAA^\mu, \partial_\mu \AAA^\la] + [\AAA_\mu , \FF^{\mu \la}]  \label{pert1.6}\\
&= 2 [ \AAA^\mu, \partial_\mu \AAA^\la] + [\partial^\la \AAA^\mu , \AAA_\mu] 
+ \big[ [\AAA^\mu, \AAA^\la], \AAA_\mu \big] \, .\notag
\end{align}
One can derive formal solutions to (\ref{pert1.6}) by means of the perturbiner ansatz 
\begin{align}
\AAA^\mu(x)&= \sum_i J_i^\mu t^{a_i} e^{k_i\cdot x} + \sum_{i,j} J_{ij}^\mu t^{a_i} t^{a_j} e^{k_{ij} \cdot x} + \sum_{i,j,l} J_{ijl}^\mu t^{a_i} t^{a_j} t^{a_l} e^{k_{ijl} \cdot x} + \ldots \notag \\
&= \sum_{P \neq \emptyset} J_P^\mu t^P e^{k_P\cdot x} \ , \ \ \ \ \te{where} \ t^{12\ldots p} = t^1 t^2 \ldots t^p \ .
\label{pert1.7}
\end{align}
The summation variables $i,j,l,\ldots=1,2,3,\ldots$ refer to external-particle labels in an unbounded range, and we 
have introduced a compact notation $\sum_{P\neq \emptyset}$ for sums over nonempty words $P=12\ldots p$
in passing to the second line.
The dependence on the spacetime coordinates $x^\mu$ enters through plane waves\footnote{The conventional
form of plane waves $e^{ik\cdot x}$ with an imaginary unit in the exponent can be recovered by redefining
the momenta in this work as $k\rightarrow ik$. The equations in the main text follow the conventions where external momenta are 
purely imaginary in order to keep factors of $i$ from proliferating.} $e^{k_P\cdot x}$, see (\ref{pert1.mand})
for the multiparticle momenta $k_P$. The color degrees of freedom in (\ref{pert1.7}) are represented 
through matrix products of the Lie-algebra generators $t^{a_i}$ whose adjoint indices $a_1,a_2,\ldots$
are associated with an unspecified gauge group. 

Upon insertion into the second line of (\ref{pert1.6}), the perturbiner ansatz (\ref{pert1.7}) can be verified
to solve the non-linear field equations $[\nabla_\mu , \FF^{\la \mu} ]=0$ if its coefficients $J^\mu_P$ obey
the Berends--Giele recursion (\ref{pert1.0}). Hence, generating series of Berends--Giele currents are formal
solutions to the field equations\footnote{Strictly speaking, contributions with several factors
of $t^{a_j}$ referring to the same external leg $j$ need to be manually suppressed by adding
nilpotent symbols to the perturbiner ansatz \cite{Rosly:1996vr}. For ease of notation, we do not include these symbols into
the equations in the main text, and all terms with repeated appearance of a given external leg are understood
to be suppressed.}. By the shuffle symmetry (\ref{pert1.3cur}) of the currents $J^\mu_P$, the 
matrix products $t^{a_i}t^{a_j}$ of the Lie-algebra generators on the right-hand side of (\ref{pert1.7}) 
conspire to nested commutators, and the perturbiner solution is guaranteed to be Lie-algebra valued \cite{Reeshuffle}.

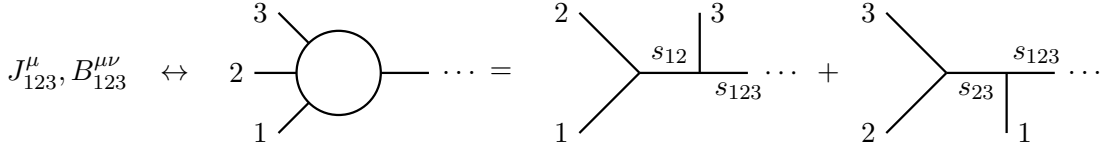
\begin{figure}[h]
\begin{center}
\begin{tikzpicture} [scale=0.8, line width=0.30mm]
\draw (-8,0) node{$J^{\mu}_{123}, B^{\mu \nu}_{123} \ \, \ \leftrightarrow  $};
\draw (-4.5,0.5) -- (-5,1) node[left]{$3$};
\draw (-4.7,0) -- (-5.4,0) node[left]{$2$};
\draw (-4.5,-0.5) -- (-5,-1) node[left]{$1$};
\draw (-4,0) circle(0.7cm);
%\draw(-4,0)node{YM};
\draw (-3.3,0)--(-2.5,0);
\draw (-2, 0) node{${\ldots} $};
\draw (-1.3,0) node{$=$};
\scope[xshift=1cm]
\draw (0,0) -- (-1,1) node[left]{$2$};
\draw (0,0) -- (-1,-1) node[left]{$1$};
\draw (0,0) -- (1.8,0);
\draw (0.5,0.3) node{$s_{12}$};
\draw (1,0) -- (1,1) node[right]{$3$};
\draw (1.65,-0.3) node{$s_{123}$};
\draw (2.7, 0) node{${\cdots} \  \ + $};
\scope[xshift=-0.4cm]
\draw (5.5,0) -- (4.5,1) node[left]{$3$};
\draw (5.5,0) -- (4.5,-1) node[left]{$2$};
\draw (5.5,0) -- (7.3,0);
\draw (6,-0.3) node{$s_{23}$};
\draw (6.5,0) -- (6.5,-1) node[right]{$1$};
\draw (7,0.3) node{$s_{123}$};
\draw (7.8, 0) node{${\ldots} $};
\endscope
\draw (5.5,1) node{};
\endscope
\end{tikzpicture}
\caption{By pairing up the two types of Berends--Giele currents $J^\mu_{12\ldots p}$ and $B^{\mu \nu}_{12\ldots p}$,
only cubic-vertex diagrams have to be considered in their recursive construction from lower-rank currents. In the depicted
example at rank $p=3$ with an additional off-shell leg $\cdots$, only two cubic diagrams of $s$-channel and $t$-channel
type contribute to the four-point amplitude obtained from $s_{123}J^\mu_{123} J^\mu_4$.}
\label{f:bg}
\end{center}
\end{figure}

As a convenient reorganization of the Berends--Giele recursion (\ref{pert1.0}), one can write the field equations as
in the first line of (\ref{pert1.6}) and insert a separate perturbiner expansion for
the non-linear field strength,
\beq
\FF^{\mu \nu}(x)= \sum_{P \neq \emptyset} B_P^{\mu \nu} t^P e^{k_P\cdot x}
\  \ \ \Rightarrow \ \ \
B^{\mu \nu}_P = k^\mu_P J^\nu_P - k^\nu_P J^\mu_P - \sum_{P=XY} (J_X^\mu J_Y^\nu - J_X^\nu J_Y^\mu ) \, .
\label{pert1.10}
\eeq
The expressions for the field-strength currents $B^{\mu \nu}_P$ in terms of $J_Q^\la$ are determined 
by the definition (\ref{pert1.4}) of $\FF^{\mu \nu}$, 
and their non-linear terms $\sum_{P=XY} J_X^{[\mu} J_Y^{\nu]}$ have already been
studied in \cite{Duhr:2006iq}. Then, inserting (\ref{pert1.7}) and (\ref{pert1.10}) into (\ref{pert1.6}) yields
a simpler but equivalent form of the recursion (\ref{pert1.0}) \cite{Mafra:2015vca}
\begin{align}
J_P^\mu = \frac{1}{2 s_P} \sum_{P=XY} \Big[ (k_Y \cdot J_X) J_Y^\mu + J_X^\nu B_Y^{\nu \mu} 
- (X\leftrightarrow Y) \Big]\, .
\label{pert1.8}
\end{align}
The trilinear term $\{J_X,J_Y,J_Z\}$ in (\ref{pert1.0B}) which represents the quartic vertex of the YM
Lagrangian has been absorbed into the non-linear part of the field-strength current $B^{\mu \nu}_P$ in (\ref{pert1.10}).
The leftover deconcatenations $P=XY$ in (\ref{pert1.8}) can be interpreted as describing cubic diagrams, see figure \ref{f:numone}.
Let us illustrate this statement with the four-point amplitude $s_{123} J^{\mu}_{123} J^\mu_4$ derived from a rank-three 
current via (\ref{pert1.1}): The two deconcatenations $(X,Y)=(12,3)$ and $(1,23)$ in the recursion (\ref{pert1.8}) 
for $J^{\mu}_{123}$ can be viewed as the two cubic diagrams in figure \ref{f:bg} where appropriate contributions 
from the quartic vertex (\ref{pert1.0B}) are automatically included.

Note that the Lorenz-gauge condition and the field equations imply the relations
\beq
k_P \cdot J_P=0 \, , \ \ \ \ \ \ k_P^\mu B_P^{\mu \nu} =\sum_{XY=P} (J_X^{\mu} B^{\mu \nu}_Y - J_Y^{\mu} B^{\mu \nu}_X )
\label{pert1.9}
\eeq
including transversality of the gluon polarizations for single-particle labels $P=i$. Moreover, the
non-linear gauge symmetry of the action (\ref{pert1.5}) under
 $\delta_{\Omega}\AAA^\mu = \partial^\mu \Omega - [\AAA^\mu,\Omega]$
and $\delta_{\Omega}\FF^{\mu \nu} =  - [\FF^{\mu \nu},\Omega]$ acts on the currents via
\beq
\delta_{\Omega} J_P^\mu = k_P^\mu \Omega_P - \sum_{XY=P} (J_X^\mu \Omega_Y - J_Y^\mu \Omega_X ) \, , 
\ \ \ \ \ \
\delta_{\Omega} B_P^{\mu \nu} =  - \sum_{XY=P} (B_X^{\mu \nu} \Omega_Y - B_Y^{\mu \nu} \Omega_X )
 \, .
\label{pert1.gauge}
\eeq
The scalar currents $\Omega_P$ are defined by the perturbiner expansion $\Omega(x)= \sum_{P\neq \emptyset} \Omega_P t^Pe^{k_P\cdot x}$ of the gauge scalar in $\delta_{\Omega}$. We will later on spell out a choice of gauge-scalar currents
$\Omega_P$ which manifests the BCJ duality at the level of Berends--Giele currents. 

Another specific choice of $\Omega_P\rightarrow \Omega^{\rm lin}_P$ allows to track the effect of linearized gauge 
transformations $e_i^\mu \rightarrow k^\mu_i$ on the $i^{\rm th}$ leg of the Berends--Giele currents in (\ref{pert1.gauge}):
One can line up the replacement $e_i^\mu \rightarrow k^\mu_i$ with a set of gauge transformations
that preserves Lorenz gauge. The condition $ \delta_{\Omega^{\rm lin}} (\partial_\mu \AAA^\mu)
=\partial_\mu( \delta_{\Omega^{\rm lin}}\AAA^\mu)=0$ then translates
into the recursion \cite{Lee:2015upy}
\beq
\Omega^{\rm lin}_P = \frac{1}{2s_{P}} \sum_{XY=P} ( (k_Y \cdot J_X) \Omega^{\rm lin}_Y - (k_X \cdot J_Y) \Omega^{\rm lin}_X )
\label{pert1.line}
\eeq
which needs to be supplemented with the initial conditions $\Omega^{\rm lin}_{j} \rightarrow \delta_{i,j}$ if the
linearized gauge transformations $e_i^\mu \rightarrow k^\mu_i$ only applies to the $i^{\rm th}$ leg. 
Precursors of the formula (\ref{pert1.line}) for linearized gauge transformations of Berends--Giele currents 
can be found in \cite{Berends:1988zn}. 

%%%%%%%%%%%%%%%%%%%%%%%%%%%%%%%%%%%%%%%%%%%%%%%%
%%%%%%%%%%%%%%%%%%%%%%%%%%%%%%%%%%%%%%%%%%%%%%%%

\subsection{Manifestly cyclic reformulation}
\label{sec:2.3}

Given that the Berends--Giele formula (\ref{pert1.1}) for color-ordered 
amplitudes ${\cal A}_{\rm YM}(1,2,\ldots,n)$ singles out the last leg $n$
which is excluded from the current $J_{12\ldots n-1}^{\mu}$, 
cyclic invariance in the external legs is obscured. We shall now review a reorganization 
of the Berends--Giele currents for YM tree amplitudes such that the $n^{\rm th}$ leg enters on completely symmetric footing. 
Moreover, the subsequent rewritings reduce $n$-point amplitudes to shorter Berends--Giele currents 
of rank $\leq \frac{n}{2}$ instead of the rank-$(n{-}2)$ currents in the recursion (\ref{pert1.0}) for $J_{12\ldots n-1}^{\mu}$.

\begin{figure}[h]
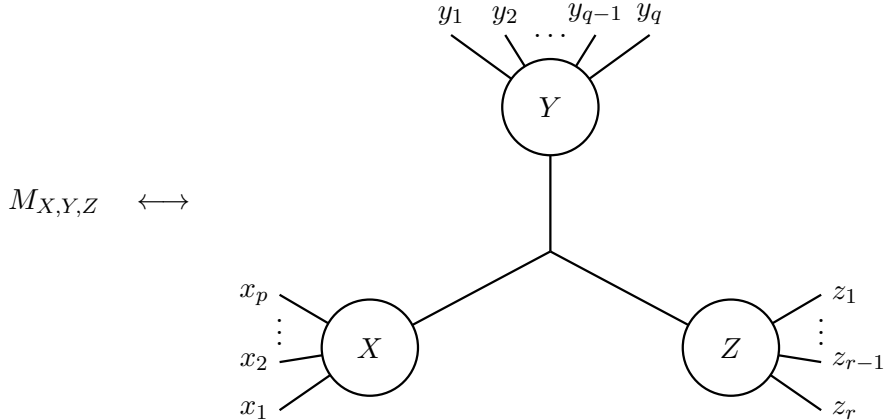

\begin{center}
\tikzpicture[xscale=0.8, yscale=0.64, line width=0.30mm]
\draw(-7.5,1)node{$M_{X,Y,Z}\ \ \ \longleftrightarrow$};
\draw(0,0) -- (0,3);
%
%\draw(0,3) -- (0,4.5);
\draw(0,3) -- (-0.75,4.5)node[above]{$y_2$};
\draw(0,3) -- (0.75,4.5)node[above]{$y_{q-1}$};
\draw(0,4.5)node{$\ldots$};
\draw(0,3) -- (-1.65,4.5)node[above]{$y_1$};
\draw(0,3) -- (1.65,4.5)node[above]{$y_q$};
\draw[fill=white](0,3) ellipse(0.8cm and 1cm);
\draw(0,3)node{$Y$};
\draw(0,0)--(3,-2);
\draw(4.5,-1.5)node{$\vdots$};
\draw(3,-2)--(4.5,-0.9)node[right]{$z_1$};
\draw(3,-2)--(4.5,-2.3)node[right]{$z_{r-1}$};
\draw(3,-2)--(4.5,-3.3)node[right]{$z_r$};
\draw[fill=white](3,-2) ellipse(0.8cm and 1cm);
\draw(3,-2)node{$Z$};
\draw(0,0)--(-3,-2);
\draw(-4.5,-1.5)node{$\vdots$};
\draw(-3,-2)--(-4.5,-0.9)node[left]{$x_p$};
\draw(-3,-2)--(-4.5,-2.3)node[left]{$x_2$};
\draw(-3,-2)--(-4.5,-3.3)node[left]{$x_1$};
\draw[fill=white](-3,-2) ellipse(0.8cm and 1cm);
\draw(-3,-2)node{$X$};
\endtikzpicture
\caption{Diagrammatic interpretation of the building block $M_{X,Y,Z}$ in (\ref{pert1.12})
with multiparticle labels $X=x_1 x_2\ldots x_p$, $Y=y_1y_2\ldots y_q$ and $Z=z_1 z_2\ldots z_r$.}
\label{f:MABC}
\end{center}
\end{figure}

The backbone of the manifestly cyclic Berends--Giele formulae is the building block \cite{Mafra:2015vca}
\beq
M_{X,Y,Z} = \frac{1}{2} \big( J_X^\mu B_Y^{\mu \nu} J_Z^\nu 
+J_Y^\mu B_Z^{\mu \nu} J_X^\nu +J_Z^\mu B_X^{\mu \nu} J_Y^\nu \big)
= \frac{1}{2}  J_X^\mu B_Y^{\mu \nu} J_Z^\nu  + {\rm cyc}(X,Y,Z)
\label{pert1.12}
\eeq
composed of three currents with multiparticle labels $X,Y,Z$ each of which represents tree-level
subdiagrams. The resulting diagrammatic interpretation of $M_{X,Y,Z}$ is depicted
in figure \ref{f:MABC}, and the definition (\ref{pert1.12}) along with $B_X^{\mu \nu}=-B_X^{\nu \mu}$
implies permutation antisymmetry $M_{X,Y,Z}=- M_{Y,X,Z}$ and $M_{X,Y,Z}=M_{Y,Z,X}$
expected from the cubic vertex in the figure.

Using $k_P\cdot J_P=0$ and $k_X+k_Y+k_Z=0$, it
was shown in \cite{Mafra:2015vca} that the $n$-point amplitude (\ref{pert1.1}) can be rewritten as
\beq
{\cal A}_{\rm YM}(1,2,\ldots,n{-}1,n) = \sum_{XY=12\ldots n-1} M_{X,Y,n} = \sum_{j=1}^{n-2} M_{12\ldots j,\, j{+}1\ldots n-1,\,n} \, .\label{pert1.13}
\eeq
As demonstrated in appendix \ref{app:A.2}, momentum conservation $k_P+k_Q=0$ and (\ref{pert1.9}) imply the following identity
\beq
\sum_{XY=P} M_{X,Y,Q} = \sum_{XY=Q} M_{P,X,Y}  \, , \label{pert1.14}
\eeq
which will be referred to as ``integration by parts''\footnote{This terminology goes back to the fact that
the building block (\ref{pert1.12}) and the amplitude representation (\ref{pert1.13}) descend from
ten-dimensional SYM \cite{Mafra:2010jq, Mafra:2015vca}: in the setup of these 
references, (\ref{pert1.14}) is a consequence of BRST integration by parts in 
pure-spinor superspace \cite{Berkovits:2000fe}.} and reads as follows in simple examples,
\begin{align}
M_{12,3,4} &= M_{1,2,34} \ , \ \ \ \ \ \ M_{123,4,5} = M_{12,3,45} + M_{1,23,45}\notag\\
M_{1234,5,6} &= M_{123,4,56} + M_{12,34,56} + M_{1,234,56}    \label{pert1.15} \\
M_{123,45,6} &+M_{123,4,56} = M_{12,3,456} +M_{1,23,456} \, . \notag
\end{align}
By repeated application to the amplitude representation (\ref{pert1.13}), one can
derive the following manifestly cyclic representations
\begin{align}
{\cal A}_{\rm YM}(1,2,3,4) &= \frac{1}{2} M_{12,3,4} + {\rm cyc}(1,2,3,4) \notag \\
{\cal A}_{\rm YM}(1,2,\ldots,5) &= M_{12,3,45} + {\rm cyc}(1,2,3,4,5) \label{pert1.16} \\
{\cal A}_{\rm YM}(1,2,\ldots,6) &= \frac{1}{3}M_{12,34,56} + \frac{1}{2}(M_{123,45,6} + M_{123,4,56}) + {\rm cyc}(1,2,\ldots,6) \notag \\
{\cal A}_{\rm YM}(1,2,\ldots,7) &= M_{123,45,67} + M_{1,234,567} + {\rm cyc}(1,2,\ldots,7) \, .\notag
\end{align}
Note in particular that the rank of the currents in the manifestly cyclic $n$-point 
amplitudes (\ref{pert1.16}) is bounded by\footnote{Earlier examples of such economic and manifestly 
cyclic Berends--Giele representations have been investigated 
in \cite{Berends:1989hf}, but the construction in the reference requires a mixture 
of quadratic, cubic and quartic combinations of Berends--Giele currents instead of a single 
building block (\ref{pert1.12}).} $\lfloor \frac{ n}{2} \rfloor$ 
rather than $n{-}2$ as expected from the recursions (\ref{pert1.0}) or (\ref{pert1.8}) for $J^\mu_{12\ldots n-1}$. In section \ref{sec:3}, 
similar expressions with manifest cyclicity and Berends--Giele currents of maximum rank $\lfloor \frac{ n}{2} \rfloor$
will be given for the deformed (\YMF) theory.

%%%%%%%%%%%%%%%%%%%%%%%%%%%%%%%%%%%%%%%%%%%%%%%%
%%%%%%%%%%%%%%%%%%%%%%%%%%%%%%%%%%%%%%%%%%%%%%%%

\subsection{BCJ duality}
\label{sec:2.4}

The organization of the Berends--Giele recursion (\ref{pert1.8}) in terms of cubic-vertex
diagrams as exemplified in figure \ref{f:bg} resonates with the BCJ duality between color and kinematics \cite{BCJ}:
According to the BCJ duality, scattering amplitudes in non-abelian gauge theories can be represented in 
a manner such that color degrees of freedom can be freely interchanged with the kinematic variables.
While ``color'' refers to contractions of structure constants $f^{a_i a_i a_k}$, polarizations and momenta
are referred to as ``kinematics'', and the notion of ``freely interchanging'' will be shortly made precise.
The three-index structure of the contracted structure constants can be visualized 
via cubic-vertex diagrams with a factor of $f^{a_i a_i a_k}$ for each vertex and contractions of the adjoint
indices along the internal edges. Similarly, the kinematic dependence 
on $e_i^\mu, k_i^\mu$ should also be organized in terms of cubic diagrams to manifest the BCJ duality. 

The non-linear extension $\sum_{XY=P} J^{[\mu}_X J^{\nu]}_Y$ of the field-strength current $B_P^{\mu \nu}$ 
in (\ref{pert1.10}) absorbs the contributions from the quartic vertex ${\rm Tr} [\AAA_\mu,\AAA_\nu][\AAA^\mu, \AAA^\nu]$ 
in the YM action (\ref{pert1.5}). This can be seen from that fact that the non-linear terms have fewer 
propagators than the rest of (\ref{pert1.8}). Hence, the use of field-strength currents
amounts to inserting $1=\frac{k_P^2}{k_P^2}$ such that a quartic vertex is ``pulled apart'' into two
cubic vertices connected by the ``fake'' propagator $k_P^2$. The choice of the channel 
$P$ in $1=\frac{k_P^2}{k_P^2}$ has to be compatible with the color dressing $ f^{ab e} f^{ecd}$ 
of the quartic vertex, where ambiguities arise from the Jacobi relations
\beq
f^{ab e} f^{ecd} + f^{ac e} f^{edb} + f^{ad e} f^{ebc} = 0\, .
\label{col:jac}
\eeq

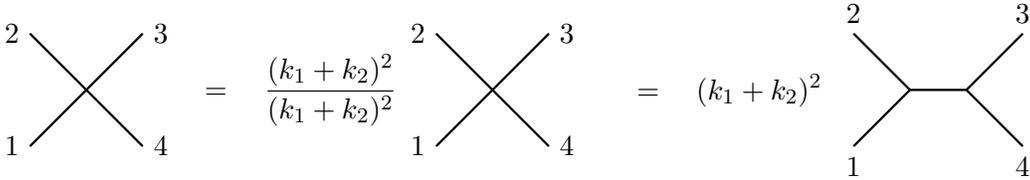
\begin{figure}[h]
\begin{center}
\begin{tikzpicture}[scale=1.5]
\scope[xshift=-3.3cm]
\draw [line width=0.30mm]  (2,0.5) -- (1.5,1) node[above,left]{$2$};
\draw [line width=0.30mm]  (2,0.5) -- (1.5,0) node[below,left]{$1$};
\draw [line width=0.30mm]  (2,0.5) -- (2.5,1) node[above,right]{$3$};
\draw [line width=0.30mm]  (2,0.5) -- (2.5,0) node[below,right]{$4$};
\endscope
\scope[xshift=0.3cm]
\draw  (0.3,0.5) node{$\displaystyle = \ \  \ \frac{(k_1+k_2)^2}{(k_1+k_2)^2}$};
\draw [line width=0.30mm]  (2,0.5) -- (1.5,1) node[above,left]{$2$};
\draw [line width=0.30mm]  (2,0.5) -- (1.5,0) node[below,left]{$1$};
\draw [line width=0.30mm]  (2,0.5) -- (2.5,1) node[above,right]{$3$};
\draw [line width=0.30mm]  (2,0.5) -- (2.5,0) node[below,right]{$4$};
\endscope
\scope[xshift=4.1cm]
\draw  (0.3,0.5) node{$\displaystyle = \ \ \ (k_1+k_2)^2 $};
\scope[xshift=-0.1cm]
\draw [line width=0.30mm]  (2,0.5) -- (1.5,1) node[above]{$2$};
\draw [line width=0.30mm]  (2,0.5) -- (1.5,0) node[below]{$1$};
\draw [line width=0.30mm]  (2,0.5) -- (2.5,0.5) ;
\draw [line width=0.30mm]  (2.5,0.5) -- (3,1) node[above]{$3$};
\draw [line width=0.30mm]  (2.5,0.5) -- (3,0) node[below]{$4$};
\endscope
\endscope
\end{tikzpicture}
%\end{center}
\caption{Quartic vertices can always be reorganized in products of cubic vertices, i.e.\ gauge-theory
amplitudes can always be parametrized in terms.}
\label{f:insert}
\end{center}
\end{figure}

In figure \ref{f:insert}, this situation is visualized in a four-point tree-level context, but there is no 
limitation to cubic-diagram parametrizations of $n$-point tree amplitudes as well as multiloop 
integrands \cite{loopBCJ, Bern:2017yxu}.  Although the BCJ duality conjecturally applies to 
loop integrands \cite{loopBCJ, Bern:2017yxu}, 
we shall focus on its well-established tree-level incarnation. 

Of course, contributions from the higher-order vertices of $({\rm Tr}\,F^3)$- and $({\rm Tr}\,F^4)$-type
can also be cast into a cubic-graph form by repeated insertions of $1=\frac{k_P^2}{k_P^2}$. For the action
(\ref{pert0.1}) of (\YMF), the color structure of the $F^3$ and $F^4$
operators also boils down to contracted structure constants \cite{Broedel:2012rc}, and the ambiguities due to 
Jacobi identities (\ref{col:jac}) arise in this situation as well. In the subsequent review of the BCJ duality,
the color-dressed tree-level amplitudes
\beq
{\cal M}_n = \sum_{\rho \in S_{n-1}} {\rm Tr}(t^{a_1} t^{a_{\rho(2)}}  t^{a_{\rho(3)}} \ldots  t^{a_{\rho(n)}} ){\cal A}(1,\rho(2),\rho(3),\ldots,\rho(n))
\label{genpara1}
\eeq
may refer to pure YM (${\cal A} \rightarrow {\cal A}_{\rm YM}$), to its $(\ap F^3+\ap^2 F^4)$-deformation
(${\cal A} \rightarrow {\cal A}_{{\rm YM}+F^3+F^4}$) or to any other generalization that obeys the BCJ duality.
Once the kinematic dependence of (\ref{genpara1}) is absorbed into cubic diagrams $I,J,K,\ldots$, one can 
choose a parametrization \cite{BCJ}
\beq
{\cal M}_n =  \sum_{I \in \Gamma_n} \frac{ C_I \, N_I }{\prod_{e \in \te{internal} \atop{\te{edges of} } \ I} s_e} \, ,
\label{genpar2}
\eeq
where $\Gamma_n$ denotes the set of cubic tree-level graphs with $n$ external legs. The color factors 
$C_I$ represent the contracted structure constants that arise from the
traces in (\ref{genpara1}). The kinematic numerators $N_I$ are combinations of
$e_i^\mu$ and $k_i^\mu$ that can be assembled from the Berends--Giele currents of the theory. 
Finally, the propagators $s_e^{-1}$ comprise Mandelstam variables (\ref{pert1.mand}) for the multiparticle
momenta in the internal edges $e$ of the graph $I$.

\begin{center}
\begin{figure}[h]
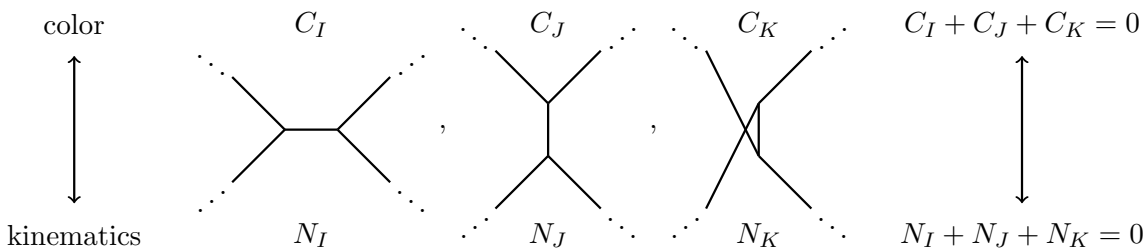

 \tikzpicture [scale=1.4]
 \draw (-2.5,-1) node {kinematics};
 \draw (-2.5,1) node {color};
 \draw [line width=0.30mm,<->] (-2.5,0.7) -- (-2.5,-0.7);
 \draw (6.5,-1) node {$N_I + N_J + N_K= 0$};
 \draw (6.5,1) node {$C_I + C_J + C_K = 0$};
 \draw [line width=0.30mm,<->] (6.5,0.7) -- (6.5,-0.7);
 % S
 \scope[yshift=-0.5cm, xshift=-2.5cm]
 \draw [line width=0.30mm]  (2,0.5) -- (1.5,1) ;
 \draw (1.3,1.2) node {$\ddots$};
 \draw [line width=0.30mm]  (2,0.5) -- (1.5,0) ;
 \draw (1.3,-0.1) node {$\iddots$};
 \draw [line width=0.30mm]  (2,0.5) -- (2.5,0.5) ;
% \draw (2.25,0.7) node {$k_{\alpha_i}$};
 \draw [line width=0.30mm]  (2.5,0.5) -- (3,1) ;
 \draw (3.2,1.2) node {$\iddots$};
 \draw [line width=0.30mm]  (2.5,0.5) -- (3,0) ;
 \draw (3.2,-0.1) node {$\ddots$};
 \draw (2.25,-0.5) node {$\displaystyle   N_I$}; 
 \draw (2.25,1.5) node {$\displaystyle   C_I$}; 
 \endscope
 % T
 \scope[xshift=-1.5cm]
 \draw (4.5,0) node{$,$};
 \draw [line width=0.30mm]  (5.5,-0.25) -- (5,0.75) ;
 \draw (4.8,0.95) node {$\ddots$};
 \draw [line width=0.30mm]  (5.5,0.25) -- (5,-0.75) ;
 \draw (4.8,-0.85) node {$\iddots$};
 \draw [line width=0.30mm]  (5.5,0.25) -- (5.5,-0.25) ;
%  \draw (5.8,0) node {$k_{\alpha_j}$};
 \draw [line width=0.30mm]  (5.5,0.25) -- (6,0.75) ;
 \draw (6.2,0.95) node {$\iddots$};
 \draw [line width=0.30mm]  (5.5,-0.25) -- (6,-0.75);
 \draw (6.2,-0.85) node {$\ddots$};
 \draw (5.5,-1) node {$\displaystyle  N_K$}; 
\draw (5.5,1) node {$\displaystyle  C_K$}; 
 \endscope
 % U
 \scope[xshift=-3.5cm]
 \draw (4.5,0) node{$,$};
 \draw [line width=0.30mm]  (5.5,0.25) -- (5,0.75) ;
 \draw (4.8,0.95) node {$\ddots$};
 \draw [line width=0.30mm]  (5.5,-0.25) -- (5,-0.75) ;
 \draw (4.8,-0.85) node {$\iddots$};
 \draw [line width=0.30mm]  (5.5,0.25) -- (5.5,-0.25) ;
% \draw (5.2,0) node {$k_{\alpha_k}$};
 \draw [line width=0.30mm]  (5.5,0.25) -- (6,0.75) ;
 \draw (6.2,0.95) node {$\iddots$};
 \draw [line width=0.30mm]  (5.5,-0.25) -- (6,-0.75);
 \draw (6.2,-0.85) node {$\ddots$};
 \draw (5.5,-1) node {$\displaystyle  N_J$}; 
 \draw (5.5,1) node {$\displaystyle  C_J$}; 
 \endscope
 \endtikzpicture
\caption{Triplets of cubic graphs $I,J,K$ whose color factors $C_{\cdot}$ and kinematic factors $N_{\cdot}$ are
both related by a Jacobi identity if the duality between color and kinematics is manifest.
The dotted lines at the corners represent arbitrary tree-level subdiagrams and are understood
to be the same for all of the three cubic graphs.}
\label{colkin}
\end{figure}
\end{center}

The parametrization (\ref{genpar2}) is said to manifest the BCJ duality if all the symmetries
of the color factors $C_I$ carry over to the kinematic numerators $N_I$. More specifically \cite{BCJ}:
\begin{itemize}
\item If two graphs $I$ and $\widehat I$ are related by a single flip of a cubic vertex, antisymmetry 
$f^{a_i a_j a_k}=f^{[a_i a_j a_k]}$ implies the color factors to have a relative minus sign.
In a duality-satisfying representation (\ref{genpar2}), the kinematic numerators exhibit the same 
antisymmetry properties under flips:
\beq
 C_{\widehat I} = - C_I \ \ \ \Longrightarrow \ \ \ N_{\widehat I} = - N_I \ .
\label{dualityA}
\eeq
\item For each triplet of graphs $I,J,K$ where the Jacobi identities (\ref{col:jac}) lead to the
vanishing of triplets $C_I+C_J+C_K$, the BCJ duality requires the corresponding triplet
of kinematic numerators to vanish as well
\beq
C_I+C_J+C_K=0 \ \ \ \Longrightarrow \ \ \ N_I+N_J+N_K=0 \ .
\label{dualityB}
\eeq
As visualized in figure \ref{colkin}, such triplets of cubic graphs only differ by a single propagator.
\end{itemize}

In later sections, we will construct local representatives of the kinematic numerators $N_I$ in (\ref{genpar2}) 
of (\YMF) which do not exhibit any poles in $s_P$ and obey the BCJ duality up to and
including the order of $\ap^2$. By the Jacobi identities (\ref{col:jac}) of the color factors, the numerators 
are still far from unique after imposing locality, and generic choices at $n\geq 5$ points will fail to obey
some of the kinematic Jacobi relations (\ref{dualityB}). Hence, finding a manifestly color-kinematics dual
parametrization (\ref{genpar2}) requires some systematics in addressing quartic and higher-order vertices
via $1= \frac{ k_P^2 }{k_P^2}$. The additional requirement of locality is particularly restrictive, and we will
see that suitable gauge transformations (\ref{pert1.gauge}) of the Berends--Giele currents in 
(\YMF) give rise to local solutions, generalizing the construction in
ten-dimensional SYM \cite{Mafra:2015vca}.

Still, the very existence of duality satisfying kinematic numerators is sufficient to
derive BCJ relations among color-ordered amplitudes \cite{BCJ}
\beq
\sum_{j=2}^{n-1} (k_{23\ldots j} \cdot k_1){\cal A}(2,3,\ldots,j,1,j{+}1,\ldots,n) = 0 \, .
\label{BCJrels}
\eeq
By combining different relabellings of (\ref{BCJrels}), any color-ordered amplitude can be
expanded in a basis of size $(n{-}3)!$. BCJ relations were shown to apply to ${\cal A} \rightarrow {\cal A}_{{\rm YM}+F^3+F^4}$
up to and including the order of $\ap^2$ \cite{Broedel:2012rc} by isolating suitable terms in the monodromy relations
of open-string tree-level amplitudes \cite{BjerrumBohr:2009rd, Stieberger:2009hq}. For a variety
of four-dimensional helicity configurations, kinematic numerators of ${\rm YM}+F^3$ subject to Jacobi relations (\ref{dualityB})
can be found in \cite{Broedel:2012rc}. We will derive generalizations to helicity-agnostic expressions in $D$ dimensions 
and include the $\ap^2$ order of (\YMF).

\begin{center}
\begin{figure}[h]
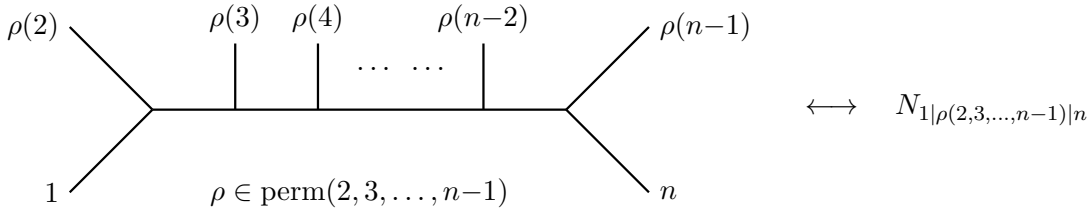

 \tikzpicture [scale=1.1, line width = 0.3mm]
 \draw(0,0) --(-1,-1) node[left]{$1$};
 \draw(0,0) --(-1,1) node[left]{$\rho(2)$};
 \draw(0,0) -- (5,0);
 \draw(2.5,-1)node{$\rho \in {\rm perm}(2,3,\ldots,n{-}1)$};
 \draw(1,0) -- (1,0.8) ;
  \draw(2,0) -- (2,0.8) ;
 \draw (2.67,0.5)node{$\ldots$};
  \draw (3.33,0.5)node{$\ldots$};
  \draw(4,0) -- (4,0.8) ;
  \draw (1,1.1)node {$\rho(3)$};
  \draw (2,1.1)node {$\rho(4)$};
  \draw (4,1.1)node {$\rho(n{-}2)$};
 \draw(5,0) --(6,-1) node[right]{$n$};
 \draw(5,0) --(6,1) node[right]{$\rho(n{-}1)$};
 %\draw(12.4,0)node{$\longleftrightarrow  \ \langle V_{12\ldots j} V_{n-1,n-2\ldots j+1} V_n \rangle $};
% \draw[<->] (7.5,0) -- (9,0);
 \draw(9.6,0)node{$\longleftrightarrow \ \ \ N_{1|\rho(2,3,\ldots,n{-}1)|n}$};
 \endtikzpicture
\caption{When the BCJ duality is manifest, the master numerators $N_{1|\rho(2,3,\ldots,n{-}1)|n}$ associated with the 
depicted $(n{-}2)!$-family of half-ladder diagrams generate all other kinematic numerators via 
Jacobi relations.}
\label{halfladders}
\end{figure}
\end{center}

Kinematic antisymmetry relations (\ref{dualityA}) and Jacobi identities (\ref{dualityB}) 
leave $(n{-}2)!$ independent instances of $N_I$. A basis of kinematic numerators under these relations can
be assembled from the ``half-ladder'' diagrams depicted in figure \ref{halfladders} which are characterized
by a fixed choice of endpoints $1$ and $n$ as well as permutations $\rho \in S_{n-2}$ of the remaining
legs $2,3,\ldots,n{-}1$. We will denote the basis numerators of the half-ladder diagrams in figure \ref{halfladders} 
by $N_{1|\rho(2,3,\ldots,n{-}1)|n}$ and refer to them as ``master numerators''.

%%%%%%%%%%%%%%%%%%%%%%%%%%%%%%%%%%%%%%%%%%%%%%%%
%%%%%%%%%%%%%%%%%%%%%%%%%%%%%%%%%%%%%%%%%%%%%%%%

\subsection{Double copy}
\label{sec:2.5}

The BCJ duality allows to convert cubic-graph parametrizations
(\ref{genpar2}) of gauge-theory amplitudes into gravitational ones:
Once the gauge-theory numerators $N_I$ satisfy the same symmetry properties
as the color factors $C_I$ (i.e.\ flip antisymmetry (\ref{dualityA}) 
and kinematic Jacobi identities (\ref{dualityB})), then the 
double-copy formula
\beq
{\cal M}^{\rm grav}_n =  \sum_{I \in \Gamma_n} \frac{  N_I  \, \tilde N_I}{\prod_{e \in \te{internal} \atop{\te{edges of} } \ I} s_e} 
\label{genpar4}
\eeq
enjoys linearized-diffeomorphism invariance. In case of undeformed YM theory,
(\ref{genpar4}) yields tree-level amplitudes of Einstein-gravity including $B$-fields, dilatons
and tentative supersymmetry partners \cite{BCJ, Bern:2010yg}. The polarizations of external
gravitons, $B$-fields or dilatons in the $j^{\rm th}$ leg are obtained by projecting the tensor 
products $e_j^\mu \tilde e_j^\nu$ in (\ref{genpar4}) to the suitable irreducible representation of the Lorentz group.

In case of (\YMF)-numerators,
the gravitational amplitudes descend from a deformation of the Einstein--Hilbert action by higher-curvature
operators of $\alpha'R^2 + \alpha'^2 R^3$ \cite{Broedel:2012rc} as seen in the low-energy effective action of the closed
bosonic string \cite{Metsaev:1986yb}, see section \ref{sec:5.4} for details. The tilde along with the second copy 
$\tilde N_I$ of the gauge-theory numerator $N_I$ indicates that the $i^{\rm th}$ external gravitational state may 
arise from the tensor product of different polarization vectors $e_i^\mu$ and $\tilde e^\mu_i$.

In the same way as (\ref{genpar4}) is obtained from gauge-theory amplitudes (\ref{genpar2}) by trading
color for kinematics, $C_I \rightarrow \tilde N_I$, one can investigate the converse replacement $N_I \rightarrow \tilde C_I$:
\beq
{\cal M}^{\rm \phi^3}_n =  \sum_{I \in \Gamma_n} \frac{ C_I \, \tilde C_I }{\prod_{e \in \te{internal} \atop{\te{edges of} } \ I} s_e} \, .
\label{genpar5}
\eeq
This double copy of color factors (with $\tilde C_I$ comprising structure constants $\tilde f^{\tilde b_i \tilde b_j \tilde b_k}$ 
of possibly different Lie algebra generators $\tilde t^{\tilde b}$) describes tree amplitudes of biadjoint 
scalars $\phi = \phi_{a|\tilde b} t^a \otimes \tilde t^{\tilde b}$ with a cubic interaction $f^{a_1a_2a_3}\tilde f^{\tilde b_1\tilde b_2
\tilde b_3}\phi_{a_1|\tilde b_1}\phi_{a_2|\tilde b_2}\phi_{a_3 |\tilde b_3} $ \cite{Cachazo:2013iea}. The two 
species $t^a$ and $\tilde t^{\tilde b}$ admit a two-fold color decomposition (\ref{genpara1}),
and we define its doubly-partial amplitudes $m(\cdot|\cdot)$ by peeling off two traces with possibly 
different orderings $\rho,\tau \in S_{n-1}$ \cite{Cachazo:2013iea},
\beq
m(1,\rho(2,\ldots,n)|1, \tau(2,\ldots,n)) = {\cal M}^{\rm \phi^3}_n \, \big|_{{\rm Tr}(t^{a_1} t^{a_{\rho(2)}} \ldots t^{a_{\rho(n)}})
{\rm Tr}(\tilde t^{\tilde b_1}  \tilde t^{\tilde b_{\tau(2)}} \ldots  \tilde t^{\tilde b_{\tau(n)}}) } \, .
\label{genpar6}
\eeq
Doubly-partial amplitudes compactly encode a solution to all the kinematic Jacobi relations: 
When reducing the gravitational amplitude (\ref{genpar4}) to the master numerators introduced 
in figure \ref{halfladders}, the coefficients are analogous $(n{-}2)! \times (n{-}2)!$ families of (\ref{genpar6})
\beq
{\cal M}_n^{\rm grav} = \! \!  \sum_{\rho,\tau \in S_{n-2}} \! \! 
N_{1|\rho(2,\ldots,n{-}1)|n} m(1, \rho(2,\ldots,n{-}1), n | 1, \tau(2,\ldots,n{-}1), n)
\tilde N_{1|\tau(2,\ldots,n{-}1)|n} \, .
 \label{genpar8}
\eeq
The gauge-theory analogue (with $C_{1|\rho(2,\ldots,n{-}1)|n} $ referring to the half-ladder diagrams as in figure \ref{halfladders}) 
\beq
{\cal M}_n = \! \! 
 \sum_{\rho,\tau \in S_{n-2}} \! \! 
C_{1|\rho(2,\ldots,n{-}1)|n} m(1, \rho(2,\ldots,n{-}1), n | 1, \tau(2,\ldots,n{-}1), n)
N_{1|\tau(2,\ldots,n{-}1)|n}
\label{fromhere}
\eeq
is equivalent to expansions of color-ordered
amplitudes in terms of master numerators \cite{Mafra:2011kj, Cachazo:2013iea}
\beq
{\cal A}(\rho(1,2,\ldots,n)) = \sum_{\tau \in S_{n-2}} m(\rho(1,2,\ldots,n) | 1, \tau(2,\ldots,n{-}1), n) \, N_{1|\tau(2,\ldots,n{-}1)|n} \, .
\label{genpar7}
\eeq
Representations of the form in (\ref{genpar8}) to (\ref{genpar7}) arise naturally from the $(\alpha'\rightarrow 0)$-limit of 
string-theory amplitudes \cite{Mafra:2011kj, Zfunctions, Stieberger:2014hba, Mafra:2016ltu} and the CHY 
formalism \cite{Cachazo:2013iea}.

By comparing the representations (\ref{genpar2}) and (\ref{fromhere}) of color-dressed gauge-theory
amplitudes, we conclude that the Jacobi relations among the cubic diagrams in figure \ref{colkin}
can be traced back to the properties of the doubly-partial amplitudes. By the symmetric role of $N_{\cdot}$ 
and $C_{\cdot}$ in (\ref{fromhere}), this applies to the Jacobi relations of both color factors and kinematic numerators.

Doubly-partial amplitudes obey BCJ relations (\ref{BCJrels}) in both of their entries and admit
bases of $(n{-}3)! \times (n{-}3)!$ elements \cite{Cachazo:2013iea}. The matrix inverse of such 
a basis appears in the more traditional formulation of the gravitational double copy at tree level:
The $(\alpha'\rightarrow 0)$ limit of the string-theory KLT relations \cite{Kawai:1985xq} yields the following manifestly 
diffeomorphism invariant rewriting of (\ref{genpar4}),
\beq
{\cal M}_n^{\rm grav} = \! \!  \sum_{\rho,\tau \in S_{n-3}} \! \! 
{\cal A}(1,\rho(2,\ldots,n{-}2),n{-}1,n)  S(\rho|\tau)_1
\tilde {\cal A}(1,\tau(2,\ldots,n{-}2),n,n{-}1) \, .
 \label{genpar9}
\eeq
The all-multiplicity form of the $(n{-}3)! \times (n{-}3)!$ KLT-matrix $S(\rho|\tau)_1$ has been studied in
\cite{Bern:1998sv, momentumKernel} and furnishes the inverse of doubly-partial 
amplitudes (\ref{genpar6}) \cite{Cachazo:2013iea},
\beq
S(\rho|\tau)_1 = - m^{-1}(1, \rho(2,\ldots,n{-}2), n{-}1,n |1, \tau(2,\ldots,n{-}2), n, n{-}1)  \, .
\label{genpar21}
\eeq
Alternatively, one can obtain the KLT matrix from the recursion \cite{momentumKernel, Carrasco:2016ldy}
\beq
S(2|2)_1 = k_1\cdot k_2 \, , \ \ \ \ \ \ S(A,j | B,j,C)_1 = k_j\cdot (k_1+k_B) S(A |B,C)_1  \, .
\label{genpar22}
\eeq
The subscript $_1$ indicates that the entries of (\ref{genpar21}) not only depend on the
momenta $k_2,\ldots,k_{n-2}$ subject to permutations $\rho,\tau$ but also on $k_1$.

Similarly, the doubly-partial amplitudes (\ref{genpar6}) can be generated from a Berends--Giele
formula analogous to (\ref{pert1.1}) \cite{Mafra:2016ltu}
\beq
m(P,n|Q,n) = s_P \phi_{P|Q}  \, ,
\label{genpar23}
\eeq
where $P$ and $Q$ are permutations of legs $1,2,\ldots,n{-}1$, and the doubly-ordered 
currents $\phi_{P|Q}$ obey the following recursion \cite{Mafra:2016ltu}:
\beq
\phi_{i|j} = \delta_{ij} \, , \ \ \ \ \ \
s_P\phi_{P|Q} = \sum_{XY=P} \sum_{AB=Q} ( \phi_{X|A} \phi_{Y|B} -  \phi_{Y|A} \phi_{X|B} )\, .
\label{genpar24}
\eeq
We will often gather rank-$r$ currents (\ref{genpar24}) in the following $(r{-}1)!\times (r{-}1)!$ 
matrix
\beq
\Phi(P|Q)_1 =  \phi_{1P|1Q} = S^{-1}(P|Q)_1 \, .
\label{genpar25}
\eeq
Examples for the output of the recursions (\ref{genpar22}) and (\ref{genpar24}) include
$\Phi(2|2)_1=s_{12}^{-1}$ and
\begin{align}
S(\rho(2,3)|\tau(2,3))_1 &= \Big( \begin{smallmatrix}
s_{12}(s_{13}{+}s_{23}) & s_{12} s_{13} \\
s_{12}s_{13} &s_{13}(s_{12}{+}s_{23}) 
\end{smallmatrix} \Big) \label{genpar26}
 \\
\Phi(\rho(2,3)|\tau(2,3))_1 &= \frac{1}{s_{123}}\Big( \begin{smallmatrix}
s_{12}^{-1}{+}s^{-1}_{23} &-s^{-1}_{23} \\
-s^{-1}_{23} &s_{13}^{-1}{+}s^{-1}_{23} 
\end{smallmatrix} \Big)
 \, .
\notag
\end{align}
We will later on use the matrices $S(P|Q)_1$ and $\Phi(P|Q)_1$ to relate shuffle
independent Berends--Giele currents to kinematic numerators subject to Jacobi identities.

%%%%%%%%%%%%%%%%%%%%%%%%%%%%%%%%%%%%%%%%%%%%%%%%
%%%%%%%%%%%%%%%%%%%%%%%%%%%%%%%%%%%%%%%%%%%%%%%%
%%%%%%%%%%%%%%%%%%%%%%%%%%%%%%%%%%%%%%%%%%%%%%%%
%%%%%%%%%%%%%%%%%%%%%%%%%%%%%%%%%%%%%%%%%%%%%%%%
%%%%%%%%%%%%%%%%%%%%%%%%%%%%%%%%%%%%%%%%%%%%%%%%

\section{Perturbiners and Berends--Giele representations for $F^3$ and $F^4$}
\label{sec:3}

In this section, we apply the Berends--Giele methods of sections \ref{sec:2.1} to \ref{sec:2.3}
to the deformed (\YMF) theory known from the low-energy regime of
open bosonic strings. The tree-level amplitudes following from the action
\begin{align}
{\cal S}_{{\rm YM}+F^3+F^4} &= \int \dd^D x \ \te{Tr} \Big\{ \, \frac{1}{4}\, \FF_{\mu \nu} \, \FF^{\mu \nu} + \frac{ 2\ap}{3} \,\FF_\mu{}^\nu \, \FF_\nu{}^\la \,\FF_{\la}{}^\mu + \frac{ \ap^2}{4}  \,  [\FF_{\mu \nu}, \FF_{\la \rho}]  [\FF^{\mu \nu}, \FF^{\la \rho}] \, \Big\} 
\label{pert0.1recap}
\end{align}
reproduce the leading orders $\ap^0, \ap^1$ in the low-energy expansion of bosonic-string amplitudes \cite{Polchinski:1998rq},
and a well-defined sector of the $\ap^2$ order: The effective action of both bosonic and supersymmetric
open strings comprises an operator $\ap^2 \zeta_2 \te{Tr} \, F^4$ which can be cleanly distinguished from
(\ref{pert0.1recap}) by its transcendental prefactor $\zeta_2=\frac{\pi^2}{6}$ \cite{Green:1981xx, Schwarz:1982jn, 
Tseytlin:1986ti}. Up to and including the order
of $\ap^2$, the amplitudes computed from (\ref{pert0.1recap}) obey BCJ relations (\ref{BCJrels})
while the $\ap^2 \zeta_2 \te{Tr} \, F^4$-operator excluded from (\ref{pert0.1recap}) is incompatible
with the BCJ duality \cite{Broedel:2012rc}. As explained in the reference, these BCJ relations to the 
order of $\ap^2$ rely on the interplay between single-insertions of the $\ap^2\te{Tr} \, F^4$-operator
in (\ref{pert0.1recap}) and double-insertions of $\ap  \te{Tr} \, F^3$.

More generally, the accompanying multiple zeta values are instrumental to
identify the $D^{2m}F^n$-operators in string effective actions that admit color-kinematics dual representations.
For instance, the entire single-trace gauge sector of the heterotic string obeys BCJ relations \cite{Stieberger:2014hba}.
The subsector of open-bosonic-string amplitudes compatible with the BCJ duality was identified in \cite{Huang:2016tag}, 
and the amplitude contributions without any zeta-value coefficient were derived from a field-theory Lagrangian \cite{Azevedo:2018dgo}.

The subsequent Berends--Giele recursions for the amplitudes of (\ref{pert0.1recap}) follow a two-fold purpose:
on the one hand, they will be used to generate economic and manifestly cyclic amplitude representations along the lines
of section \ref{sec:2.3}. On the other hand, they set the stage for
\begin{itemize}
\item an off-shell realization of the BCJ duality in section \ref{sec:4}
\item a kinematic proof of the BCJ relations in section \ref{sec:5.2}
\item a construction of manifestly local gauge-theory numerators subject to kinematic Jacobi relations
in section \ref{sec:5.3}.
\end{itemize}
All of these results hold to the order of $\ap^2$ and are based on a non-linear gauge 
transformation of the generating series of Berends--Giele currents similar to (\ref{pert1.gauge}).

%%%%%%%%%%%%%%%%%%%%%%%%%%%%%%%%%%%%%%%%%%%%%%%%
%%%%%%%%%%%%%%%%%%%%%%%%%%%%%%%%%%%%%%%%%%%%%%%%

\subsection{Berends--Giele recursions for $F^3$ and $F^4$}
\label{sec:3.1}

Our Berends--Giele approach to (\YMF) follows the lines of
section \ref{sec:2.2} to derive recursions for the currents from the non-linear equations of motion. 
The field variation of the action (\ref{pert0.1recap}) is given by\footnote{It is convenient to use
$\frac{ \delta   }{\delta \AAA_\la} \te{Tr}(\FF_{\mu \nu} X)= \delta^\la_\mu [\nabla_\nu,X]
- \delta^\la_\nu [\nabla_\mu,X]$ in intermediate steps of deriving (\ref{pert3.2}). The extensions of this lemma
to $\FF_{\mu \nu}$-dependent quantities $X$ follows straightforwardly from the Leibniz rule.} 
\begin{align}
\frac{ \delta {\cal S}_{{\rm YM}+F^3+F^4}  }{\delta \AAA_\la} = [\nabla_\mu , \FF^{\la \mu} ] + 2\ap [\nabla_\mu , [\FF^{ \mu \nu} , \FF_\nu{}^\la]] + 2\ap^2   \Big[ \nabla_\mu , \big[ [ \FF^{\mu \la} , \FF_{\rho \sigma}] , \FF^{\rho\sigma} \big] \Big]
\label{pert3.2}
\end{align}
and augments (\ref{pert1.5}) by $\ap$-corrections. In Lorenz gauge 
$\partial_\mu \AAA^\mu=0$, setting (\ref{pert3.2}) to zero amounts to a wave equation analogous to (\ref{pert1.6}),
\begin{align}
\Box \AAA^\la &= [\AAA^\mu, \partial_\mu \AAA^\la] + [\AAA_\mu , \FF^{\mu \la}]
+ 2\ap \big\{  [ \nabla_\mu \FF^{ \mu \nu} , \FF_\nu{}^\la] +  [ \FF^{ \mu \nu} , \nabla_\mu \FF_\nu{}^\la] \big\}
\label{pert3.3}
\\
&+ 2\ap^2 \Big\{   \big[ [ \nabla_\mu \FF^{\mu \la} , \FF_{\rho \sigma}] , \FF^{\rho\sigma} \big]  
+  \big[ [ \FF^{\mu \la} , \nabla_\mu \FF_{\rho \sigma}] , \FF^{\rho\sigma} \big]  
+  \big[ [ \FF^{\mu \la} , \FF_{\rho \sigma}] , \nabla_\mu  \FF^{\rho\sigma} \big]  \Big\} \, . \notag
\end{align}
Since we will only be interested in the amplitude contributions up to the order of $\ap^2$,
we can simplify (\ref{pert3.3}) by dropping terms of order $\alpha'^{3}$ and higher. At
the first order in $\ap$, this allows to replace $\nabla_\mu \FF^{ \mu \nu}= 
2\ap [\FF^{\rho \sigma} , \nabla_\rho \FF_{\sigma}{}^\nu] + {\cal O}(\ap^2)= -\ap
 [\FF^{\rho \sigma} , \nabla^\nu \FF_{\rho \sigma}] + {\cal O}(\ap^2)$ such that
\begin{align}
\Box \AAA^\la &= [\AAA^\mu, \partial_\mu \AAA^\la] + [\AAA_\mu , \FF^{\mu \la}]
+ 2\ap    [ \FF^{ \mu \nu} , \nabla_\mu \FF_\nu{}^\la] 
+ 4\ap^2 \big[  [\FF^{\mu \la} , \FF^{\rho \sigma} ] , \nabla_\mu \FF_{\rho \sigma} \big]+ {\cal O}(\ap^3)\, .
\label{pert3.4}
\end{align}
This form of the field equations gives rise to an efficient Berends--Giele recursion:
We will study formal solutions of (\ref{pert3.4}) modulo $\ap^3$ that descend from a perturbiner ansatz
\beq
\AAA^{\mu} = \sum_{P \neq \emptyset} A_P^{\mu} t^P e^{k_P\cdot x}
\ , \ \ \ \ \ \
\FF^{\mu \nu} = \sum_{P \neq \emptyset} F_P^{\mu \nu} t^P e^{k_P\cdot x} \, ,
\label{pert3.5}
\eeq
where Lorenz gauge and the definition $\FF^{\mu \nu}= - [\nabla^\mu, \nabla^\nu]$ of the field strength imply
\beq
k_P \cdot A_P= 0
\ , \ \ \ \ \ \
F_P^{\mu \nu} =  k^\mu_P A^\nu_P - k^\nu_P A^\mu_P - \sum_{P=XY} (A_X^\mu A_Y^\nu - A_X^\nu A_Y^\mu )  \, .
\label{pert3.6}
\eeq
In comparison to the perturbiners (\ref{pert1.7}) and (\ref{pert1.10}) of undeformed YM theory, the 
Berends--Giele currents have been renamed
as $J_P^\mu \rightarrow A_P^\mu$ and $B^{\mu \nu}_P \rightarrow F_P^{\mu \nu}$ in order to
distinguish these $\ap$-dependent quantities from the YM currents in (\ref{pert1.8}) and (\ref{pert1.10}) ,
\beq
J_P^{\mu} = \lim_{\ap \rightarrow0} A_P^{\mu} \ , \ \ \ \ \ \
B_P^{\mu \nu} = \lim_{\ap \rightarrow0} F_P^{\mu \nu}  \, .
\label{pert3.5FT}
\eeq
In the same way as the Berends--Giele recursion of undeformed YM theory
benefits from field-strength currents $B_P^{\mu \nu}$, the perturbiner solutions to (\ref{pert3.4})
are conveniently expressed in terms of the additional auxiliary currents
\beq
\nabla^\mu \FF^{\nu \la} =  \sum_{P \neq \emptyset} F_P^{\mu| \nu \la} t^P e^{k_P\cdot x} \ , \ \ \ \ \ \
[\FF^{\mu \nu} , \FF^{\la \rho}] =   \sum_{P \neq \emptyset} G_P^{\mu \nu | \la \rho} t^P e^{k_P\cdot x}\, .
\label{pert3.5aux}
\eeq
By their definition in (\ref{pert3.5aux}), the auxiliary currents are determined by 
$A_P^\mu$ and $F_P^{\mu \nu}$,
\begin{align}
F_P^{\mu| \nu \la} &= k_P^\mu F_P^{\nu \la} - \sum_{P=XY} ( A_X^\mu F_Y^{\nu \la} - A_Y^\mu F_X^{\nu \la}) \notag \\
G_P^{\mu \nu | \la \rho} &= \sum_{P=XY} (F_X^{\mu \nu} F_Y^{\la \rho} - F_Y^{\mu \nu} F_X^{\la \rho}) \, ,
\label{pert3.8}
\end{align}
in the same way as the $F_P^{\mu \nu}$ in (\ref{pert3.6}) boil down to the elementary currents $A_P^\mu$.
All the above currents can be shown to obey shuffle symmetry
\beq
A_{P\shuffle Q}^\mu
=F_{P\shuffle Q}^{ \mu \nu }
=F_{P\shuffle Q}^{\mu| \nu \la} 
=G_{P\shuffle Q}^{\mu \nu | \la \rho}  = 0 \ \forall \ P,Q\neq \emptyset
\label{pert3.shuffle}
\eeq
by repeating the arguments for the currents $J_P^\mu$ and $F_P^{\mu \nu}$ of pure YM 
theory \cite{Berends:1988zn, Lee:2015upy}.

With the above definitions, the Berends--Giele recursion induced by the field equation (\ref{pert3.4}) 
takes the simple form
\begin{align}
A_i^\mu &= e_i^\mu
\label{pert3.7}
\\
A_P^\mu &= \frac{1}{2 s_P} \sum_{P=XY} \Big[ (k_Y \cdot A_X) A_Y^\mu + A_X^\nu F_Y^{\nu \mu} + 2\ap F^{\nu \la}_X F_Y^{\nu|\la}{}^{\mu} + 4\ap^2 G_X^{\nu \mu| \rho \sigma} F_Y^{\nu|\rho \sigma}
- (X\leftrightarrow Y) \Big] \, .
\notag
\end{align}
In analogy to (\ref{pert1.1}), the leading orders $\alpha'^{\leq 2}$ of the tree amplitudes 
resulting from the action (\ref{pert0.1recap}) are then given by
\beq
{\cal A}_{{\rm YM}+F^3+F^4}(1,2,\ldots,n{-}1,n) = s_{12\ldots n-1} A^\mu_{12\ldots n-1} A^\mu_n + {\cal O}(\ap^3) \, .
\label{pert3.1}
\eeq
For instance, the rank-two current due to (\ref{pert3.7}) with $X=1$ and $Y=2$ and the resulting
three-point amplitude read
\begin{align}
s_{12} A_{12}^\mu &= (k_2\cdot e_1) e_2^\mu -  (k_1\cdot e_2) e_1^\mu + \frac{1}{2}(k_1^\mu - k_2^\mu)(e_1\cdot e_2) \notag \\
& \ \ \ \
 + \ap(k_1^\mu - k_2^\mu) \big[ (k_1\cdot k_2) (e_1\cdot e_2) - (k_1\cdot e_2) (k_2\cdot e_1) \big]
\label{pert3.1threept}\\
{\cal A}_{{\rm YM}+F^3+F^4}(1,2,3) &= \big[(k_2 \cdot e_1) (e_2\cdot e_3) + {\rm cyc}(1,2,3) \big] + 2\ap (k_2 \cdot e_1)(k_3 \cdot e_2)(k_1 \cdot e_3)
\, .
\notag
\end{align}
The last term of the three-point function illustrates a fundamental difference between the tensor structure
of YM amplitudes and their $F^3$ corrections: Contractions of the type $(k\cdot e)^n$ do not occur in $n$-point
amplitudes of YM \cite{Barreiro:2012aw, Barreiro:2013dpa, Boels:2016xhc}. Hence, the expressions for 
${\cal A}_{{\rm YM}+F^3+F^4}$ in this work do not belong to the class of $D^{2m}F^n$ amplitudes that are 
accessible from the open superstring through a combination of color-ordered SYM trees \cite{Mafra:2011nv}.

By the shuffle symmetry (\ref{pert3.shuffle}) of the currents, the amplitude representation (\ref{pert3.1})
can be used to demonstrate ${\cal A}_{{\rm YM}+F^3+F^4}(1,2,\ldots,n)$ to also obey the Kleiss--Kuijf 
relations (\ref{pert1.3}). 
The $\ap^1$-order of our results up to and including $n=6$ points has
been checked to match the $D$-dimensional CHY formulae of\footnote{We are grateful to Song He
and Yong Zhang for providing the analytic expressions.} \cite{He:2016iqi}. The
$D=4$ helicity components of the CHY expressions\footnote{See
\cite{Zhang:2016rzb} for a systematic study of the reduction of CHY formulae to
$D=4$ dimensions.} in turn have been verified to 
agree with the results of \cite{Dixon:1993xd, Dixon:2004za, Cohen:2010mi}.

We emphasize that the form of the recursion in (\ref{pert3.7}) only involves deconcatenations $P=XY$ into two words $X,Y$
rather than three-word expressions with $P=XYZ$ as seen in (\ref{pert1.0}). Hence, the amplitudes (\ref{pert3.1}) naturally arise
in a cubic-graph parametrization (\ref{genpar2}) as visualized in figure \ref{f:bgf3f4}. The cubic-graph organization extends
to the order of $\ap^2 $, although each term of the $ {\rm Tr} \, F^4$ vertex in (\ref{pert0.1recap}) involves at least 
four powers of the $\AAA^\mu$ field.
Still, the quartic-vertex origin of the terms $ G_X^{\nu \mu| \rho \sigma} F_Y^{\nu|\rho \sigma}$ in (\ref{pert3.7})
is visible through the absence of single-particle currents $G_i^{\mu \nu | \la \rho}=0$ since there is no deconcatenation
$XY=i$ in (\ref{pert3.8}). Like this, 
the part $ G_X^{\nu \mu| \rho \sigma} F_Y^{\nu|\rho \sigma}$ of the recursion (\ref{pert3.7}) can only contribute
at minimum length $|X|+|Y|=3$, i.e.\ to amplitudes (\ref{pert3.1}) at multiplicity $n\geq 4$. 

\begin{figure}[h]
\begin{center}
\begin{tikzpicture} [scale=0.8, line width=0.30mm]
\draw (-9.0,0.5) node{$A^{\mu}_{123}, F^{\mu \nu}_{123},$};
\draw (-9.0,-0.5) node{$F^{\mu| \nu\la}_{123},G^{\mu \nu|\la\rho}_{123}$};
\draw (-8.0,0) node{$\phantom{A^{\mu}_{123}, F^{\mu \nu}_{123}} \ \ \ \ \ \leftrightarrow  $};
\draw(-7.2,0)node{$\Bigg\}$};
\draw (-4.5,0.5) -- (-5,1) node[left]{$3$};
\draw (-4.7,0) -- (-5.4,0) node[left]{$2$};
\draw (-4.5,-0.5) -- (-5,-1) node[left]{$1$};
\draw (-4,0) circle(0.7cm);
\draw(-4,0.3)node{\tiny YM};
\draw(-4,0)node{\tiny $+F^3$};
\draw(-4,-0.3)node{\tiny$+F^4$};
\draw (-3.3,0)--(-2.5,0);
\draw (-2, 0) node{${\ldots} $};
\draw (-1.3,0) node{$=$};
\scope[xshift=1cm]
\draw (0,0) -- (-1,1) node[left]{$2$};
\draw (0,0) -- (-1,-1) node[left]{$1$};
\draw (0,0) -- (1.8,0);
\draw (0.5,0.3) node{$s_{12}$};
\draw (1,0) -- (1,1) node[right]{$3$};
\draw (1.65,-0.3) node{$s_{123}$};
\draw (2.7, 0) node{${\cdots} \  \ + $};
\scope[xshift=5cm]
\draw (0,0) -- (-1,1) node[left]{$2$};
\draw (0,0) -- (-1,-1) node[left]{$1$};
\draw (0,0) -- (1.8,0);
\draw[fill=white](0,0)circle(0.3cm);
%\draw (0.5,0.3) node{$s_{12}$};
\draw (1,0) -- (1,1) node[right]{$3$};
%\draw (1.65,-0.3) node{$s_{123}$};
\draw (2.7, 0) node{${\cdots} \  \ \phantom{+} $};
\endscope
\scope[xshift=-5cm, yshift=-3cm]
\draw(-2.0,0)node{$+$};
\draw (0,0) -- (-1,1) node[left]{$2$};
\draw (0,0) -- (-1,-1) node[left]{$1$};
\draw (0,0) -- (1.8,0);
%\draw (0.5,0.3) node{$s_{12}$};
\draw (1,0) -- (1,1) node[right]{$3$};
\draw[fill=white](1,0)circle(0.3cm);
%\draw (1.65,-0.3) node{$s_{123}$};
\draw (2.7, 0) node{${\cdots} \  \  +$};
\endscope
\scope[xshift=0cm, yshift=-3cm]
\draw (0,0) -- (-1,1) node[left]{$2$};
\draw (0,0) -- (-1,-1) node[left]{$1$};
\draw (0,0) -- (1.8,0);
\draw[fill=black](0,0)circle(0.3cm);
%\draw (0.5,0.3) node{$s_{12}$};
\draw (1,0) -- (1,1) node[right]{$3$};
%\draw[fill=white](1,0)circle(0.3cm);
%\draw (1.65,-0.3) node{$s_{123}$};
\draw (2.7, 0) node{${\cdots} \  \ +$};
%\draw(4.7,0)node{$(1\leftrightarrow 3) \ \ +$};
\endscope
\scope[xshift=5cm, yshift=-3cm]
\draw (0,0) -- (-1,1) node[left]{$2$};
\draw (0,0) -- (-1,-1) node[left]{$1$};
\draw (0,0) -- (1.8,0);
%\draw (0.5,0.3) node{$s_{12}$};
\draw (1,0) -- (1,1) node[right]{$3$};
\draw[fill=black](1,0)circle(0.3cm);
%\draw (1.65,-0.3) node{$s_{123}$};
\endscope
\scope[xshift=-5cm, yshift=-6cm]
\draw(-2.0,0)node{$+$};
\draw (0,0) -- (-1,1) node[left]{$2$};
\draw (0,0) -- (-1,-1) node[left]{$1$};
\draw (0,0) -- (1.8,0);
\draw[fill=white](0,0)circle(0.3cm);
%\draw (0.5,0.3) node{$s_{12}$};
\draw (1,0) -- (1,1) node[right]{$3$};
\draw[fill=white](1,0)circle(0.3cm);
%\draw (1.65,-0.3) node{$s_{123}$};
\draw (2.7, 0) node{${\cdots} \  \ +$};
\draw(4.7,0)node{$(1\leftrightarrow 3) \ \ +$};
\draw(6.9,0)node{${\cal O}(\ap^3)$};
\endscope
\draw (5.5,1) node{};
\endscope
\end{tikzpicture}
\caption{Berends--Giele currents $A^\mu_{12\ldots p},F^{\mu \nu}_{12\ldots p},F^{\mu |\nu\la}_{12\ldots p}$ 
and $G^{\mu \nu|\la \rho}_{12\ldots p}$ of rank $p$ combine the diagrams and propagators expected in a color-ordered 
$(p{+}1)$-point tree amplitude of (\YMF) with an off-shell leg $\cdots$. Vertices 
marked with a white and black dot represent the first and second order in $\ap$ on the right hand side of (\ref{pert3.7}),
i.e.\ the cubic-graph parametrization of $\alpha' F^3$ and $\alpha'^2F^4$ insertions.}
\label{f:bgf3f4}
\end{center}
\end{figure}

Note that the equation of motion (\ref{pert3.2}) translates into the following expression for the tensor divergence
of $F_P^{\mu \nu}$:
\beq
k_P^\la F_P^{\la \mu} = \sum_{P=XY} \Big[  A_X^\la F_Y^{\la \mu} + 2\ap F^{\nu \la}_X F_Y^{\nu|\la}{}^{\mu} + 4\ap^2 G_X^{\nu \mu| \rho \sigma} F_Y^{\nu|\rho \sigma}
- (X\leftrightarrow Y) \Big] + {\cal O}(\ap^3)\, .
\label{pert3.9}
\eeq
%

%%%%%%%%%%%%%%%%%%%%%%%%%%%%%%%%%%%%%%%%%%%%%%%%
%%%%%%%%%%%%%%%%%%%%%%%%%%%%%%%%%%%%%%%%%%%%%%%%

\subsection{Manifestly cyclic Berends--Giele representations}
\label{sec:3.2}

\begin{figure}[h]
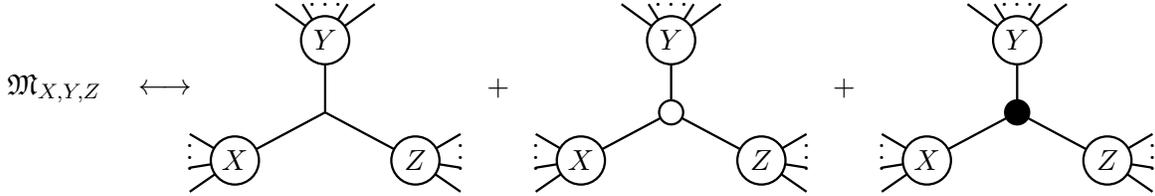

\begin{center}
\tikzpicture[xscale=0.4, yscale=0.32, line width=0.30mm]
\draw(-7.5,1)node{$\Mfrak_{X,Y,Z}\ \ \ \longleftrightarrow$};
\draw(5.75,1)node{$+$};
\draw(17.25,1)node{$+$};
%%%%%
\draw(0,0) -- (0,3);
\draw(0,3) -- (-0.75,4.5);
\draw(0,3) -- (0.75,4.5);
\draw(0,4.5)node{$\ldots$};
\draw(0,3) -- (-1.65,4.5);
\draw(0,3) -- (1.65,4.5);
\draw[fill=white](0,3) ellipse(0.8cm and 1cm);
\draw(0,3)node{$Y$};
\draw(0,0)--(3,-2);
\draw(4.5,-1.5)node{$\vdots$};
\draw(3,-2)--(4.5,-0.9);
\draw(3,-2)--(4.5,-2.3);
\draw(3,-2)--(4.5,-3.3);
\draw[fill=white](3,-2) ellipse(0.8cm and 1cm);
\draw(3,-2)node{$Z$};
\draw(0,0)--(-3,-2);
\draw(-4.5,-1.5)node{$\vdots$};
\draw(-3,-2)--(-4.5,-0.9);
\draw(-3,-2)--(-4.5,-2.3);
\draw(-3,-2)--(-4.5,-3.3);
\draw[fill=white](-3,-2) ellipse(0.8cm and 1cm);
\draw(-3,-2)node{$X$};
%%%%
%%%%
%%%%
\scope[xshift=11.5cm]
\draw(0,0) -- (0,3);
\draw(0,3) -- (-0.75,4.5);
\draw(0,3) -- (0.75,4.5);
\draw(0,4.5)node{$\ldots$};
\draw(0,3) -- (-1.65,4.5);
\draw(0,3) -- (1.65,4.5);
\draw[fill=white](0,3) ellipse(0.8cm and 1cm);
\draw(0,3)node{$Y$};
\draw(0,0)--(3,-2);
\draw(4.5,-1.5)node{$\vdots$};
\draw(3,-2)--(4.5,-0.9);
\draw(3,-2)--(4.5,-2.3);
\draw(3,-2)--(4.5,-3.3);
\draw[fill=white](3,-2) ellipse(0.8cm and 1cm);
\draw(3,-2)node{$Z$};
\draw(0,0)--(-3,-2);
\draw(-4.5,-1.5)node{$\vdots$};
\draw(-3,-2)--(-4.5,-0.9);
\draw(-3,-2)--(-4.5,-2.3);
\draw(-3,-2)--(-4.5,-3.3);
\draw[fill=white](-3,-2) ellipse(0.8cm and 1cm);
\draw(-3,-2)node{$X$};
\draw[fill=white](0,0) ellipse(0.4cm and 0.5cm);
\endscope
%%%%
%%%%
%%%%
\scope[xshift=23cm]
\draw(0,0) -- (0,3);
\draw(0,3) -- (-0.75,4.5);
\draw(0,3) -- (0.75,4.5);
\draw(0,4.5)node{$\ldots$};
\draw(0,3) -- (-1.65,4.5);
\draw(0,3) -- (1.65,4.5);
\draw[fill=white](0,3) ellipse(0.8cm and 1cm);
\draw(0,3)node{$Y$};
\draw(0,0)--(3,-2);
\draw(4.5,-1.5)node{$\vdots$};
\draw(3,-2)--(4.5,-0.9);
\draw(3,-2)--(4.5,-2.3);
\draw(3,-2)--(4.5,-3.3);
\draw[fill=white](3,-2) ellipse(0.8cm and 1cm);
\draw(3,-2)node{$Z$};
\draw(0,0)--(-3,-2);
\draw(-4.5,-1.5)node{$\vdots$};
\draw(-3,-2)--(-4.5,-0.9);
\draw(-3,-2)--(-4.5,-2.3);
\draw(-3,-2)--(-4.5,-3.3);
\draw[fill=white](-3,-2) ellipse(0.8cm and 1cm);
\draw(-3,-2)node{$X$};
\draw[fill=black](0,0) ellipse(0.4cm and 0.5cm);
\endscope
\endtikzpicture
\caption{In the diagrammatic interpretation of the building block $\Mfrak_{X,Y,Z}$ in (\ref{pert3.12}), the central
vertex can either be of YM type (first term), of $(\ap F^3)$-type (white dot) or of $(\ap^2 F^4)$-type (black dot).
The blobs labelled by $X,Y,Z$ represent currents of (\YMF).}
\label{f:MABC2}
\end{center}
\end{figure}

This section is dedicated to a reformulation of the Berends--Giele formula (\ref{pert3.1}) for ${\cal A}_{{\rm YM}+F^3+F^4}$ 
such as to manifest cyclicity and to reduce the maximum 
rank of the Berends--Giele constituents (\ref{pert3.7}) on the right-hand side. This amounts to identifying a deformation
of the cyclic $M_{X,Y,Z}$ building block in (\ref{pert1.12}) that preserves the structure of the economic and manifestly cyclic
amplitude representations (\ref{pert1.16}) up to and including the order of $\ap^2$.
The desired cyclic building block analogous to $M_{X,Y,Z}$ reads
\begin{align}
\Mfrak_{X,Y,Z} &= \frac{1}{2} \big( A_X^\mu F_Y^{\mu \nu} A_Z^\nu  + {\rm cyc}(X,Y,Z) \big) -2\ap F_X^{\mu \nu} F_Y^{\nu \la} F_Z^{\la \mu} \notag \\
& + \Big( \frac{\ap}{2}  F_X^{\mu | \nu \la} F_Y^{ \nu \la} A_Z^\mu + 2 \ap^2  G_X^{\mu \nu | \la \rho} F_Y^{ \mu| \la \rho} A_Z^\nu  \pm {\rm perm}(X,Y,Z) \Big) \label{pert3.12}\\
&+ \Big( \frac{\ap^2}{2}   G_X^{\mu \nu | \la \rho} F_Y^{\mu \nu} F_Z^{\la \rho} - 2\ap^2   F_X^{\mu \nu} F_Y^{\mu| \la \rho} F_Z^{\nu| \la \rho} + {\rm cyc}(X,Y,Z) \Big)\,,
\notag
\end{align}
where the notation $\pm {\rm perm}(X,Y,Z)$ instructs to add five permutations with alternating signs
and enforces permutation antisymmetry $\Mfrak_{X,Y,Z}=-\Mfrak_{Y,X,Z}=\Mfrak_{Y,Z,X}$.
The diagrammatic interpretation of the building block (\ref{pert3.12}) can be found in figure \ref{f:MABC2}.

In analogy with (\ref{pert1.13}), the amplitude formula (\ref{pert3.1}) can be rewritten as
\begin{align}
{\cal A}_{{\rm YM}+F^3+F^4}(1,2,\ldots,n{-}1,n)  
&= \sum_{XY=12\ldots n-1} \Mfrak_{X,Y,n} + {\cal O}(\ap^3) \label{pert3.13} \\
& = \sum_{j=1}^{n-2} \Mfrak_{12\ldots j, \, j{+}1\ldots n-1,\, n} + {\cal O}(\ap^3)  \, .
\notag
\end{align}
To first order in $\ap$, the equivalence with (\ref{pert3.1}) is proven to all multiplicities in
appendix \ref{app:A.1}. At second order in $\ap^2$, (\ref{pert3.13}) has been checked analytically to multiplicity
$n=6$ and numerically up to and including $n=8$. In the same way as we are only interested in the
orders $\ap^{\leq 2}$ of the amplitudes (\ref{pert3.13}), we will consistently drop terms at the orders $\ap^{\geq 3}$
in later equations of this work and skip the disclaimer ${\cal O}(\ap^3)$ for ease of notation.

Amplitude representations with manifest cyclicity and lower-rank Berends--Giele currents can be
obtained by an integration-by-parts property that takes the same form as (\ref{pert1.14}),
\beq
\sum_{XY=A} \Mfrak_{X,Y,B} = \sum_{XY=B} \Mfrak_{A,X,Y} \, ,  \label{pert3.14}
\eeq
starting with $\Mfrak_{12,3,4}=\Mfrak_{1,2,34}$, see (\ref{pert1.15}) for higher-point examples. At the order of $\ap$, a general 
proof of (\ref{pert3.14}) can be found in appendix \ref{app:A.2}. At the order of $\ap^2$, (\ref{pert3.14})
has been checked analytically at $|A|+|B|\leq 6$ as well as numerically at $|A|+|B|\leq 8$ and is conjectural
at higher multiplicity.

By applying (\ref{pert3.14}) to the amplitude representation  (\ref{pert3.13}), one gets
manifestly cyclic expressions analogous to (\ref{pert1.16}),
\begin{align}
{\cal A}_{{\rm YM}+F^3+F^4}(1,2,3,4) &= \frac{1}{2} \Mfrak_{12,3,4} + {\rm cyc}(1,2,3,4) \notag \\
{\cal A}_{{\rm YM}+F^3+F^4}(1,2,\ldots,5) &= \Mfrak_{12,3,45} + {\rm cyc}(1,2,3,4,5) \label{pert13.16} \\
{\cal A}_{{\rm YM}+F^3+F^4}(1,2,\ldots,6) &= \frac{1}{3}\Mfrak_{12,34,56} + \frac{1}{2}(\Mfrak_{123,45,6} + \Mfrak_{123,4,56}) + {\rm cyc}(1,2,\ldots,6) \notag \\
{\cal A}_{{\rm YM}+F^3+F^4}(1,2,\ldots,7) &= \Mfrak_{123,45,67} + \Mfrak_{1,234,567} + {\rm cyc}(1,2,\ldots,7) \, ,\notag
\end{align}
as well as
\begin{align}
{\cal A}_{{\rm YM}+F^3+F^4}(1,2,\ldots,8) &=\frac{1}{2}(\Mfrak_{1234,567,8}+\Mfrak_{1234,56,78}+\Mfrak_{1234,5,678})  \notag \\
& \ \ \ +  \Mfrak_{123,456,78}  + {\rm cyc}(1,2,\ldots,8)  \notag \\
{\cal A}_{{\rm YM}+F^3+F^4}(1,2,\ldots,9) &=   \Mfrak_{1234,567,89}+\Mfrak_{1234,56,789}+\Mfrak_{1234,5678,9}  \notag \\
&\ \ \ + \frac{1}{3}\Mfrak_{123,456,789}+ {\rm cyc}(1,2,\ldots,9)  \label{pert23.16} \\
{\cal A}_{{\rm YM}+F^3+F^4}(1,2,\ldots,0) &= \frac{1}{2}(\Mfrak_{12345,6789,0}+\Mfrak_{12345,678,90}+\Mfrak_{12345,67,890}+\Mfrak_{12345,6,7890}) \notag \\
& \ \ \ + \Mfrak_{1234,567,890} + \Mfrak_{1234,5678,90}  + {\rm cyc}(1,2,\ldots,0) \, . \notag
\end{align}
The rational prefactors of non-prime 
multiplicities $n=4,6,8,9$ avoid overcounting of cubic diagrams when combinations of currents
are invariant under less than $n$ cyclic shifts $i\rightarrow i{+}1$. Integration by parts (\ref{pert3.14}) 
can be used to bypass such prefactors in expressions like
\begin{align}
{\cal A}_{{\rm YM}+F^3+F^4}(1,2,3,4) &=  \Mfrak_{12,3,4} +  \Mfrak_{23,4,1} \notag \\
{\cal A}_{{\rm YM}+F^3+F^4}(1,2,\ldots,6) &=  \Mfrak_{12,34,56} + \Mfrak_{23,45,61} + \Mfrak_{123,45,6} + \Mfrak_{123,4,56}   \label{pert13.16a} \\
&+ \Mfrak_{234,56,1} + \Mfrak_{234,5,61}+ \Mfrak_{345,61,2} + \Mfrak_{345,6,12} \, ,\notag
\end{align}
also see \cite{Mafra:2010ir, Mafra:2010jq, Mafra:2015vca} for the antecedents of these 
representations in ten-dimensional SYM. Similarly, the all-multiplicity series of cyclic representations  
\beq
{\cal A}_{{\rm YM}+F^3+F^4}(1,2,\ldots,n) = \frac{1}{2(n{-}3)} \sum_{j=2}^{n-2} \sum_{l=j+1}^{n-1}\Mfrak_{12\ldots j,\, j+1 \ldots l, \, l+1 \ldots n} + {\rm cyc}(1,2,\ldots,n) 
\label{pert13.37}
\eeq
can be imported from ten-dimensional SYM \cite{Mafra:2011nv}.

%%%%%%%%%%%%%%%%%%%%%%%%%%%%%%%%%%%%%%%%%%%%%%%%
%%%%%%%%%%%%%%%%%%%%%%%%%%%%%%%%%%%%%%%%%%%%%%%%

\subsection{Gauge algebra of $F^3+F^4$ building blocks}
\label{sec:3.3}

The action of non-linear gauge transformations  $\delta_{\Omega}\AAA^\mu = \partial^\mu \Omega - [\AAA^\mu,\Omega]$
is not altered by the higher-mass-dimension operators in the action (\ref{pert0.1recap}). Hence, given perturbiner
components $\Omega_P$ for the gauge scalars $\Omega$, the $\alpha'$-deformed
currents of the previous subsection follow the transformations of the YM currents (\ref{pert1.gauge}),
\begin{align}
\delta_{\Omega} A_P^\mu &= k_P^\mu \Omega_P - \sum_{XY=P} (A_X^\mu \Omega_Y - A_Y^\mu \Omega_X ) \, , 
\ \ \ \
\delta_{\Omega} F_P^{\mu \nu} =  - \sum_{XY=P} (F_X^{\mu \nu} \Omega_Y - F_Y^{\mu \nu} \Omega_X ) 
\label{pert1.gau}
\\
\delta_{\Omega} F_P^{\mu|\nu \la} &=- \sum_{XY=P} (F_X^{\mu|\nu \la}  \Omega_Y - F_Y^{\mu|\nu \la} \Omega_X ) \, , 
\ \ \ \
\delta_{\Omega} G_P^{\mu \nu|\la \rho} =  - \sum_{XY=P} (G_X^{\mu \nu|\la \rho} \Omega_Y - G_Y^{\mu \nu|\la \rho} \Omega_X )
 \, .
\notag
\end{align}
One can therefore verify non-linear gauge invariance of the amplitude formula (\ref{pert3.1}) 
by repeating the arguments of the undeformed gauge theory: Among the three terms in the gauge variation
\begin{align}
\delta_\Omega(s_{12\ldots n-1} A_{12\ldots n-1} \cdot A_n) &= s_{12\ldots n-1} \Big\{ (A_{12\ldots n-1}\cdot k_n) \Omega_n + \Omega_{12\ldots n-1} (k_{12\ldots n-1}\cdot A_n) \notag \\
& \ \ \ \ \ \ \ \ \ \ \ \ \ \ \ \ \ \ + \sum_{12\ldots n{-}1 = XY} (\Omega_Y A_X^\mu - \Omega_X A_Y^\mu) A_n^\mu\Big\} \, ,
\label{pert3.1gvar}
\end{align}
the first one vanishes by the Lorenz-gauge condition $k_{12\ldots n-1}\cdot A_{12\ldots n-1}=0$ and
the second one due to transversality $k_n\cdot A_n=0$ (using momentum conservation $ k^\mu_{12 \ldots n-1}= -k^\mu_n  $
in both cases). The currents in the second line of (\ref{pert3.1gvar}) have multiplicity
$|X|,|Y|\leq n{-}2$ and are therefore regular as $s_{12\ldots n-1}\rightarrow 0$, so multiplication with
$s_{12\ldots n-1}$ causes this term to vanish as well. This rests on the reasonable assumption that
the gauge scalars $\Omega_P$ descend from a perturbiner and can only have poles in $s_Q$ 
for subsets $Q \subseteq P$.

By the Leibniz property of $\delta_{\Omega}$, momentum conservation and the expression (\ref{pert3.9}) 
for contractions of the form $k_P^\mu F_P^{\mu \nu}$, one can infer the non-linear gauge transformation
of the building blocks (\ref{pert3.12}). The result is most conveniently expressed in terms of a scalar quantity
$\Omega_{X,Y,Z,W}=\Omega_{[X,Y,Z,W]}$ which is totally antisymmetric in four multiparticle labels $X,Y,Z,W$,
\begin{align}
\delta_\Omega  \Mfrak_{X,Y,Z} &= \sum_{X=PQ} \Omega_{P,Q,Y,Z}  + \sum_{Y=PQ} \Omega_{P,Q,Z,X} + \sum_{Z=PQ} \Omega_{P,Q,X,Y}   \label{pert3.51}
\\
\Omega_{X,Y,Z,W} &= \Omega_X \Mfrak_{Y,Z,W}-  \Omega_Y \Mfrak_{Z,W,X} 
+  \Omega_Z \Mfrak_{W,X,Y} -  \Omega_W \Mfrak_{X,Y,Z} \, .
\notag
\end{align}
While the $\alpha' \rightarrow 0$ limit of (\ref{pert3.51}) descends from BRST variations of superspace 
building blocks in ten-dimensional SYM theory \cite{Mafra:2015vca}, a proof to the first order in $\ap$
is given in appendix \ref{app:A.3}. At the order of $\ap^2$, we have tested (\ref{pert3.51}) to the order of $|X|+|Y|+|Z|=6$.

Based on the gauge algebra (\ref{pert3.51}) and permutation antisymmetry $\Omega_{X,Y,Z,W}=\Omega_{[X,Y,Z,W]}$,
one can check the non-linear gauge invariance of the manifestly cyclic amplitude 
representations in (\ref{pert13.16}), (\ref{pert23.16}) and (\ref{pert13.37}). In particular,
this will be exploited in later sections to evaluate the $\Mfrak_{X,Y,Z}$ in BCJ gauge which
is tailored to manifest the BCJ duality via local numerators.

%%%%%%%%%%%%%%%%%%%%%%%%%%%%%%%%%%%%%%%%%%%%%%%%
%%%%%%%%%%%%%%%%%%%%%%%%%%%%%%%%%%%%%%%%%%%%%%%%
%%%%%%%%%%%%%%%%%%%%%%%%%%%%%%%%%%%%%%%%%%%%%%%%
%%%%%%%%%%%%%%%%%%%%%%%%%%%%%%%%%%%%%%%%%%%%%%%%
%%%%%%%%%%%%%%%%%%%%%%%%%%%%%%%%%%%%%%%%%%%%%%%%

\section{Kinematic Jacobi identities in off-shell diagrams}
\label{sec:4}

\begin{figure}[h]
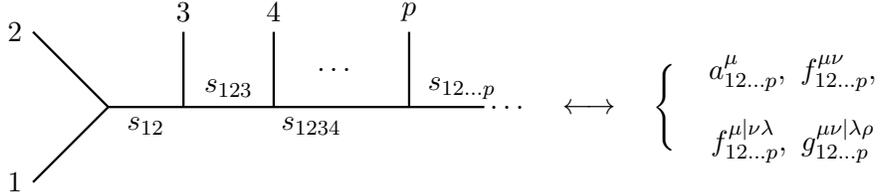

\begin{center}
\tikzpicture[line width=0.30mm]
\draw (0,0) -- (-1,1) node[left]{$2$};
\draw (0,0) -- (-1,-1) node[left]{$1$};
\draw (0,0) -- (5,0);
\draw (0.5,-0.25) node{$s_{12}$};
\draw (1,0) -- (1,1) node[above]{$3$};
\draw (1.6,0.25) node{$s_{123}$};
\draw (2.2,0) -- (2.2,1) node[above]{$4$};
\draw (2.7,-0.25) node{$s_{1234}$};
\draw (3, 0.5) node{$\ldots$};
%\draw (3, -0.2) node{$s_{i_1i_2i_3i_4}$};
\draw (4,0) -- (4,1) node[above]{$p$};;
\draw (4.7,0.25) node{$s_{12...p}$};
\draw (5.3,0) node{$\ldots$};
\scope[xshift=-0.4cm]
\draw (7,0) node{$\quad \longleftrightarrow \quad \Bigg\{$};
\draw(9.5,0.45)node{$ a_{12\ldots p}^\mu , \ f_{12\ldots p}^{\mu \nu},$};
\draw(9.5,-0.45)node{$ f_{12\ldots p}^{\mu|\nu \la} , \ g_{12\ldots p}^{\mu \nu|\la\rho}$};
\endscope
\endtikzpicture
\caption{This section is dedicated to constructing local and Jacobi-satisfying kinematic representatives for the depicted
cubic diagram of (\YMF). 
The notation $a_{12\ldots p}^\mu$, $f_{12\ldots p}^{\mu \nu}$, $ f_{12\ldots p}^{\mu|\nu \la} $ and 
$g_{12\ldots p}^{\mu \nu|\la\rho}$ will be introduced in subsection \ref{sec:4.1} and refers to four types 
of such solutions at different mass dimensions.}
\label{f:localdiag}
\end{center}
\end{figure}

The purpose of this section is to manifest the BCJ duality between color and kinematics
in off-shell diagrams of (\YMF).
We will construct local solutions to the kinematic Jacobi relations (\ref{dualityB}) in the 
subdiagram with an off-shell leg drawn in figure \ref{f:localdiag}. This amounts to
assigning kinematic numerators to the cubic-vertex diagram in the figure 
which share the symmetries of the associated color factors
\beq
C^b_{123\ldots p} = f^{a_1 a_2 c} f^{c a_3 d} f^{d a_4 e} \ldots f^{y a_{p-1} z} f^{z a_p b} \, .
\label{pert4.1}
\eeq
The adjoint indices $a_1,a_2,\ldots ,a_p$ refer to $p$ on-shell legs, and an off-shell leg
is associated with a free adjoint index $b$ carried by the rightmost factor in (\ref{pert4.1}). 

When identifying the dotted off-shell line in figure \ref{f:localdiag} with an external on-shell leg,
we recover the half-ladder diagrams of figure \ref{halfladders} that define the master numerators 
at $n=p{+}1$ points. Accordingly, permutations of figure \ref{f:localdiag} in $2,3,\ldots,p$ will be 
associated with the master numerators in an off-shell setup: Cubic diagrams which are not of 
half-ladder form or do not have leg $1$ and the off-shell leg $\cdots$ at their endpoints can be reached from $(p{-}1)!$
permutations of figure \ref{f:localdiag} through a sequence of Jacobi identities.

In specific examples of (\ref{pert4.1}) at rank $p\leq 5$, antisymmetry $f^{a_i a_j a_k} = f^{[a_i a_j a_k]} $ and Jacobi 
identities (\ref{col:jac}) imply the so-called {\it Lie symmetries} for the color factors,
\begin{align}
0 &= C^b_{12\ldots }+ C^b_{21\ldots} \, , \ \ \ \ \ \ 0 = C^b_{123\ldots} +C^b_{231\ldots} +C^b_{312\ldots} 
\notag \\
0 &= C^b_{1234\ldots} - C^b_{1243\ldots} + C^b_{3412\ldots} - C^b_{3421\ldots} \label{pert4.2}\\
0 &= C^b_{12345\ldots} -  C^b_{12354\ldots} -  C^b_{12453\ldots} +  C^b_{12543\ldots}
+ C^b_{45321\ldots} - C^b_{45312\ldots} \, .
\notag
\end{align}
The ellipsis in the subscript of each term indicates that lower-rank symmetries in the first labels extend to
higher rank. For instance, $C_{12}^b = f^{a_1 a_2 b} = - f^{a_2 a_1 b} = -C_{21}^b$ can be shown to
persist at any rank $p>2$ by contraction with $ f^{b a_3 c} f^{c a_4 d} \ldots f^{x a_{p-1} y} f^{y a_p z}$ which yields 
$C_{123\ldots p}^z =  -C_{213\ldots p}^z$. The generalization of the Lie symmetries (\ref{pert4.2}) to higher rank 
will be spelt out in (\ref{pert4.2LIE}) and can be checked to leave $(p{-}1)!$ independent permutations 
of $C_{123\ldots p}^b$ at rank $p$. 

We will now describe the construction of local kinematic factors for (\YMF)
with the same Lie symmetries of (\ref{pert4.2}) which imply kinematic Jacobi relations. The recursive procedure
presented in this section closely follows the steps of \cite{Mafra:2014oia}, where local superspace building
blocks with Lie symmetries have been constructed for ten-dimensional SYM.

%%%%%%%%%%%%%%%%%%%%%%%%%%%%%%%%%%%%%%%%%%%%%%%%
%%%%%%%%%%%%%%%%%%%%%%%%%%%%%%%%%%%%%%%%%%%%%%%%

\subsection{Local multiparticle polarizations up to rank three}
\label{sec:4.1}

As we already saw for the Berends--Giele currents of the previous section, each
cubic vertex of (\YMF) may introduce powers 
of $\ap^0,\ap^1$ or $\ap^2$ into the kinematic factors. 
We no longer distinguish these contributions from the individual vertices (as done by the white and black
circles in figure \ref{f:bgf3f4} and \ref{f:MABC2}) and collectively refer to all contributions at orders $\ap^{\leq2}$ 
through the off-shell diagram in figure \ref{f:localdiag}.
We will start from the numerators in the Berends--Giele recursion (\ref{pert3.7}) to construct solutions
to the kinematic Jacobi identities -- i.e.\ realizations of the Lie symmetries in (\ref{pert4.2}) -- up
to the order of $\ap^2$. 

Kinematic representatives for the diagram in figure \ref{f:localdiag} with Lie symmetries will 
be referred to as {\it multiparticle polarizations} and denoted by
lowercase parental letters $a_{12\ldots p}^\mu$,
$f_{12\ldots p}^{\mu \nu}$, $ f_{12\ldots p}^{\mu|\nu \la} $ and $g_{12\ldots p}^{\mu \nu|\la\rho}$.
This notation will help to distinguish the local multiparticle polarizations from the Berends--Giele 
currents $A_{P}^\mu$, $F_{P}^{\mu \nu}$, $ F_{P}^{\mu|\nu \la} $ and $G_{P}^{\mu \nu|\la\rho}$
with kinematic poles. In the same way as all the four species of Berends--Giele currents enter the
cyclic building block $\Mfrak_{X,Y,Z}$ in (\ref{pert3.12}), we will later on see that the analogous four species
of multiparticle polarizations can be combined to Jacobi-satisfying kinematic numerators $N_I$ in the sense of 
section \ref{sec:2.4}.

\begin{figure}[h]
\begin{center}
\begin{tikzpicture} [scale=0.8, line width=0.30mm]
\draw (-4.9,0.05) node{$a^{\mu}_{12},f^{\mu \nu}_{12}, f_{12}^{\mu| \nu \la}, g_{12}^{\mu \nu | \la \rho} \  \ \  \longleftrightarrow  $};
\draw (0,0) -- (-1,1) node[left]{$2$};
\draw (0,0) -- (-1,-1) node[left]{$1$};
\draw (0,0) -- (0.8,0);
\draw (0.5,0.3) node{$s_{12}$};
%\draw (1,0) -- (1,1) node[right]{$3$};
%\draw (1.65,-0.3) node{$s_{123}$};
\draw (1.5, 0) node{${\cdots} \  \ , $};
\end{tikzpicture}
\caption{Diagrammatic interpretation of two-particle polarizations $a^{\mu}_{12},  f^{\mu \nu}_{12}, f_{12}^{\mu| \nu \la}, g_{12}^{\mu \nu | \la \rho}$.}
\label{f:rank2}
\end{center}
\end{figure}
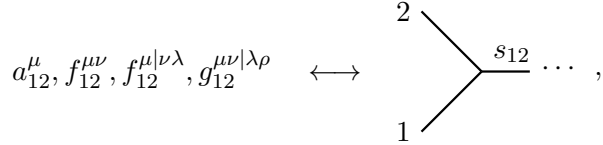

At rank one, the local multiparticle polarizations are defined to match their Berends--Giele counterparts
which include the transverse polarization vectors $e_i^\mu$ and do not exhibit any kinematic poles,
\begin{align}
a_i^\mu &= e_i^\mu = A_i^\mu \ , \ \ \ \ \ \
f_i^{\mu \nu} = k_i^\mu e_i^\nu - k_i^\nu e_i^\mu = F_i^{\mu \nu} \notag \\
f_i^{\mu|\nu \la} &= k_i^\mu f^{\nu \la}_i = F^{\mu|\nu \la}_i \ , \ \ \ \ \ \
g^{\mu\nu| \la \rho}_i=0= G^{\mu\nu| \la \rho}_i \, .
\label{pert4.3}
\end{align}
The simplest multiparticle polarization $a_{12}^\mu$ at rank two is defined by
isolating the numerator in the Berends--Giele current (\ref{pert3.1threept}),
\begin{align}
a_{12}^\mu &= \frac{1}{2 }\Big[ (k_2 {\cdot} a_{1}) a_2^\mu-(k_{1} {\cdot} a_{2}) a_{1}^\mu
 + a_{1}^\nu f_2^{\nu \mu}
 - a_{2}^\nu f_{1}^{\nu \mu} 
 + 2\ap( f^{\nu \la}_{1} f_2^{\nu|\la \mu}  -  f^{\nu \la}_{2} f_{1}^{\nu|\la \mu} )
\Big] \, ,
\label{pert4.4}
\end{align}
where the absence of contributions at order $\ap^2$ is plausible by the valence
of the Feynman vertices from $\ap^2 F^4$. The alternative presentation of (\ref{pert4.4})
as $a_{12}^\mu =s_{12}A_{12}^\mu$ generalizes to the following
two-particle polarizations at higher mass dimension,
\begin{align}
f_{12}^{\mu \nu} &=  k_{12}^\mu a_{12}^\nu - k_{12}^\nu a_{12}^\mu - (k_1\cdot k_2) (a_1^\mu a_2^\nu-a_1^\nu a_2^\mu)  =s_{12}F_{12}^{\mu \nu} \notag \\
f_{12}^{\mu| \nu \la} &=  k_{12}^\mu f_{12}^{\nu\la} - (k_1\cdot k_2)(a_1^\mu f_2^{\nu \la} - a_2^\mu f_1^{\nu \la})  =s_{12}F_{12}^{\mu| \nu \la} 
\label{pert4.5} \\
g_{12}^{\mu \nu | \la \rho} &= (k_1\cdot k_2) (f_1^{\mu \nu} f_2^{\la \rho} - f_2^{\mu \nu} f_1^{\la \rho}) = s_{12} G_{12}^{\mu \nu | \la \rho}\, .
\notag
\end{align}
The local multiparticle polarizations are still proportional to their Berends--Giele counterparts
since the latter only describe a single cubic diagram, see figure \ref{f:rank2}.
By the shuffle symmetry $A_{12}^\mu=-A_{21}^\mu$ of Berends--Giele currents or
the antisymmetry $C_{12}^b=-C_{21}^b$ of the dual color factors (\ref{pert4.1}), we have
\beq
a_{12}^\mu=-a_{21}^\mu \, , \ \ \ \ f_{12}^{\mu \nu} = - f_{21}^{\mu \nu} 
\, , \ \ \ \ f_{12}^{\mu| \nu \la} = - f_{21}^{\mu| \nu \la}
\, , \ \ \ \ g_{12}^{\mu \nu | \la \rho} = - g_{21}^{\mu \nu | \la \rho}\, .
\label{pert4.6}
\eeq

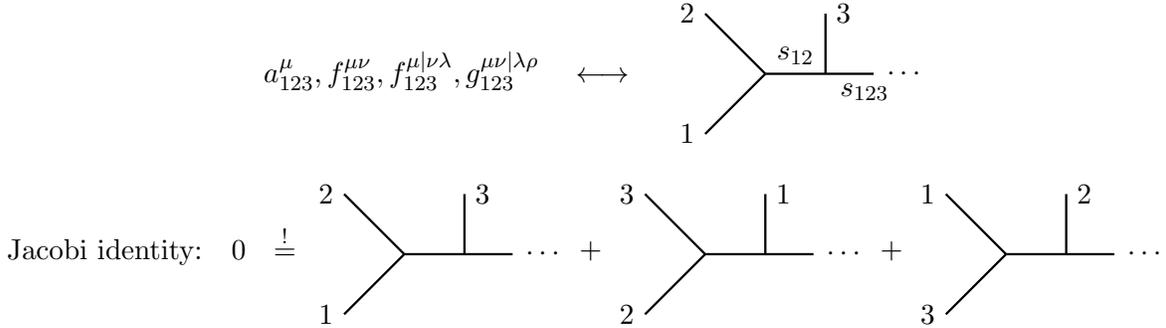
\begin{figure}[h]
\begin{center}
\begin{tikzpicture} [scale=0.8, line width=0.30mm]
\draw (-5.3,0.05) node{$a^{\mu}_{123},f^{\mu \nu}_{123}, f_{123}^{\mu| \nu \la}, g_{123}^{\mu \nu | \la \rho} \ \ \ \longleftrightarrow   $};
\draw (0,0) -- (-1,1) node[left]{$2$};
\draw (0,0) -- (-1,-1) node[left]{$1$};
\draw (0,0) -- (1.8,0);
\draw (0.5,0.3) node{$s_{12}$};
\draw (1,0) -- (1,1) node[right]{$3$};
\draw (1.65,-0.3) node{$s_{123}$};
\draw (2.3, 0) node{${\cdots}  $};
%%%%
\draw(-10.2,-2.85)node{Jacobi identity:$\ \ \ 0\ \ \stackrel{!}{=}$};
\draw(-2.9,-2.95)node{$+$};
\draw(2.1,-2.95)node{$+$};
%%%%
\scope[yshift=-3cm,xshift=-6cm]
\draw (0,0) -- (-1,1) node[left]{$2$};
\draw (0,0) -- (-1,-1) node[left]{$1$};
\draw (0,0) -- (1.8,0);
%\draw (0.5,0.3) node{$s_{12}$};
\draw (1,0) -- (1,1) node[right]{$3$};
%\draw (1.65,-0.3) node{$s_{123}$};
\draw (2.3, 0) node{${\cdots}  $};
\endscope
\scope[yshift=-3cm,xshift=-1cm]
\draw (0,0) -- (-1,1) node[left]{$3$};
\draw (0,0) -- (-1,-1) node[left]{$2$};
\draw (0,0) -- (1.8,0);
%\draw (0.5,0.3) node{$s_{12}$};
\draw (1,0) -- (1,1) node[right]{$1$};
%\draw (1.65,-0.3) node{$s_{123}$};
\draw (2.3, 0) node{${\cdots}  $};
\endscope
\scope[yshift=-3cm,xshift=4cm]
\draw (0,0) -- (-1,1) node[left]{$1$};
\draw (0,0) -- (-1,-1) node[left]{$3$};
\draw (0,0) -- (1.8,0);
%\draw (0.5,0.3) node{$s_{12}$};
\draw (1,0) -- (1,1) node[right]{$2$};
%\draw (1.65,-0.3) node{$s_{123}$};
\draw (2.3, 0) node{${\cdots}  $};
\endscope
\end{tikzpicture}
\caption{Diagrammatic interpretation of three-particle polarizations $a^{\mu}_{123},  f^{\mu \nu}_{123}, f_{123}^{\mu| \nu \la},$ and $g_{123}^{\mu \nu | \la \rho}$ subject to kinematic Jacobi relations such as $a^{\mu}_{123}+{\rm cyc}(1,2,3)=0$.}
\label{f:rankthree}
\end{center}
\end{figure}

Starting from rank three, Berends--Giele currents involve multiple cubic diagrams. 
Multiparticle polarizations for the individual diagrams can be built by isolating one of 
the two deconcatenations $(X,Y)=(12,3)$ and $(X,Y)=(1,23)$ in (\ref{pert3.7}) that 
contribute to $A_{123}^\mu$.
The numerator w.r.t.\ $s_{12}^{-1} s_{123}^{-1}$ stems from $(X,Y)=(12,3)$ and reads
\begin{align}
\widehat{a}_{123}^\mu &= \frac{1}{2 }\Big[ (k_3 \cdot a_{12}) a_3^\mu
-(k_{12} \cdot a_{3}) a_{12}^\mu
 + a_{12}^\nu f_3^{\nu \mu}
 - a_{3}^\nu f_{12}^{\nu \mu} \label{pert4.7} \\
& \ \ \ \ \ + 2\ap( f^{\nu \la}_{12} f_3^{\nu|\la \mu}  -  f^{\nu \la}_{3} f_{12}^{\nu|\la \mu}  )
 + 4\ap^2 g_{12}^{\nu \mu| \rho \sigma} f_3^{\nu|\rho \sigma}
%-g_{3}^{\nu \mu| \rho \sigma} f_{12}^{\nu|\rho \sigma})
\Big] \, ,\notag
\end{align}
where a formal antisymmetry under exchange of labels $12\leftrightarrow 3$ can be manifested
by adding $0=-2\ap^2 g_{3}^{\nu \mu| \rho \sigma} f_{12}^{\nu|\rho \sigma}$.
The Berends--Giele numerator $\widehat{a}_{123}^\mu$ should ideally share the symmetries (\ref{pert4.2}) of the 
color factor $C_{123}^b$. Indeed, antisymmetry $\widehat{a}_{123}^\mu=-\widehat{a}_{213}^\mu$ in the first two
indices is inherited from the property (\ref{pert4.6}) of the rank-two input. However, the first non-trivial kinematic Jacobi identity 
for the triplet of cubic diagrams in figure \ref{f:rankthree}
requires $\widehat{a}_{123}^\mu + \widehat{a}_{231}^\mu + \widehat{a}_{312}^\mu$ to vanish, which is not the case.
Still, the obstruction takes a special form, where one can factor out the overall momentum $k_{123}^\mu$
and isolate a scalar quantity $h_{123}$ that captures the deviation from the Lie symmetries
\beq
\widehat{a}_{123}^\mu + \widehat{a}_{231}^\mu + \widehat{a}_{312}^\mu = 3 k_{123}^\mu h_{123}\,.
\label{pert4.8}
\eeq
Amusingly, the explicit form of
\begin{align}
6h_{123} &=  \Big( \frac{1}{2} a_1^\mu f_2^{\mu \nu} a_3^\nu -  2 \ap^2 f_1^{\mu \nu} f_2^{\mu|\la \rho} f_3^{\nu|\la \rho}
+ {\rm cyc}(1,2,3) \Big)
\notag \\
& \ \ \ \ \ \ - 2\ap  f_{1}^{\mu \nu} f_2^{\nu \la} f_3^{\la \mu} + \frac{\ap}{2} \big(
f_1^{\mu| \nu \la} f_2^{\nu \la} a_3^\mu \pm {\rm perm}(1,2,3)
\big)
\label{pert4.8scal}
\\
&=   \Mfrak_{1,2,3}
\notag
\end{align}
can be reproduced from the cyclic building block of (\ref{pert3.12}).
Already the left-hand side of (\ref{pert4.8}) implies permutation-antisymmetry $h_{123}=h_{[123]}$,
so a redefinition of the Berends--Giele numerator (\ref{pert4.7}) via
\beq
a_{123}^\mu = \widehat{a}_{123}^\mu -  k_{123}^\mu h_{123}
\label{pert4.9}
\eeq
yields the desired Lie symmetries of the color factors,
\beq
a_{123}^\mu = - a_{213}^\mu  \, , \ \ \ \ \ \ a_{123}^\mu +a_{231}^\mu +a_{312}^\mu =0\, .
\label{pert4.9scal}
\eeq
As we will see, the appearance of the overall momentum $k_{123}^\mu$ in the correction (\ref{pert4.9})
to $\widehat{a}_{123}^\mu$ is essential to absorb the analogous improvements of Berends--Giele 
currents into a non-linear gauge transformation (\ref{pert1.gau}). 

Given a multiparticle polarization $a_{12\ldots p}^\mu$ at rank $p$,
the construction of its analogues $f_{12\ldots p}^{\mu \nu}$, $ f_{12\ldots p}^{\mu|\nu \la} $ and $g_{12\ldots p}^{\mu \nu|\la\rho}$ at higher mass dimension involves contact terms $\sim s_{ij}$ that preserve
the Lie symmetries. For instance, the rank-three generalizations of (\ref{pert4.5})
\begin{align}
f_{123}^{\mu \nu} &= k_{123}^\mu a_{123}^\nu - (k_{12}\cdot k_3) a_{12}^\mu a_3^\nu  -(k_1\cdot k_2) (a_{1}^\mu a_{23}^\nu - a_{2}^\mu a_{13}^\nu) - (\mu \leftrightarrow \nu)
\notag \\
f_{123}^{\mu| \nu\la } &= k_{123}^\mu f_{123}^{\nu\la} - (k_{12}\cdot k_3) (a_{12}^\mu f_3^{\nu\la} -f_{12}^{\nu \la} a_3^\mu ) 
\label{pert4.10} \\
& \ \ \ \ \ \ - (k_1\cdot k_2) (a_{1}^\mu f_{23}^{\nu \la}  - a_{23}^\mu f_{1}^{\nu \la} - a_{2}^\mu f_{13}^{\nu \la} + a_{13}^\mu f_{2}^{\nu \la})
\notag \\
g_{123}^{\mu \nu | \la \rho}&= (k_{12}\cdot k_3) (f_{12}^{\mu \nu} f_3^{\la \rho} -f_{12}^{ \la \rho} f_3^{\mu \nu}) 
 + (k_1\cdot k_2) (f_{1}^{\mu \nu}f_{23}^{ \la \rho}  - f_{23}^{\mu \nu} f_{1}^{ \la \rho} 
- f_{2}^{\mu \nu} f_{13}^{\la \rho} + f_{13}^{\mu \nu} f_{2}^{\la \rho}) 
\notag
\end{align}
are easily checked to reproduce the symmetries (\ref{pert4.9scal}) of $a_{123}^\mu$.
These contact terms are the local equivalents of the deconcatenation terms in
the Berends--Giele currents $F_P^{\mu \nu}$, $F_P^{\mu|\nu \la}$ and $G_P^{\mu \nu| \la \rho}$
in (\ref{pert3.6}) and (\ref{pert3.8}), see section \ref{sec:5.1} for more details and \cite{Mafra:2014oia} 
for superspace analogues.
While the difference between $a_{123}^\mu$ and $\widehat{a}_{123}^\mu$ in (\ref{pert4.9})
drops out from the definition (\ref{pert4.10}) of $f_{123}^{\mu \nu}$, it will be crucial at higher
rank to always build $f_{12\ldots p}^{\mu \nu} ,f_{12\ldots p}^{\mu| \nu\la } $ and 
$g^{\mu \nu | \la\rho}_{12\ldots p}$ from the redefined fields subject to Lie symmetries.

%%%%%%%%%%%%%%%%%%%%%%%%%%%%%%%%%%%%%%%%%%%%%%%%
%%%%%%%%%%%%%%%%%%%%%%%%%%%%%%%%%%%%%%%%%%%%%%%%

\subsection{Local multiparticle polarizations at rank four and five}
\label{sec:4.2}

The numerators of the Berends--Giele recursion (\ref{pert3.7}) serve as a starting point
to construct higher-rank multiparticle polarizations for the diagrams in figure \ref{f:localdiag} 
that satisfy the Lie symmetries (\ref{pert4.2}) of the dual color factors. The higher-rank systematics 
of the redefinition in (\ref{pert4.9}) is most conveniently illustrated via examples at $p=4,5$.

Given the multiparticle polarizations at rank three in (\ref{pert4.9}) and (\ref{pert4.10}),
their Lie symmetries imply that the rank-four object
\begin{align}
\widehat{a}_{1234}^\mu &= \frac{1}{2 }\Big[ (k_4 \cdot a_{123}) a_4^\mu
-(k_{123} \cdot a_{4}) a_{123}^\mu
 + a_{123}^\nu f_4^{\nu \mu}
 - a_{4}^\nu f_{123}^{\nu \mu} \notag \\
& \ \ \ \ \   + 2\ap( f^{\nu \la}_{123} f_4^{\nu|\la \mu}  -  f^{\nu \la}_{4} f_{123}^{\nu|\la \mu} )
 + 4\ap^2 g_{123}^{\nu \mu| \rho \sigma} f_4^{\nu|\rho \sigma}
\Big]  \label{pert4.11}
\end{align}
obeys $\widehat a_{1234}^\mu = - \widehat a_{2134}^\mu$ and $\widehat a_{1234}^\mu +\widehat a_{2314}^\mu
 +\widehat a_{3124}^\mu =0$.  However, the Lie symmetry at rank four is not yet satisfied by (\ref{pert4.11}), and in 
contrast to (\ref{pert4.8}), it is {\it not} possible to factorize $k_{1234}^\mu$ from 
$\widehat a^\mu_{1234} - \widehat a^\mu_{1243} + \widehat a^\mu_{3412} - \widehat a^\mu_{3421} $. Instead, we 
will need redefinitions $\widehat{a}_{1234}^\mu  \rightarrow a'{}^{\mu}_{\! \! 1234}  \rightarrow a_{1234}^\mu $ in two 
steps, where an intermediate object $a'{}^{\mu}_{\! \! 1234} $ is built from permutations of the scalar $h_{ijk}$ in the 
rank-three redefinition, see (\ref{pert4.8scal}),
\begin{align}
a'{}^{\mu}_{\! \! 1234} &= \widehat{a}_{1234}^\mu - (k_{12}\cdot k_3) a_3^\mu h_{124} - (k_{1}\cdot k_2)(a_2^\mu h_{134} - a_1^\mu h_{234}) \, . \label{pert4.12} 
\end{align}
The pattern of subtractions in (\ref{pert4.12}) has been inferred by mimicking BRST transformations in ten-dimensional
pure-spinor superspace \cite{Mafra:2014oia}, and it should be possible to give a similar motivation
from a study of linearized gauge variations. The key benefit of the redefinition in (\ref{pert4.12}) is that the deviation from
the rank-four Lie symmetry now takes a factorized form
\begin{align}
a'{}^{\mu}_{\! \! 1234}  - a'{}^{\mu}_{\! \! 1243} + a'{}^{\mu}_{\! \! 3412} - a'{}^{\mu}_{\! \! 3421}&= 4k_{1234}^\mu h_{1234} \, ,\label{pert4.13} 
\end{align}
see (\ref{pert4.23}) for convenient representations of the scalar $h_{1234}$. The left-hand side of (\ref{pert4.13})
along with $a'{}^{\mu}_{\! \! 1234}=-a'{}^{\mu}_{\! \! 2134}$ and $a'{}^{\mu}_{\! \! 1234}+a'{}^{\mu}_{\! \! 2314}
+a'{}^{\mu}_{\! \! 3124}=0$ imply the symmetries $h_{1234}= -h_{2134}=h_{3412}= - h_{3421}$ 
and $h_{1234}+h_{2314}+h_{3124}=0$. Like this, the redefinition
\begin{align}
a^{\mu}_{1234} = a'{}^{\mu}_{\! \! 1234} - k_{1234}^\mu h_{1234} \label{pert4.14} 
\end{align}
leads to the desired Lie symmetries
\begin{align}
a^{\mu}_{1234} = - a^{\mu}_{1234}  \, , \ \ \ \
a^{\mu}_{1234} + a^{\mu}_{2314} +a^{\mu}_{3124} =0 \, , \ \ \ \ 
a^\mu_{1234} -  a^\mu_{1243} + a^\mu_{3412} -  a^\mu_{3421}=0 \, .\label{pert4.15} 
\end{align}
This final form of the multiparticle polarization $a^{\mu}_{1234}$ can be used to construct 
its counterparts at higher mass dimensions, where the Lie-symmetry preserving contact 
terms in (\ref{pert4.10}) generalize to\footnote{As an example for the antisymmetrization 
prescriptions in (\ref{pert4.16}), the subtraction of $(a^\mu_P f^{\nu \la}_Q \leftrightarrow a^\mu_Q f^{\nu \la}_P )$ 
extends the term $a_{123}^\mu f^{\nu \la}_{4}$ to 
the combination $a_{123}^\mu f^{\nu \la}_{4} - a_{4}^\mu f^{\nu \la}_{123}$.}
\begin{align}
f_{1234}^{\mu \nu} &= k_{1234}^\mu a_{1234}^\nu - (k_{123}\cdot k_4)a_{123}^\mu a_4^\nu   - (k_{12}\cdot k_3) (a_{12}^\mu a_{34}^\nu + a_{124}^\mu a_{3}^\nu)\notag\\
& \ \ \ - (k_{1}\cdot k_2) (a_{13}^\mu a_{24}^\nu + a_{14}^\mu a_{23}^\nu + a_{134}^\mu a_{2}^\nu - a_{234}^\mu a_{1}^\nu) - (\mu \leftrightarrow \nu) \notag \\
%%%%
f_{1234}^{\mu| \nu \la} &= k_{1234}^\mu f_{1234}^{\nu\la} - \big[ (k_{123}\cdot k_4) a_{123}^\mu f_4^{\nu \la}   + (k_{12}\cdot k_3) (a_{12}^\mu f_{34}^{\nu \la}  + a_{124}^\mu f_{3}^{\nu\la} ) \label{pert4.16}  \\
& \ \ \ + (k_{1}\cdot k_2)(a_{13}^\mu f_{24}^{\nu \la}  
+ a_{14}^\mu f_{23}^{\nu \la} 
 + a_{134}^\mu f_{2}^{\nu \la}   
 - a_{234}^\mu f_{1}^{\nu \la} )  - (a^\mu_P f^{\nu \la}_Q \leftrightarrow a^\mu_Q f^{\nu \la}_P ) \big] \notag \\
 %%%%
g_{1234}^{\mu \nu| \la\rho } &= (k_{123}\cdot k_4)f_{123}^{\mu \nu} f_4^{\la \rho}  + (k_{12}\cdot k_3)(f_{12}^{\mu \nu} f_{34}^{\la \rho}  + f_{124}^{\mu \nu} f_{3}^{\la \rho} )\notag\\
& \ \ \ + (k_{1}\cdot k_2) (f_{13}^{\mu \nu} f_{24}^{\la \rho} 
+ f_{14}^{\mu \nu} f_{23}^{\la \rho} 
 + f_{134}^{\mu \nu} f_{2}^{\la \rho}  
 - f_{234}^{\mu \nu} f_{1}^{\la \rho} )  - (f^{\mu\nu}_P f^{\la \rho}_Q \leftrightarrow f^{\mu\nu}_Q f^{\la \rho}_P ) \, .\notag 
\end{align}
At higher rank, analogous redefinitions in two steps $\widehat{a}_{123\ldots p}^\mu  \rightarrow a'{}^{\mu}_{\! \! 123\ldots p}  
\rightarrow a_{123\ldots p}^\mu $ will be sufficient to attain Lie symmetries, i.e.\ there are no additional intermediate
steps at $p>4$. For instance, the Lie-symmetry satisfying multiparticle
polarizations (\ref{pert4.14}) and (\ref{pert4.16}) at rank four can be used to recursively construct a rank-five
quantity 
\begin{align}
\widehat{a}_{12345}^\mu &= \frac{1}{2 }\Big[ (k_5 \cdot a_{1234}) a_5^\mu
-(k_{1234} \cdot a_{5}) a_{1234}^\mu
 + a_{1234}^\nu f_5^{\nu \mu}
 - a_{5}^\nu f_{1234}^{\nu \mu} \notag \\
&\ \ \ \ \  + 2\ap( f^{\nu \la}_{1234} f_5^{\nu|\la \mu}  -  f^{\nu \la}_{5} f_{1234}^{\nu|\la \mu}  )
 + 4\ap^2 g_{1234}^{\nu \mu| \rho \sigma} f_5^{\nu|\rho \sigma}
\Big]  \label{pert4.17}
\end{align}
subject to the symmetries (\ref{pert4.15}) in its first four labels. The rank-five Lie symmetry
can be enforced by first performing subtractions analogous to (\ref{pert4.12}) and \cite{Mafra:2014oia},
\begin{align}
a'{}^{\mu}_{\! \! 12345} &= \widehat{a}_{12345}^\mu -  (k_{123}\cdot k_4)  a_4^\mu h_{1235} -  (k_{12}\cdot k_3)  (a_3^\mu h_{1245} + a_{34}^\mu h_{125} - a_{12}^\mu h_{345})
\notag \\
&\ \ -  (k_{1}\cdot k_2) (a_2^\mu h_{1345} + a_{23}^\mu h_{145} + a_{24}^\mu h_{135}
- a_1^\mu h_{2345} - a_{13}^\mu h_{245} - a_{14}^\mu h_{235}) \, , \label{pert4.18} 
\end{align}
and then identifying a rank-five scalar $h_{12345}$ along the lines of (\ref{pert4.13}),
\begin{align}
a'{}^{\mu}_{\! \! 12345}  - a'{}^{\mu}_{\! \! 12354} 
+ a'{}^{\mu}_{\! \! 45123} - a'{}^{\mu}_{\! \! 45213}
- a'{}^{\mu}_{\! \! 45312} + a'{}^{\mu}_{\! \! 45321}
&= 5 k_{12345}^\mu h_{12345} \, . \label{pert4.19} 
\end{align}
Note that $a'{}^{\mu}_{\! \! 12345} $ only satisfies Lie symmetries in its first four labels, as 
one can check via $h_{1234}=-h_{2134}$ and  $h_{1234}+h_{2314}+h_{3124}=0$ as well as $h_{123}=h_{[123]}$.
This endows the resulting $h_{12345}$ on the right-hand side with the same Lie symmetries in its first four legs
and an additional reflection property $h_{12345}+h_{45312}=0$. On these grounds, the redefinition
\begin{align}
a^{\mu}_{12345} = a'{}^{\mu}_{\! \! 12345} - k_{12345}^\mu h_{12345} \label{pert4.20} 
\end{align}
leads to the desired Lie symmetries in all the five labels
\begin{align}
&a^{\mu}_{12345} = - a^{\mu}_{12345}  \, , \ \ \ \
a^{\mu}_{12345} + a^{\mu}_{23145} +a^{\mu}_{31245} =0 \, , \ \ \ \ 
a^\mu_{12345} -  a^\mu_{12435} + a^\mu_{34125} -  a^\mu_{34215}=0 \notag \\
&\ \ \ \ \ \ a^{\mu}_{12345}  - a^{\mu}_{12354} 
+ a^{\mu}_{ 45123} - a^{\mu}_{ 45213}
- a^{\mu}_{ 45312} + a^{\mu}_{ 45321} =0 \, .\label{pert4.21} 
\end{align}
The remaining multiparticle polarizations are given by
\begin{align}
f_{12345}^{\mu\nu} &= k_{12345}^\mu a^{\nu}_{12345} - (k_{1234} \cdot k_5) a_{1234}^\mu a_5^\nu - (k_{123} \cdot k_4 )(a_{1235}^\mu a_4^\mu + a_{123}^\mu a_{45}^\nu) \notag \\
&- (k_{12}\cdot k_3)(a_{1245}^\mu a_3^\nu + a_{124}^\mu a_{35}^\nu + a_{125}^\mu a_{34}^\nu + a_{12}^\mu a_{345}^\nu) \notag \\
&- (k_1\cdot k_2)( a_{1345}^\mu a_2^\nu + a_{134}^\mu a_{25}^\nu + a_{135}^\mu a_{24}^\nu + a_{145}^\mu a_{23}^\nu \label{pert4.22} \\
& \ \ \ \ \ \ + a_{13}^\mu a_{245}^\nu + a_{14}^\mu a_{235}^\nu + a_{15}^\mu a_{234}^\nu + a_1^\mu a_{2345}^\nu)
-(\mu \leftrightarrow \nu)\, , \notag
\end{align}
and similar expressions for $f_{12345}^{\mu|\nu \la}$ and $g_{12345}^{\mu \nu | \la\rho}$ can be
inferred by analogy with (\ref{pert4.16}) or from the all-rank formula (\ref{pert4.gh}).

We emphasize that the expressions for the scalars $h_{12\ldots p}$ result from a fully
constructive procedure, i.e.\ they can be read off from (\ref{pert4.8}), (\ref{pert4.13}) and (\ref{pert4.19}) after factoring
out $k_{12\ldots p}^\mu$. Similar to (\ref{pert4.8scal}), one can use the cyclic building block (\ref{pert3.12}) to rewrite
\begin{align}
h_{1234} &= \frac{1}{24} (2 s_{12} \Mfrak_{12,3,4} +  s_{13} \Mfrak_{13,2,4} -  s_{14} \Mfrak_{14,2,3}  
- s_{23} \Mfrak_{23,1,4} +  s_{24} \Mfrak_{24,1,3} +  2s_{34} \Mfrak_{34,1,2} )\notag \\
&= + \frac{1}{48} \big( (k_{123}\cdot a_4) \Mfrak_{1,2,3}
- (k_{234}\cdot a_1) \Mfrak_{2,3,4} + ( k_{134}\cdot a_2) \Mfrak_{1,3,4} - ( k_{124}\cdot a_3) \Mfrak_{1,2,4} \big) \notag \\
& \ \ \ \ +\frac{ 1}{8}( s_{12} \Mfrak_{12,3,4} + s_{34} \Mfrak_{34,1,2} )
\label{pert4.23}
\end{align}
and similar expressions for $h_{12345}$ are spelt out in appendix \ref{app:D.1}.

%%%%%%%%%%%%%%%%%%%%%%%%%%%%%%%%%%%%%%%%%%%%%%%%
%%%%%%%%%%%%%%%%%%%%%%%%%%%%%%%%%%%%%%%%%%%%%%%%

\subsection{Local multiparticle polarizations at higher rank}
\label{sec:4.3}

The recursive construction of multiparticle polarizations will now be summarized in terms
of all-rank formulae that closely follow their superspace antecedents \cite{Mafra:2014oia}
but incorporate $\ap$-corrections. The Lie symmetries of $a_{12\ldots q}^\mu$,
$f_{12\ldots q}^{\mu \nu}$, $ f_{12\ldots q}^{\mu|\nu \la} $ and $g_{12\ldots q}^{\mu \nu|\la\rho}$ at lower rank
$q=p{-}1$ propagate to the first $p{-}1$ labels of the following quantity:
\begin{align}
\widehat{a}_{12\ldots p}^\mu &= \frac{1}{2 }\Big[ (k_p \cdot a_{12\ldots p-1}) a_p^\mu
-(k_{12\ldots p-1} \cdot a_{p}) a_{12\ldots p-1}^\mu
 + a_{12\ldots p-1}^\nu f_p^{\nu \mu}
 - a_{p}^\nu f_{12\ldots p-1}^{\nu \mu} \notag \\
& \ \ \ \ \  + 2\ap( f^{\nu \la}_{12\ldots p-1} f_p^{\nu|\la \mu} -  f^{\nu \la}_{p} f_{12\ldots p-1}^{\nu|\la \mu} )
 + 4\ap^2 g_{12\ldots p-1}^{\nu \mu| \rho \sigma} f_p^{\nu|\rho \sigma}
\Big]  \label{pert4.ab}\, .
\end{align}
When reinstating the vanishing term $0=-2\ap^2g_{p}^{\nu \mu| \rho \sigma} f_{12\ldots p-1}^{\nu|\rho \sigma}$,
this expression exhibits formal antisymmetry under exchange of labels $12\ldots p{-}1\leftrightarrow p$. 
In order to isolate the deviations from the Lie symmetries at rank $p$, one first has to subtract\footnote{OS is
grateful to Carlos Mafra for identifying the property $j{+}1,j{+}2 \ldots p{-}1 =X\shuffle Y$ of the words $X,Y$ in
the second sum of (\ref{pert4.cd}).}
\begin{align}
a'{}^{\mu}_{\! \! 12\ldots p} &= \widehat{a}_{12\ldots p}^\mu - \sum_{j=2}^{p-1} (k_{12\ldots j-1} \cdot k_j) 
\sum_{j+1,j+2 \ldots p-1 \atop{=X\shuffle Y} } \big( h_{12\ldots (j-1)Yp} a_{jX}^\mu -  h_{jYp} a_{12\ldots (j-1) X}^\mu \big)  \label{pert4.cd} \, ,
\end{align}
with $h_{i}=h_{ij}=0$ as well as $h_{ijk}$ and $h_{ijkl}$ defined in (\ref{pert4.8scal}) and (\ref{pert4.23}), respectively.
These subtractions vanish at rank $p\leq 3$, and their instances at $p=4,5$ are spelt out in (\ref{pert4.12}) and (\ref{pert4.18}), 
respectively. These equations might be helpful to see an example of the summation prescription of the form
$a_1 a_2\ldots a_k =X\shuffle Y$ in (\ref{pert4.cd}): For a given $k$-particle label $a_1 a_2\ldots a_k$, the 
sum runs over all the $2^k$ pairs of words $X$ and $Y$ whose shuffle product contains $a_1 a_2\ldots a_k$,
for instance\footnote{As a rank-two example of the above summation prescription, $a_1 a_2 =X\shuffle Y$
allows for the four choices of $(X,Y)$, namely $(a_1a_2 ,\emptyset),(\emptyset,a_1a_2),(a_1 ,a_2),(a_2 ,a_1)$.} $(X,Y)=(\emptyset,\emptyset)$ in case of $k=0$ as well as $(X,Y)=(a_1,\emptyset)$ 
and $(X,Y)=(\emptyset,a_1)$ in case of $k=1$.

The outcome of (\ref{pert4.cd}) still obeys the Lie symmetries in the first $p{-}1$ labels
and is claimed to factorize $k_{12\ldots p}^\mu$ when probing the rank-$p$ Lie symmetry:
\begin{align}
\begin{array}{cl}
a'{}^{\mu}_{\! \! 12\ldots n+1[n+2[\ldots  [2n-1[2n,2n+1]] \ldots ]]}
- a'{}^{\mu}_{\! \! 2n+1,2n,\ldots n+2[n+1[\ldots  [3[21]] \ldots ]]} &:\, p=2n{+}1 \ \te{odd} \\
a'{}^{\mu}_{\! \! 12\ldots n[n+1[\ldots  [2n-2[2n-1,2n]] \ldots ]]}
+ a'{}^{\mu}_{\! \! 2n,2n-1,\ldots n+1[n[\ldots  [3[21]] \ldots ]]} &: \, p=2n \ \te{even}
\end{array} 
\bigg\} &= p k_{12\ldots p}^\mu h_{12\ldots p}
\label{pert4.2LIE} \, .
\end{align}
The symmetries of the scalar $h_{12\ldots p}$ induced by the left-hand side ensure that the final form
\begin{align}
a^{\mu}_{12\ldots p} &= a'{}^{\mu}_{\! \! 12\ldots p}  - k_{12\ldots p}^\mu h_{12\ldots p} \label{pert4.ef} 
\end{align}
of the multiparticle polarizations obeys all the Lie symmetries of the color factor (\ref{pert4.1}),
\begin{align}
0&= \bigg\{
\begin{array}{cl}
a^{\mu}_{12\ldots n+1[n+2[\ldots  [2n-1[2n,2n+1]] \ldots ]]}
- a^{\mu}_{ 2n+1,2n,\ldots n+2[n+1[\ldots  [3[21]] \ldots ]]} &:\, p=2n{+}1 \ \te{odd} \\
a^{\mu}_{12\ldots n[n+1[\ldots  [2n-2[2n-1,2n]] \ldots ]]}
+ a^{\mu}_{ 2n,2n-1,\ldots n+1[n[\ldots  [3[21]] \ldots ]]} &: \, p=2n \ \te{even}
\end{array} 
\label{pert4.2LIEfinal} \, .
\end{align}
Hence, when interpreted as the kinematic numerator of the off-shell diagram in figure \ref{f:localdiag},
the multiparticle polarization $a^{\mu}_{12\ldots p} $ in (\ref{pert4.ef}) obeys kinematic Jacobi identities. 

The remaining multiparticle polarizations $f_{12\ldots p}^{\mu \nu}$, $ f_{12\ldots p}^{\mu|\nu \la} $ 
and $g_{12\ldots p}^{\mu \nu|\la\rho}$ of higher mass dimension are obtained by the following generalization of (\ref{pert4.10}),
(\ref{pert4.16}) and (\ref{pert4.22}) 
\begin{align}
f_{12\ldots p}^{\mu \nu} &= k_{12\ldots p}^\mu a_{12\ldots p}^\nu - 
\sum_{j=2}^p(k_{12\ldots j-1} \cdot k_j) \sum_{j{+}1,j{+}2\ldots p \atop{= X\shuffle Y}} a^\mu_{12\ldots j-1X} a^\nu_{jY}
-(\mu \leftrightarrow \nu )\notag \\
f_{12\ldots p}^{\mu|\nu \la}  &= k_{12\ldots p}^\mu f^{\nu \la}_{12\ldots p} - \Big[
\sum_{j=2}^p(k_{12\ldots j-1} \cdot k_j) \sum_{j{+}1,j{+}2\ldots p \atop{= X\shuffle Y}} a^\mu_{12\ldots j-1X} f^{\nu \la}_{jY}
- (a^\mu_P f^{\nu \la}_Q \leftrightarrow a^\mu_Q f^{\nu \la}_P) \Big]  \label{pert4.gh}
\\
g_{12\ldots p}^{\mu \nu|\la\rho} &=\sum_{j=2}^p(k_{12\ldots j-1} \cdot k_j) \sum_{j{+}1,j{+}2\ldots p \atop{= X\shuffle Y}} f^{\mu \nu}_{12\ldots j-1X} f^{\la \rho}_{jY}  - (f^{\mu\nu}_P f^{\la \rho}_Q \leftrightarrow f^{\mu\nu}_Q f^{\la \rho}_P) 
\, , \notag
\end{align}
see the explanation above for the summation prescription $j{+}1,j{+}2\ldots p=X\shuffle Y$.
The pattern of contact terms on the right-hand side preserves the Lie symmetries in all the $p$ labels
and will be connected with Berends--Giele currents in section \ref{sec:5.1}.
In the next section, these four Jacobi-satisfying kinematic representatives (\ref{pert4.ef}) and (\ref{pert4.gh}) 
of the off-shell diagram in figure \ref{f:localdiag} will be combined to on-shell numerators of (\YMF).

%%%%%%%%%%%%%%%%%%%%%%%%%%%%%%%%%%%%%%%%%%%%%%%%
%%%%%%%%%%%%%%%%%%%%%%%%%%%%%%%%%%%%%%%%%%%%%%%%
%%%%%%%%%%%%%%%%%%%%%%%%%%%%%%%%%%%%%%%%%%%%%%%%
%%%%%%%%%%%%%%%%%%%%%%%%%%%%%%%%%%%%%%%%%%%%%%%%
%%%%%%%%%%%%%%%%%%%%%%%%%%%%%%%%%%%%%%%%%%%%%%%%

\section{BCJ gauge and BCJ numerators for (\YMF)}
\label{sec:5}

%%%%%%%%%%%%%%%%%%%%%%%%%%%%%%%%%%%%%%%%%%%%%%%%
%%%%%%%%%%%%%%%%%%%%%%%%%%%%%%%%%%%%%%%%%%%%%%%%

\subsection{Berends--Giele currents in BCJ gauge}
\label{sec:5.1}

In this section, we relate the Jacobi-satisfying numerators for cubic off-shell diagrams 
as constructed in the previous section to gauge-transformed Berends--Giele currents.
The idea is to compare the Lorenz-gauge currents $A_P^\mu,F_P^{\mu \nu},F_P^{\mu|\nu\la}$
and $G_P^{\mu \nu|\la\rho}$ of section \ref{sec:3.1} with alternative currents 
$A_P^{\mu, \, {\rm BCJ}},\ldots ,G_P^{\mu \nu|\la\rho, \, {\rm BCJ}}$ obtained from
multiparticle polarizations. More precisely, these alternative currents are defined by
combining cubic diagrams in the usual color-ordered manner and dressing them with
multiparticle polarizations and propagators. The simplest examples are $A^{\mu, \, {\rm BCJ}}_{1}=e_1^\mu$ 
as well as
\begin{align}
A^{\mu, \, {\rm BCJ}}_{12} &= \frac{ a_{12}^{\mu} }{s_{12}} \ , \ \ \ \ \ \ A^{\mu, \, {\rm BCJ}}_{123} = \frac{ a_{123}^{\mu} }{s_{12}s_{123}} + \frac{ a_{321}^{\mu} }{s_{23}s_{123}} \label{xyzabc}
\\
A_{1234}^{\mu, \, {\rm BCJ}} &= \frac{1}{s_{1234}} \Big\{ \frac{ a_{1234}^\mu }{s_{12} s_{123}} 
+ \frac{ a_{3214}^\mu }{s_{23} s_{123}} + \frac{ a_{1234}^\mu - a_{1243}^\mu }{s_{12} s_{34}} 
- \frac{ a_{4321}^\mu }{s_{34} s_{234}} - \frac{ a_{2341}^\mu }{s_{23} s_{234}}  \Big\} \, , \notag
\end{align}
and similar definitions apply to $F_{\ldots}^{\mu \nu, \, {\rm BCJ}}, F_{\ldots}^{\mu|  \nu\la, \, {\rm BCJ}},
G_{\ldots}^{\mu \nu|\la\rho, \, {\rm BCJ}}$ with $a^\mu_{\ldots} \rightarrow f^{\mu \nu}_{\ldots}, 
f^{\mu| \nu \la}_{\ldots}, g^{\mu \nu|\la\rho}_{\ldots}$ on the right-hand side of (\ref{xyzabc}).
The rank-two currents are degenerate with $A^{\mu, \, {\rm BCJ}}_{12} =A^{\mu}_{12} $, while the
redefinitions of numerators $a^\mu_{12\ldots p}$ at ranks $p\geq 3$ by $h_{12\ldots p}$ introduce differences 
between $A_P^\mu$ from the recursion (\ref{pert3.7}) and the alternative currents in (\ref{xyzabc}). 

\begin{figure}[h]
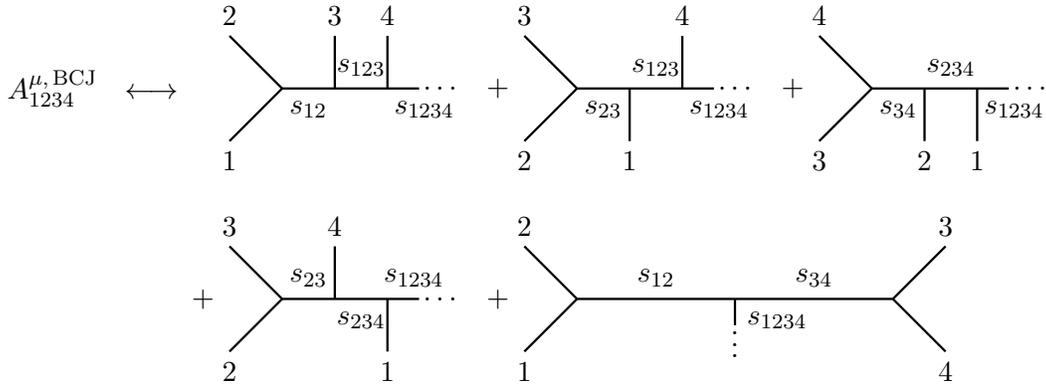

\begin{center}
\tikzpicture [scale=1.4,line width=0.30mm]
\draw (-1.8,2) node {$A_{1234}^{\mu, \, {\rm BCJ}}  \ \ \longleftrightarrow $};
\scope[yshift=2cm]
\draw (0,0) -- (-0.5,0.5) node[above]{$2$};
\draw (0,0) -- (-0.5,-0.5) node[below]{$1$};
%\draw (-0.75,0) node{$+$};
\draw (0,0) -- (1.3,0);
\draw (0.25,-0.2) node{$s_{12}$};
\draw (0.5,0) -- (0.5,0.5) node[above]{$3$};
\draw (0.75,0.2) node{$s_{123}$};
\draw (1,0) -- (1,0.5) node[above]{$4$};
\draw (1.35,-0.2) node{$s_{1234}$};
\draw (1.5, 0) node{$\ldots  $};
%\draw (2.5, 0) node{ $=\displaystyle {T_{1234}\over s_{12}s_{123}s_{1234}}$};
\endscope
\scope[xshift=2.8cm]
\draw (0,2) -- (-0.5,2.5) node[above]{$3$};
\draw (0,2) -- (-0.5,1.5) node[below]{$2$};
\draw (-0.75,2) node{$+$};
\draw (0,2) -- (1.3,2);
\draw (0.25,1.8) node{$s_{23}$};
\draw (0.5,2) -- (0.5,1.5) node[below]{$1$};
\draw (0.75,2.2) node{$s_{123}$};
\draw (1,2) -- (1,2.5) node[above]{$4$};
\draw (1.35,1.8) node{$s_{1234}$};
\draw (1.5,2) node{$\ldots  $};
%\draw (2.6, 2) node{ $=\displaystyle {T_{3214}\over s_{23}s_{123}s_{1234}}$};
\endscope
\scope[xshift=1.4cm]
\draw (4.2, 2) -- (3.7, 2.5) node[above]{$4$};
\draw (4.2, 2) -- (3.7, 1.5) node[below]{$3$};
\draw (3.45,2) node{$+$};
\draw (4.2, 2) -- (5.5, 2);
\draw (4.45, 1.8) node{$s_{34}$};
\draw (4.7, 2) -- (4.7, 1.5) node[below]{$2$};
\draw (4.95, 2.2) node{$s_{234}$};
\draw (5.2, 2) -- (5.2, 1.5) node[below]{$1$};
\draw (5.55, 1.8) node{$s_{1234}$};
\draw (5.7, 2) node{$\ldots  $};
%\draw (6.7, 2) node{$=\displaystyle {T_{3421} \over s_{34}s_{234}s_{1234}}$};
\endscope
\scope[xshift=-4.2cm]
\draw (4.2, 0) -- (3.7, 0.5) node[above]{$3$};
\draw (4.2, 0) -- (3.7, -0.5) node[below]{$2$};
\draw (3.45,0) node{$+$};
\draw (4.2, 0) -- (5.5, 0);
\draw (4.45,0.2) node{$s_{23}$};
\draw (4.7, 0) -- (4.7, 0.5) node[above]{$4$};
\draw (4.95, -0.2) node{$s_{234}$};
\draw (5.2, 0) -- (5.2, -0.5) node[below]{$1$};
\draw (5.45, 0.2) node{$s_{1234}$};
\draw (5.7, 0) node{$\ldots  $};
%\draw (6.8, 0) node{$=\displaystyle {T_{3241} \over s_{23}s_{234}s_{1234}}$};
\endscope
\scope[xshift=1.8cm,yshift=1.5cm]
  \draw (1, -1.5) -- (4.0, -1.5);
  \draw (1, -1.5) -- (0.5, -1.0) node[above]{$2$};
  \draw (1, -1.5) -- (0.5, -2.0) node[below]{$1$};
  \draw (0.25,-1.5) node{$+$};
  \draw (4.0, -1.5) -- (4.5, -1.0) node[above]{$3$};
  \draw (4.0, -1.5) -- (4.5, -2.0) node[below]{$4$};
  %off-shell leg
  \draw (2.5, -1.5) -- (2.5, -1.75);
  \draw (2.5, -1.85) node{$\vdots$};
  \draw (1.75, -1.3) node{$s_{12}$};
  \draw (3.25, -1.3) node{$s_{34}$};
  \draw (2.9, -1.7) node{$s_{1234}$};
 % \draw (5.2, -1.5) node{$= \displaystyle {2T_{12[34]}\over s_{12}s_{34}s_{1234}}$};
\endscope
\endtikzpicture
\end{center}
\caption{Cubic-diagram expansion of the rank-four Berends--Giele current $A_{1234}^{\mu, \, {\rm BCJ}}$
built from multiparticle polarizations.}
\label{fig:M1to4}
\end{figure}

At rank three, there are two cubic diagrams contributing to 
$A^{\mu, \, {\rm BCJ}}_{123}$ after dropping the distinction between order-$\ap^0,\ap^1,\ap^2$ vertices in
figure \ref{f:bgf3f4}, and the five cubic diagrams at rank four are depicted in figure \ref{fig:M1to4}.
The numerator for the last cubic diagram of $A^{\mu, \, {\rm BCJ}}_{1234}$ in the figure with 
propagators $(s_{12}s_{34} s_{1234})^{-1}$ is defined
to be $a_{1234}^\mu - a_{1243}^\mu$ by its relation to half-ladder numerators via Jacobi identities\footnote{The 
Lie symmetry $a_{1234}^\mu - a_{1243}^\mu = - (a_{3412}^\mu - a_{3421}^\mu)$ ensures that 
this numerator changes sign when trading $12\leftrightarrow 34$ by a flip of the central cubic 
vertex and therefore obeys (\ref{dualityA}).}. 

More generally, each cubic diagram contributing to $A_{12\ldots p}^{\mu, \, {\rm BCJ}}$ can be derived from the
half-ladder topology via kinematic Jacobi relations. The half-ladder numerators at rank $p$ in turn can be expanded
in the $(p{-}1)!$-element basis $\{a^{\mu}_{1\rho(23\ldots p)}, \, \rho \in S_{p-1} \}$ by their Lie symmetries
(\ref{pert4.2LIEfinal}). As already noted in a superspace context \cite{Zfunctions, Mafra:2014oia},
currents and master numerators are related by the inverse of the $(p{-}1)!\times (p{-}1)!$ KLT matrix in (\ref{genpar25}),
\begin{align}
A^{\mu, \, {\rm BCJ}}_{1\rho(23\ldots p)} &= \!  \sum_{\sigma \in S_{p-1}} \Phi(\rho|\sigma)_1  \, a_{1\sigma(23\ldots p)}^\mu \, , \ \ \ \ \ \ \ \; \!
F^{\mu \nu, \, {\rm BCJ}}_{1\rho(23\ldots p)} = \! \sum_{\sigma \in S_{p-1}}  \! \Phi (\rho|\sigma)_1  \, f_{1\sigma(23\ldots p)}^{\mu \nu} \label{pert4.ij} \\
F^{\mu|\nu \la, \, {\rm BCJ}}_{1\rho(23\ldots p)} &= \!  \sum_{\sigma \in S_{p-1}} \Phi(\rho|\sigma)_1  \, f_{1\sigma(23\ldots p)}^{\mu|\nu \la} \, , \ \ \ \ \ 
G^{\mu \nu|\la \rho, \, {\rm BCJ}}_{1\rho(23\ldots p)} = \! \sum_{\sigma \in S_{p-1}}  \! \Phi (\rho|\sigma)_1  \, g_{1\sigma(23\ldots p)}^{\mu \nu|\la \rho} \, .
\notag
\end{align}
The cubic-graph expansion endows the alternative currents 
$A_P^{\mu \, {\rm BCJ}},\ldots ,G_P^{\mu \nu|\la\rho, \, {\rm BCJ}}$ with the
same shuffle relations  (\ref{pert3.shuffle}) as the Lorenz-gauge currents
$A_P^{\mu},\ldots ,G_P^{\mu \nu|\la\rho}$. Hence, the former also descend from
Lie-algebra valued perturbiners \cite{Reeshuffle} such as
\beq
\AAA^{\mu, \, {\rm BCJ}} =  \sum_{P \neq \emptyset} A_P^{\mu, \, {\rm BCJ}} t^P e^{k_P\cdot x}
\ , \ \ \ \ \ \
\FF^{\mu \nu, \, {\rm BCJ}} = \sum_{P \neq \emptyset} F_P^{\mu \nu, \, {\rm BCJ}} t^P e^{k_P\cdot x} \, .
\label{pert4.kl}
\eeq
By direct comparison of the currents $A_P^{\mu},F_P^{\mu \nu}$ and $A_P^{\mu, \, {\rm BCJ}},F_P^{\mu \nu, \, {\rm BCJ}}$, 
the redefinitions of the multiparticle polarizations via $h_{12\ldots p}$ conspire to shuffle symmetric 
scalars $H_{12\ldots p}$,
\begin{align}
A_{123}^{\mu, \, {\rm BCJ}} &= A_{123}^{\mu} + k_{123}^\mu H_{123} \, , \ \ \ \ \ \ F_{123}^{\mu \nu, \, {\rm BCJ}} = F_{123}^{\mu \nu} \notag\\
A_{1234}^{\mu, \, {\rm BCJ}} &= A_{1234}^{\mu} - A_1^\mu H_{234} + H_{123} A_4^\mu +k_{1234}^\mu H_{1234} \label{pert4.mn} \\
F_{1234}^{\mu \nu, \, {\rm BCJ}} &= F_{1234}^{\mu \nu} - F_1^{\mu \nu} H_{234} + H_{123} F_4^{\mu  \nu}\, ,
\notag
\end{align}
for instance 
\begin{align}
H_{123} &=  \frac{h_{123}}{s_{123}}  \Big( \frac{1}{s_{23}} - \frac{ 1}{s_{12}} \Big)
=  \frac{\Mfrak_{1,2,3}}{6s_{123}}  \Big( \frac{1}{s_{23}} - \frac{ 1}{s_{12}} \Big) \label{pert4.op}
\\
H_{1234} &= \frac{1}{s_{1234}} \Big\{ h_{1234} \Big(   \frac{1}{s_{34}s_{234}} - \frac{1}{s_{12}s_{123}} \Big) - \frac{h_{3214}}{s_{23}} \Big(\frac{1}{s_{123}}-\frac{1}{s_{234}} \Big) + \frac{1}{4} \Big( \frac{\mathfrak{M}_{12,3,4}}{s_{34}} - \frac{\mathfrak{M}_{34,1,2}}{s_{12}} \Big)  \notag \\
&+\frac{1}{12} \Big( \big(k_{123} \cdot a_4 \big) \frac{ \mathfrak{M}_{1,2,3}}{s_{123}} \Big(\frac{1}{s_{12}}-\frac{1}{s_{23}} \Big) + \big( k_{234} \cdot a_1 \big) \frac{ \mathfrak{M}_{2,3,4}}{s_{234}} \Big( \frac{1}{s_{34}}-\frac{1}{s_{23}} \Big) \Big) \label{rk4H}  \\
& +  \frac{1}{24s_{12}s_{34}} \big[ \mathfrak{M}_{1,2,3} (k_{123} \cdot a_4) -\mathfrak{M}_{1,2,4} (k_{124} \cdot a_3) -\mathfrak{M}_{1,3,4} (k_{134} \cdot a_2)+\mathfrak{M}_{2,3,4} (k_{234} \cdot a_1) \big]\Big\} \, .\notag
\end{align}
An alternative expression for $H_{1234}$ can be found in appendix \ref{app:D.0}.

Given that $H_{i}=H_{ij}=0$, the redefinitions (\ref{pert4.mn}) up to rank four
line up with the general form of a non-linear gauge transformation (\ref{pert1.gau})
\begin{align}
A_{P}^{\mu, \, {\rm BCJ}}  &= A_{P}^{\mu} +k_P^\mu H_P- \sum_{XY=P} ( A_X^\mu H_Y -  A_Y^\mu H_X)  \notag \\
F_{P}^{\mu \nu, \, {\rm BCJ}}  &= F_{P}^{\mu \nu}- \sum_{XY=P} ( F_X^{\mu \nu} H_Y -   F_Y^{\mu \nu} H_X) \label{pert4.qr} \\
F_{P}^{\mu| \nu \la, \, {\rm BCJ}}  &= F_{P}^{\mu| \nu \la} -\sum_{XY=P} ( F_X^{\mu| \nu \la} H_Y -   F_Y^{\mu |\nu \la} H_X ) 
\notag \\
G_{P}^{\mu \nu| \la \rho, \, {\rm BCJ}}  &= G_{P}^{\mu \nu |\la \rho}- \sum_{XY=P} ( G_X^{\mu \nu| \la \rho}  H_Y -  G_Y^{\mu \nu | \la \rho} H_X)
\notag
\end{align}
with gauge parameters $\Omega_P \rightarrow H_P$. These transformations are checked to apply to all 
currents up to and including rank five, see appendix \ref{app:D.2} for the explicit form of $H_{12345}$, and 
the existence of suitable $H_{12\ldots p}$ is conjectural at higher rank $p\geq 6$.

As the punchline of (\ref{pert4.qr}), the local Jacobi-satisfying numerators
for off-shell diagrams are related to Lorenz-gauge currents through a non-linear gauge transformation generated by
\beq
\Omega = \sum_{i,j,l} H_{ijl} t^i t^j t^l e^{k_{ijl} \cdot x} +  \sum_{i,j,l,m} H_{ijlm} t^i t^j t^l t^m e^{k_{ijlm} \cdot x}  + \ldots
= \sum_{|P|\geq 3}  H_P t^P e^{k_P\cdot x} \, .
\label{pert4.st}
\eeq
In the next sections, the local multiparticle polarizations will be used to 
manifest the BCJ duality between color and kinematics in tree-level amplitudes of (\YMF).
Hence, the transformed currents $A_P^{\mu \, {\rm BCJ}},\ldots ,G_P^{\mu \nu|\la\rho, \, {\rm BCJ}}$
related by (\ref{pert4.ij}) are said to be in {\it BCJ gauge} \cite{Lee:2015upy, Mafra:2015vca}.
Note that the first perturbiner solutions to the field equations of ten-dimensional 
SYM were actually constructed in BCJ gauge \cite{Mafra:2015gia}.

%%%%%%%%%%%%%%%%%%%%%%%%%%%%%%%%%%%%%%%%%%%%%%%%
%%%%%%%%%%%%%%%%%%%%%%%%%%%%%%%%%%%%%%%%%%%%%%%%

\subsection{Kinematic derivation of the BCJ relations}
\label{sec:5.2}

As a first application of BCJ-gauge currents, we derive the BCJ relations (\ref{BCJrels}) of
${\cal A}_{{\rm YM}+F^3+F^4}$ amplitudes by inverting their correspondence (\ref{pert4.ij})
with multiparticle polarizations. The same sequence of arguments has been applied to derive
BCJ relations for tree amplitudes of ten-dimensional SYM from superspace currents in BCJ
gauge \cite{Mafra:2015vca}, and we adapt the reasoning of the reference to the bosonic amplitudes
up to the $\ap^2$-order.

At rank $p$, inversion of (\ref{pert4.ij}) relates
multiparticle polarizations to BCJ-gauge currents via
\begin{align}
a_{1\sigma(23\ldots p)}^\mu&= \!  \sum_{\tau \in S_{p-1}} S(\sigma|\tau)_1  \,  A^{\mu, \, {\rm BCJ}}_{1\tau(23\ldots p)}  \, , \ \ \ \ \ \, 
 f_{1\sigma(23\ldots p)}^{\mu \nu}  = \! \sum_{\tau \in S_{p-1}}  S(\sigma|\tau)_1  \, F^{\mu \nu, \, {\rm BCJ}}_{1\tau(23\ldots p)}  \label{pert4.uv} \\
f_{1\sigma(23\ldots p)}^{\mu|\nu \la}    &= \!  \sum_{\tau \in S_{p-1}} S(\sigma|\tau)_1  \, F^{\mu|\nu \la, \, {\rm BCJ}}_{1\tau(23\ldots p)} \, , \ \ \ \ \ 
g_{1\sigma(23\ldots p)}^{\mu \nu|\la \tau}  = \! \sum_{\tau \in S_{p-1}}  S(\sigma|\tau)_1  \,  G^{\mu \nu|\la \tau, \, {\rm BCJ}}_{1\tau(23\ldots p)}  \, ,
\notag
\end{align}
where $S(\sigma|\rho)_1$ denotes the KLT matrix defined in (\ref{genpar22}). The
Lie symmetries of the numerators of the BCJ-gauge currents ensure that the
Mandelstam invariants from $S(\sigma|\rho)_1$ cancel all of their kinematic poles on
the right-hand sides of (\ref{pert4.uv}). However, when repeating these matrix 
multiplications with Lorenz-gauge currents $A^\mu_P$, some of the kinematic poles
in $(m \geq 3)$-particle channels persist\footnote{For instance, the product $\sum_{\tau \in S_2} S(2,3|\tau)_1 A^\mu_{1\tau(23)}
= s_{12}(s_{23}A^\mu_{123} - s_{13} A_{213}^\mu)$ can be thought of as a Lorenz-gauge 
analogue of $a_{123}^\mu$ and exhibits a pole in $s_{123}$ with residue $\sim s_{12} k_{123}^\mu h_{123}$.} in $\sum_{\tau \in S_{p-1}} S(\sigma|\tau)_1 
A^{\mu}_{1\tau(23\ldots p)} $ with $p\geq 3$. Hence, it is a peculiarity of BCJ-gauge 
currents that {\it local} objects are obtained from matrix multiplication with $S(\sigma|\rho)_1$.
Similarly, the pole $s^{-1}_{12\ldots p}$ in the $p$-particle channel
drops out from the following rank-$p$ combinations of BCJ-gauge currents,
\begin{align}
a_{12}^\mu &= s_{12} A_{12}^{\mu , \, {\rm BCJ}} \, , \ \ \ \
\frac{ a_{123}^\mu }{s_{12}}=  s_{23} A_{123}^{\mu , \, {\rm BCJ}}  - s_{13} A_{213}^{\mu , \, {\rm BCJ}}  \label{bcjgauge2} \\
\frac{ a_{1234}^\mu }{s_{12} s_{123}} + \frac{ a_{3214}^\mu }{s_{23} s_{123}} &= s_{34} A_{1234}^{\mu , \, {\rm BCJ}} -s_{24}(A_{1324}^{\mu , \, {\rm BCJ}}+A_{3124}^{\mu , \, {\rm BCJ}}) + s_{14} A_{3214}^{\mu , \, {\rm BCJ}}\, ,
\notag
\end{align}
i.e.\ they are non-singular as $s_{12\ldots p} \rightarrow 0$ but obviously
exhibit poles in lower-multiplicity channels such as $s_{12\ldots p-1}^{-1}$.
The same is true for the rank-five expression
\begin{align}
\frac{1}{s_{1234}} \Big( \frac{ a_{12345}^\mu }{s_{12} s_{123}} &+ \frac{ a_{32145}^\mu }{s_{23} s_{123}} 
- \frac{ a_{43215}^\mu }{s_{34} s_{234}} - \frac{ a_{23415}^\mu }{s_{23} s_{234}}  + \frac{ a_{12345}^\mu - a_{12435}^\mu }{s_{12} s_{34}} \Big) \label{bcjgauge2a} \\
&\! \! \! \! \! \! \! \! \! \! \! \! \! \! \! = s_{45} A_{12345}^{\mu , \, {\rm BCJ}} - s_{35} A_{(12\shuffle 4)35}^{\mu , \, {\rm BCJ}}+s_{25} A_{(43\shuffle 1)25}^{\mu , \, {\rm BCJ}} - s_{15} A_{43215}^{\mu , \, {\rm BCJ}} \, . \notag
\end{align}
This can be used to derive BCJ relations among color-ordered (\YMF) amplitudes. 
We exploit that the amplitude formula (\ref{pert3.1}) is invariant under non-linear gauge transformations -- see 
section \ref{sec:3.3} -- and can therefore be written in terms of BCJ-gauge currents,
\beq
{\cal A}_{{\rm YM}+F^3+F^4}(1,2,\ldots,n{-}1,n) = s_{12\ldots n-1} A^{\mu , \, {\rm BCJ}}_{12\ldots n-1} A^\mu_n   \, .
\label{pert3.1BCJ}
\eeq
The right-hand side is nonzero and finite by the interplay of 
the vanishing Mandelstam invariant $s_{12\ldots n-1}$ and the compensating $(n{-}1)$-particle pole 
of $A^{\mu , \, {\rm BCJ}}_{12\ldots n-1}$. If the propagator $s^{-1}_{12\ldots n-1}$ cancels in a linear 
combination of currents, then multiplication with $s_{12\ldots n-1}$ yields vanishing expressions in the
$n$-particle momentum phase space. For instance, since $s_{23} A_{123}^{\mu , \, {\rm BCJ}}  
- s_{13} A_{213}^{\mu , \, {\rm BCJ}}=a_{123}^\mu/s_{12}$ does not have the pole $s_{123}^{-1}$ of the
individual currents, the quantity $s_{123} a_{123}^\mu/s_{12}= s_{123}(s_{23} A_{123}^{\mu , \, {\rm BCJ}}  
- s_{13} A_{213}^{\mu , \, {\rm BCJ}})$ vanishes by four-particle momentum conservation.
In combination with the amplitude formula (\ref{pert3.1BCJ}), this implies the four-point BCJ relation (\ref{BCJrels})
\begin{align}
0 &= \frac{s_{123} a_{123}^\mu}{s_{12}} 
= s_{123}( s_{23} A_{123}^{\mu , \, {\rm BCJ}}  - s_{13} A_{213}^{\mu , \, {\rm BCJ}})  \notag \\
&= s_{23}  {\cal A}_{{\rm YM}+F^3+F^4}(1,2,3,4)  - s_{13} {\cal A}_{{\rm YM}+F^3+F^4}(2,1,3,4) \, .
\label{bcjgauge4}
\end{align}
Similarly, the rank-$(p{\leq} 5)$ combinations in (\ref{bcjgauge2}) and (\ref{bcjgauge2a}) with 
regular $s_{12\ldots p}\rightarrow0$ limit imply the following five- and six-point BCJ relations 
after multiplication with the vanishing $(p=n{-}1)$-point 
Mandelstam invariant $s_{12\ldots p}$,
\begin{align}
0 &= s_{1234} \Big( \frac{ a_{1234}^\mu }{s_{12} s_{123}} + \frac{ a_{3214}^\mu }{s_{23} s_{123}} \Big)
= s_{34} {\cal A}_{{\rm YM}+F^3+F^4}(1,2,3,4,5)  \notag \\
& \ \ \ \
-s_{24}({\cal A}_{{\rm YM}+F^3+F^4}(1,3,2,4,5)+{\cal A}_{{\rm YM}+F^3+F^4}(3,1,2,4,5)) 
+ s_{14} {\cal A}_{{\rm YM}+F^3+F^4}(3,2,1,4,5) 
\notag \\
0 &= \frac{s_{12345}}{s_{1234}} \Big( \frac{ a_{12345}^\mu }{s_{12} s_{123}} + \frac{ a_{32145}^\mu }{s_{23} s_{123}} 
- \frac{ a_{43215}^\mu }{s_{34} s_{234}} - \frac{ a_{23415}^\mu }{s_{23} s_{234}}  + \frac{ a_{12345}^\mu - a_{12435}^\mu }{s_{12} s_{34}} \Big) \label{bcjgauge5} \\
&= s_{45} {\cal A}_{{\rm YM}+F^3+F^4}(1,2,3,4,5,6) - s_{35} {\cal A}_{{\rm YM}+F^3+F^4}((1,2\shuffle 4),3,5,6) \notag \\
& \ \ \ \ +s_{25} {\cal A}_{{\rm YM}+F^3+F^4}((4,3\shuffle 1),2,5,6) - s_{15} {\cal A}_{{\rm YM}+F^3+F^4}(4,3,2,1,5,6)\, .
\notag
\end{align}
This calls for an all-multiplicity formula for analogous combinations with regular 
behaviour as $s_{12\ldots p}\rightarrow 0$: The right-hand sides of (\ref{bcjgauge2})
can be generated through the $S$-map \cite{Mafra:2014oia, Mafra:2015vca}
\beq
A^{\mu , \, {\rm BCJ}}_{S[P,Q]}= 
\sum_{i=1}^{|P|} \sum_{j=1}^{|Q|}(-1)^{i-j+|P|-1} s_{p_i q_j} A^{\mu , \, {\rm BCJ}}_{(p_1 p_2\ldots p_{i-1} \shuffle p_{|P|} p_{|P|-1} \ldots p_{i+1}) p_i q_j(q_{j-1}\ldots q_2 q_1 \shuffle q_{j+1} \ldots q_{|Q|})   }
\label{bcjgauge3}
\eeq
involving words $P=p_1 p_2\ldots p_{|P|}$ and $Q=q_1 q_2 \ldots q_{|Q|}$. BCJ gauge
of the currents implies that the $S$-map defined in (\ref{bcjgauge3}) removes the pole in $s_{PQ}$
\cite{Mafra:2014oia} and therefore paves the way for the following form of the BCJ relations \cite{Mafra:2015vca}
\begin{align}
0 &= (-1)^{|P|-1} s_{PQ} A^{\mu , \, {\rm BCJ}}_{S[P,Q]} A^\mu_n\label{bcjgauge6} \\
&=\sum_{i=1}^{|P|} \sum_{j=1}^{|Q|}(-1)^{i-j} s_{p_i q_j} {\cal A}_{{\rm YM}+F^3+F^4}((p_1 p_2\ldots p_{i-1} \shuffle p_{|P|}  \ldots p_{i+1}) ,p_i, q_j,(q_{j-1}\ldots q_1 \shuffle q_{j+1} \ldots q_{|Q|}),n) \, .
\notag
\end{align}
Suitable choices of $P$ and $Q$ in (\ref{bcjgauge6}) reproduce
various representations of the BCJ relations \cite{BCJ, Stieberger:2009hq, BjerrumBohr:2009rd, Chen:2011jxa}. 
Setting $P=1$ and $Q=23\ldots n{-}1$, for instance, one recovers a form of the BCJ relations
\begin{align}
0 &=\sum_{j=2}^{n-1}(-1)^{j} s_{1j} {\cal A}_{{\rm YM}+F^3+F^4}(1,j,(j{-}1,j{-}2,\ldots ,3,2 \shuffle j{+}1, \ldots ,n{-}1),n) 
\label{bcjgauge60}
\end{align}
which is equivalent to (\ref{BCJrels}) by the KK relations ${\cal A}_{{\rm YM}+F^3+F^4}((X\shuffle Y),n)=0
\ \forall \ X,Y\neq \emptyset$.

%%%%%%%%%%%%%%%%%%%%%%%%%%%%%%%%%%%%%%%%%%%%%%%%
%%%%%%%%%%%%%%%%%%%%%%%%%%%%%%%%%%%%%%%%%%%%%%%%

\subsection{Local Jacobi-satisfying numerators}
\label{sec:5.3}

In this section, we will exploit the multiparticle polarizations of (\YMF) to 
construct local and Jacobi-satisfying cubic-diagram numerators. The most direct approach is to 
expand the BCJ-gauge current in the amplitude representation (\ref{pert3.1BCJ}) via (\ref{pert4.ij}),
\beq
{\cal A}_{{\rm YM}+F^3+F^4}(1,\tau(2,\ldots,n{-}1),n) = \sum_{\rho \in S_{n-2}} s_{12\ldots n-1} \Phi(\tau|\rho)_1 a^\mu_{1\rho(23\ldots n-1)} e^\mu_n \, , \ \ \ \ \tau \in S_{n-2} \, ,
\label{bcjgauge61}
\eeq
where the formally vanishing Mandelstam invariant in $s_{12\ldots n-1} \Phi(\tau|\rho)_1$ cancels
in each entry of the inverse KLT matrix (see the recursion in (\ref{genpar24}) and (\ref{genpar25})). 
From the remaining propagators in $\Phi(\tau|\rho)_1$, the expressions $a^\mu_{1\rho(23\ldots n-1)} e^\mu_n$ 
will be shown below to take the role of master numerators of the half-ladder diagrams depicted in 
figure \ref{halfladders}. The $(n{-}2)!$ KK-independent permutations of ${\cal A}_{{\rm YM}+F^3+F^4}$ 
in (\ref{bcjgauge61}) incorporate each cubic diagram at least once and therefore define all of the numerators. 

In order to demonstrate that the numerators in (\ref{bcjgauge61}) obey kinematic Jacobi identities, 
we bring it into the form of the general amplitude representation (\ref{genpar7}) with manifest
color-kinematics duality,
\beq
{\cal A}_{{\rm YM}+F^3+F^4}(\sigma(1,2,\ldots,n)) = \! \sum_{\rho \in S_{n-2}}  \!
m(\sigma | 1,\rho(2,\ldots,n{-}1),n) a^\mu_{1\rho(23\ldots n-1)} e^\mu_n \, , \ \ \ \ \sigma \in S_{n} \, .
\label{bcjgauge62}
\eeq
Consistency with (\ref{bcjgauge61}) can be conveniently checked by expressing $\Phi(\tau|\rho)_1$ with $\tau,\rho \in S_{n-2}$
as a putative $(n{+}1)$-point doubly-partial amplitude $-m(1,\tau,n,n{+}1|1,\rho,n{+}1,n)$ via (\ref{genpar21}) 
and (\ref{genpar25}). By its Berends--Giele representation (\ref{genpar23}) \cite{Mafra:2016ltu}, the 
latter can be written as
\beq
\Phi(\tau|\rho)_1 = - s_{12\ldots n} \phi_{1\tau n| n 1\rho} = - \sum_{XY=1\tau n} \sum_{AB=n1\rho} ( \phi_{X|A} \phi_{Y|B} -  \phi_{Y|A} \phi_{X|B})\, .
\label{bcjgauge63}
\eeq
Since $\phi_{P|Q}$ vanishes unless $P$ is a permutation of $Q$, the only contribution 
arises from the deconcatenations with $A=n$ and $Y=n$ leading to
\beq
s_{12\ldots n-1}\Phi(\tau|\rho)_1 =  - s_{12\ldots n-1} (\phi_{1\tau|n} \phi_{n|1\rho} -  \phi_{n|n} \phi_{1\tau|1\rho}) = m(1,\tau,n|1,\rho,n)\, .
\label{bcjgauge64}
\eeq
Hence, (\ref{bcjgauge62}) at $\sigma=(1,\tau,n)$ reduces to (\ref{bcjgauge61}).
For other choices of $\sigma$ in turn, validity of (\ref{bcjgauge62}) follows from the KK relations of both sides.
Hence, by the discussion around (\ref{genpar7}), the cubic-diagram numerators of (\ref{bcjgauge61}) are 
composed from the masters $a^\mu_{1\rho(23\ldots n-1)} e^\mu_n$ as dictated by Jacobi identities.

Note that the cubic-diagram numerators in (\ref{bcjgauge61}) and (\ref{bcjgauge62}) are not crossing 
symmetric, i.e.\ their functional form in terms of polarizations and momenta
depends on the position of the singled-out legs $1$ and $n$ in the diagram.

\begin{figure}[h]
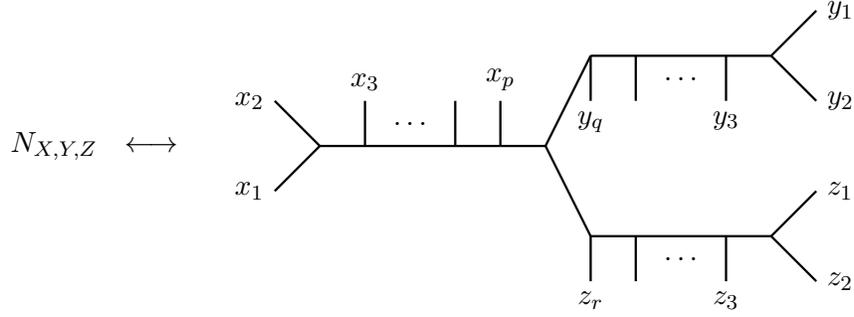

\begin{center}
 \tikzpicture [scale=1.2,line width=0.30mm]
\draw (-4,0.5)node{$N_{X,Y,Z} \ \ \longleftrightarrow$};
\draw (1,0.5) -- (1.5,-0.5) ;
\draw (1,0.5) -- (1.5,1.5) ;
\draw (1,0.5) -- (0.5,0.5)  ;
\draw (1.5,1.5) -- (1.5,1) node[below]{$y_q$};
\draw (2,1.5) -- (2,1) ;
\draw (2.5,1.25) node{$\ldots$};
\draw (3,1.5) -- (3,1) node[below]{$y_{3}$};
\draw (1.5,1.5) -- (3.5,1.5);
\draw (3.5,1.5) -- (4,1) node[right]{$y_2$};
\draw (3.5,1.5) -- (4,2) node[right]{$y_1$};
\scope[xshift=1cm]
\draw (-0.5,0.5) -- (-0.5,1)node[above]{$x_p$};
\draw (-1,0.5) -- (-1,1);
\draw (-1.5,0.75) node{$\ldots$};
\draw (-2,0.5) -- (-2,1) node[above]{$x_{3}$};
\draw (-0.5,0.5) -- (-2.5,0.5);
\draw (-2.5,0.5) -- (-3,0) node[left]{$x_1$};
\draw (-2.5,0.5) -- (-3,1) node[left]{$x_2$};
\endscope
\draw (1.5,-0.5) -- (1.5,-1) node[below]{$z_r$};
\draw (2,-0.5) -- (2,-1) ;
\draw (2.5,-0.75) node{$\ldots$};
\draw (3,-0.5) -- (3,-1) node[below]{$z_{3}$};
\draw (1.5,-0.5) -- (3.5,-0.5);
\draw (3.5,-0.5) -- (4,-1) node[right]{$z_2$};
\draw (3.5,-0.5) -- (4,0) node[right]{$z_1$};
\endtikzpicture 
\caption{Diagrammatic interpretation of the local building block 
$N_{X,Y,Z}$ with multiparticle labels $X=x_1x_2\ldots x_p, \ Y=y_1y_2\ldots y_q$
and $Z=z_1 z_2 \ldots z_r$ referring to three off-shell half-ladder diagrams that are connected
by the central vertex. Note that
we no longer distinguish between the orders of $\alpha'$ carried by the individual 
cubic vertices and therefore suppress the white and black dots of figure \ref{f:MABC2}.}
\label{f:localMABC}
\end{center}
\end{figure}

In the same way as the manifestly cyclic representations of section \ref{sec:3.2} assemble $n$-point amplitudes 
from Berends--Giele currents of maximum rank $\lfloor \frac{ n}{2} \rfloor$, we will next spell out alternative 
numerators in terms of lower-rank multiparticle polarizations. In analogy to the cyclic building block $\Mfrak_{X,Y,Z}$ 
in (\ref{pert3.12}) composed of Lorenz-gauge currents, we define the local combination
\begin{align}
N_{X,Y,Z} &= \frac{1}{2} \big( a_X^\mu f_Y^{\mu \nu} a_Z^\nu  + {\rm cyc}(X,Y,Z) \big) -2\ap f_X^{\mu \nu} f_Y^{\nu \la} f_Z^{\la \mu} \notag \\
& + \Big( \frac{\ap}{2}  f_X^{\mu | \nu \la} f_Y^{ \nu \la} a_Z^\mu + 2 \ap^2  g_X^{\mu \nu | \la \rho} f_Y^{ \mu| \la \rho} a_Z^\nu  \pm {\rm perm}(X,Y,Z) \Big) \label{pert3.12loc}\\
&+ \Big( \frac{\ap^2}{2}   g_X^{\mu \nu | \la \rho} f_Y^{\mu \nu} f_Z^{\la \rho} - 2\ap^2   f_X^{\mu \nu} f_Y^{\mu| \la \rho} f_Z^{\nu| \la \rho} + {\rm cyc}(X,Y,Z) \Big)
\notag
\end{align}
to describe the cubic diagram in figure \ref{f:localMABC} (see figure \ref{f:MABC2} for
the analogous diagrammatic interpretation of $\Mfrak_{X,Y,Z}$). 

\begin{figure}[h]
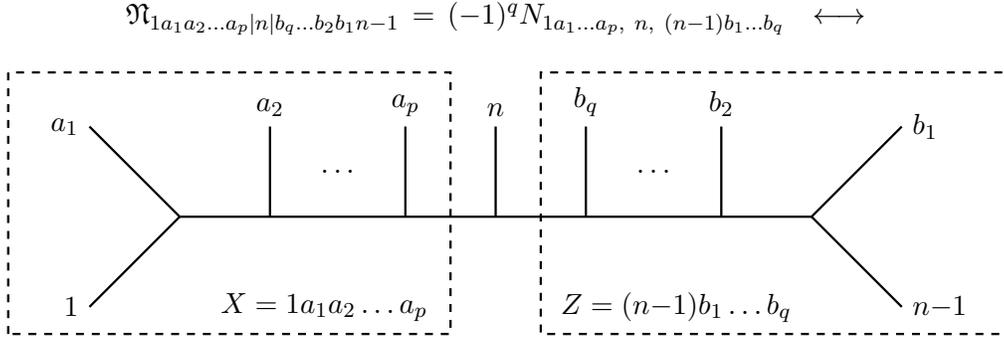

\begin{center}
 \tikzpicture [scale=1.2,line width=0.30mm]
\draw(0,0) --(-1,-1) node[left]{$1$};
\draw(0,0) --(-1,1) node[left]{$a_1$};
\draw(0,0) -- (7,0);
\draw(1,0) -- (1,1) node[above]{$a_2$};
\draw (1.75,0.5)node{$\ldots$};
\draw (2.5,0)--(2.5,1) node[above]{$a_p$};
\draw (3.5,0)--(3.5,1) node[above]{$n$};
\draw (4.5,0)--(4.5,1) node[above]{$b_q$};
\draw (5.25,0.5)node{$\ldots$};
\draw(6,0) -- (6,1) node[above]{$b_2$};
\draw(7,0) --(8,-1) node[right]{$n{-}1$};
\draw(7,0) --(8,1) node[right]{$b_1$};
\draw[dashed] (-1.9,-1.3) rectangle (3,1.6);
\draw[dashed] (9.2,-1.3) rectangle (4.0,1.6);
%
%\draw(1.8,-1)node{$X=12\ldots j$};
%\draw(5.85,-1)node{$Z=(n{-}1)(n{-}2)\ldots j{+}1$};
\draw(1.6,-1)node{$X=1a_1a_2\ldots a_p$};
\draw(5.5,-1)node{$Z=(n{-}1) b_1 \ldots b_{q}$};
%
%\draw(3.5,2.2)node{$\Nfrak_{12\ldots j|n| j+1 \ldots n-2,n-1} \, = \, (-1)^{n-j} N_{12\ldots j ,\ n, \ (n{-}1)(n{-}2)\ldots j{+}1}  \ \ \longleftrightarrow$};
\draw(3.5,2.2)node{$\Nfrak_{1a_1a_2\ldots a_p|n|b_q  \ldots b_{2} b_1 n-1} \, = \, (-1)^{q} N_{1a_1\ldots a_p , \ n ,\  (n-1)b_{1}\ldots b_q}  \ \ \longleftrightarrow$};
\endtikzpicture 
\caption{An alternative choice of master numerators composed of multiparticle polarizations of smaller rank 
as compared to (\ref{bcjgauge62}). The external legs $a_1,a_2,\ldots,a_p,b_1,b_2,\ldots,b_q$ are permutations
of $2,3,\ldots,n{-}2$ where $p{+}q=n{-}3$ and $p=0,1,\ldots,n{-}3$.}
\label{f:BCJnums}
\end{center}
\end{figure}

There is an ambiguity in relating cubic diagrams to the combinations $N_{X,Y,Z}$
in (\ref{pert3.12loc}): Each of the $n{-}2$ cubic vertices may be associated with the 
central vertex in figure \ref{f:localMABC}, e.g.\ all of $N_{123,4,5}$, $N_{12,3,45}$ and $N_{1,2,543}$ 
describe the same cubic diagram. A valid $(n{-}2)!$-set of $N_{X,Y,Z}$ to serve as the master
numerators for half-ladder diagrams is given by
\beq
\Nfrak_{1a_1a_2\ldots a_p|n|b_q  \ldots b_{2} b_1 n-1} = (-1)^{q} N_{1a_1a_2\ldots a_p , \ n ,\  (n-1)b_{1}b_2\ldots b_q} \, .
\label{bcjgauge65}
\eeq
As a defining property of these master numerators $\Nfrak_{\ldots}$, the central vertex of figure \ref{f:localMABC} is always 
chosen to be adjacent to leg $n$ which therefore enters in a single-particle slot.
As depicted in figure \ref{f:BCJnums}, the numerators in (\ref{bcjgauge65}) describe half-ladder diagrams 
with endpoints $1$ and $n{-}1$, where the location of leg $n$ decides about the partition into the three
subdiagrams associated with the slots of $N_{X,Y,Z}$. The remaining labels $a_1,a_2,\ldots,a_p,b_1,b_2,\ldots,b_q$
are a permutation of $2,3,\ldots,n{-}2$ with $p{+}q=n{-}3$. Together with the $n{-}2$ different choices of 
$p=0,1,\ldots,n{-}3$, this exhausts the total of $(n{-}2)!$ permutations of the larger set $2,3,\ldots,n{-}2,n$.

The collection of $\Nfrak_{1a_1a_2\ldots a_p|n|b_q  \ldots b_{2} b_1 n-1}$ in (\ref{bcjgauge65}) and figure \ref{f:BCJnums}
can be used as an alternative to the master numerators $a^\mu_{1\rho(23\ldots n-1)}e_n^\mu$ in (\ref{bcjgauge62}).
As a practical advantage of the $\Nfrak_{\ldots}$, their constituents in (\ref{pert3.12loc}) only require multiparticle 
polarizations of maximal rank $n{-}2$ instead of the rank-$(n{-}1)$ quantities $a^\mu_{1\rho(23\ldots n-1)}$. 
As demonstrated in appendix \ref{app:E}\footnote{Also see \cite{Mafra:2011kj, Mafra:2016ltu} for Jacobi-satisfying
superspace numerators in ten-dimensional SYM with the same combinatorial structure 
as (\ref{pert3.12loc}) and (\ref{bcjgauge65}).}, 
they yield Jacobi-satisfying amplitude representations of the form (\ref{genpar7}),
\begin{align}
&{\cal A}_{{\rm YM}+F^3+F^4}(\sigma(1,2,\ldots,n))  = \sum_{j=1}^{n-2} \sum_{\rho \in S_{n-3}}  \Nfrak_{1\rho(23\ldots j)|n| \rho(j+1\ldots n-2)n-1}
\notag \\
& \ \ \ \ \ \ \ \ \ \ \ \ \ \  \times m(\sigma| 1,\rho(2,3,\ldots,j),n,\rho(j{+}1,\ldots,n{-}2),n{-}1) \,.\label{bcjgauge66}
\end{align}
Note that the $\alpha' \rightarrow 0$ order of (\ref{bcjgauge66}) follows from the field-theory
limit of the pure-spinor superstring based on the amplitude representations of \cite{Mafra:2011kj, Mafra:2016ltu}
and the superspace gauge described in \cite{Lee:2015upy}.

Moreover, one can further reduce the maximum rank of the multiparticle polarizations by 
a generalization of the integration-by-parts relation (\ref{pert3.14}). The latter still holds 
when the $\Mfrak_{A,B,C}$ are constructed from BCJ gauge currents\footnote{This follows from
the fact that the difference of the left- and right-hand side of (\ref{pert3.14}) is invariant under non-linear
gauge transformations (\ref{pert3.51}).} and multiplied by
KLT matrices for the slots $A,B,C$, so the rank-$(n{-}2)$ cases in (\ref{bcjgauge66}) can be reduced as follows
\begin{align}
\Nfrak_{123|5|4}&= N_{123,5,4} = \frac{1}{s_{45}} 
 \big[ (k_{12}\cdot k_3) N_{12,3,54} + (k_1\cdot k_2)(N_{1,23,54} + N_{13,2,54}) \big] \notag \\
 \Nfrak_{1234|6|5}&= N_{1234,6,5} = \frac{1}{s_{56}} 
 \big[ (k_{123} \cdot k_4) N_{123,4,65} + (k_{12}\cdot k_3) (N_{124,3,65} +N_{12,34,65} )\label{bcjgauge72} \\
& \ \ \  \ \ \  \ \ \  \ \ \  \ \ \  \ \ \  \ \ \  + (k_1\cdot k_2)(N_{1,234,65} + N_{134,2,65}+N_{13,24,65} + N_{14,23,65}) \big] \, . \notag
\end{align}
The $n$-point generalization involves the summation prescription of the 
form $a_1 a_2\ldots a_p = X\shuffle Y$ that has been introduced in section \ref{sec:4.3}
\begin{align}
\Nfrak_{12\ldots n-2|n|n-1}&= N_{12\ldots n-2,n,n-1} = \frac{1}{s_{n-1,n}} 
\sum_{j=2}^{n-2}(k_{12\ldots j-1} \cdot k_j) \sum_{j{+}1,j{+}2\ldots n{-}2 \atop{= X\shuffle Y}} N_{12\ldots j-1X,jY,n(n-1)}
 \,. \label{bcjgauge71}
\end{align}
The right-hand sides of (\ref{bcjgauge72}) and (\ref{bcjgauge71}) at $n\geq 5$ 
can be assembled from multiparticle polarizations of maximum rank $n{-}3$, and the
spurious poles in $s_{n-1,n}$ cancel after combining all the terms. The same strategy applies
to permutations of $\Nfrak_{12\ldots n-2|n|n-1}$ and $\Nfrak_{1|n|23\ldots (n-2)(n-1)}$ in $2,3,\ldots,n{-}2$.
Like this, the $n$-point amplitude representation (\ref{bcjgauge66}) with manifest BCJ duality
and local numerators is completely determined by multiparticle polarizations of rank $n{-}3$. For instance,
the explicit construction of multiparticle polarizations up to rank five in section \ref{sec:4}
is sufficient to pinpoint all the eight-point numerators in (\ref{bcjgauge66}).

Note that the special footing of legs $1,n{-}1$ and $n$ in (\ref{bcjgauge66}) breaks the crossing symmetry
even more heavily than the numerators in (\ref{bcjgauge62}). Still, one can restore crossing symmetry
by averaging over all choices of singling out legs $i,j,k \in \{1,2,\ldots,n\}$ instead of $1,n{-}1$ and $n$.

%%%%%%%%%%%%%%%%%%%%%%%%%%%%%%%%%%%%%%%%%%%%%%%%
%%%%%%%%%%%%%%%%%%%%%%%%%%%%%%%%%%%%%%%%%%%%%%%%

\subsection{Relation to string-theory and gravity amplitudes}
\label{sec:5.4}

A major motivation for the construction of (\YMF)
numerators with manifest locality stems from their connection with gravitational quantities
through the double copy. Following the lines of \cite{Broedel:2012rc}, the double copy of 
${\cal A}_{{\rm YM}+F^3+F^4}$ to amplitudes from higher-curvature operators 
$\alpha'R^2 + \alpha'^2 R^3$ can be extracted from the string-theory KLT
relations \cite{Kawai:1985xq} (also see \cite{BjerrumBohr:2003vy, BjerrumBohr:2003af}):
The leading $\ap$-orders of the open-bosonic-string amplitudes
\beq
{\cal A}_{{\rm open} \atop{\rm bosonic}}(\sigma) = {\cal A}_{{\rm YM}+F^3+F^4}(\sigma) + \zeta_2 {\cal A}_{{\rm super}-F^4}(\sigma) + {\cal O}(\ap^3)
\label{R2R3.1}
\eeq
comprise our results for ${\cal A}_{{\rm YM}+F^3+F^4}$ (and the aforementioned contribution from the 
supersymmetrizable $F^4$-operator which is incompatible with the BCJ duality
and can be distinguished by its coefficient $\zeta_2$ \cite{Broedel:2012rc}). By the interplay with the
trigonometric factors in the KLT formula, both copies of ${\cal A}_{{\rm super}-F^4}(\sigma)$ 
drop out from relevant orders of the closed bosonic string \cite{Huang:2016tag},
\beq
{\cal M}_{{\rm closed} \atop{\rm bosonic}}  = \! \! \!
 \sum_{\rho,\tau \in S_{n-3}}  \! \! \!
{\cal A}_{{\rm YM}+F^3+F^4}(1,\rho,n{-}1,n)  S(\rho|\tau)_1
\tilde {\cal A}_{{\rm YM}+F^3+F^4}(1,\tau,n,n{-}1)  + {\cal O}(\ap^3)\, ,
\label{R2R3.2}
\eeq
where $S(\rho|\tau)_1$ is the field-theory KLT matrix defined in (\ref{genpar22}) and the permutations $\rho,\tau$ act on 
$2,3,\ldots,n{-}2$. Hence, to the orders considered, the right-hand side of (\ref{R2R3.2})
describes amplitudes from the low-energy effective action of the closed bosonic string \cite{Metsaev:1986yb}
\begin{align}
{\cal S}_{{\rm closed} \atop{\rm bosonic}}   &\sim \int {\rm d}^D x \, \sqrt{g}  \Big\{
R - 2 (\partial_\mu \varphi)^2 - \frac{1}{12} H^2  + \frac{\ap}{4} e^{-2\varphi} \big[ R_{\mu \nu \la \rho} R^{\mu \nu \la \rho} - 4 R_{\mu \nu} R^{\mu \nu}+R^2 \big]\notag \\
& \ \ \ \ +\ap^2  e^{-4\varphi}  \big[ \tfrac{1}{16} R^{\mu \nu}{}_{\alpha \beta} R^{\alpha \beta}{}_{\la \rho} R^{\la \rho}{}_{\mu \nu}
- \tfrac{1}{12} R^{\mu \nu}{}_{\alpha \beta} R^{\nu \la}{}_{\beta \rho} R^{\la \mu}{}_{\rho \alpha} \big]
+ {\cal O}(\ap^3)  \Big\}  \,  ,  \label{R2R3.4}
\end{align}
where $\varphi$ denotes the dilaton and $H= \dd B$ is the field strength of the $B$-field. In spite of the dilaton
admixtures via $e^{-2\varphi}, e^{-4\varphi}$, the operators along with the first and second order of $\ap$ are
collectively referred to as $R^2$ and $R^3$. While the $R^2$ operator can be reconciled with up to sixteen
supercharges \cite{Gross:1986mw}, the $R^3$ operator is not supersymmetrizable \cite{Grisaru:1976nn}.

Given the multitude of propagators in KLT formulae of the form (\ref{R2R3.2}), the 
locality properties of gravity amplitudes are more transparent in representations
involving Jacobi-satisfying numerators as in (\ref{genpar8}). For instance, our master numerators 
for (\YMF) in (\ref{bcjgauge62}) and (\ref{bcjgauge66}) 
admit a realization of the double-copy structure via
\begin{align}
&{\cal M}_{{\rm GR}+ R^2 +   R^3}  =  \! \sum_{\rho \in S_{n-2}}  \!
( a^\mu_{1\rho(23\ldots n-1)} e^\mu_n)  \, \tilde {\cal A}_{{\rm YM}+F^3+F^4}(1,\rho(2,3,\ldots,n{-}1),n)
 + {\cal O}(\ap^3) \notag \\
&=  \sum_{j=1}^{n-2} \sum_{\rho \in S_{n-3}}  \! \Nfrak_{1\rho(2\ldots j)|n| \rho(j+1\ldots n-2)n-1}\,
 \tilde {\cal A}_{{\rm YM}+F^3+F^4}( 1,\rho(2,\ldots,j),n,\rho(j{+}1,\ldots,n{-}2),n{-}1) \notag \\
 & \ \ \ \ \ + {\cal O}(\ap^3) 
\, ,
\label{R2R3.3}
\end{align}
where each term has the propagator structure of cubic diagrams.
The subscript ${\rm GR}+R^2 + R^3$ is just a schematic shorthand for the amplitudes
generated by the action (\ref{R2R3.4}) to the orders of $\alpha'^2$. As emphasized in \cite{Broedel:2012rc},
the $\alpha'^2$-order of (\ref{R2R3.3}) receives contribution from both single-insertions of $R^3$ operators
and double-insertions of $R^2$ operators. 

In $D=4$ spacetime dimensions, the $R^2$ contribution to (\ref{R2R3.4}) is the topological Gauss--Bonnet term.
The components at the first order in $\alpha'$ of (\ref{R2R3.3}) with graviton helicities are therefore guaranteed to vanish.
Still, the double insertions of $R^2$ contribute to the $\alpha'^2$-order of graviton components in four dimensions
since the prefactor of $e^{-2\varphi}$ in (\ref{R2R3.4}) allows for dilaton exchange \cite{Broedel:2012rc}.

On the right-hand side of (\ref{R2R3.2}) or (\ref{R2R3.3}), the $\alpha'^2$ order receives both symmetric 
and asymmetric contributions: Terms of the form ${\cal A}_{{\rm YM}+F^3+F^4} \big|_{\ap}
 \tilde {\cal A}_{{\rm YM}+F^3+F^4} \big|_{\ap}$ where both gauge-theory halves contribute a factor of
 $\ap$ have been carefully analyzed in $D=4$ helicity components \cite{Broedel:2012rc}.
Our results on the $\ap^2$-order of ${\cal A}_{{\rm YM}+F^3+F^4} $ and its master numerators additionally deliver the 
 contributions to $D$-dimensional amplitudes of ${\rm GR}+R^2 + R^3$, where both powers of $\ap$ stem from the same 
 gauge-theory factor. 
These contributions involving ${\cal A}_{{\rm YM}+F^3+F^4} \big|_{\ap^2}$ explain\footnote{We are grateful to Johannes Br\"odel for helpful discussions on this point and checking a representative four-dimensional helicity example.} the departure of ${\cal M}_{{\rm GR}+ R^2 +   R^3} \big|_{\ap^2}$ from the
double copy of the first $\ap$-order ${\cal A}_{{\rm YM}+F^3+F^4} \big|_{\ap}$ which has already been
observed in certain $D=4$ helicity components \cite{Broedel:2012rc}.
Hence, there is no need to consider higher orders from the $\ap$-expansion of the 
$\sin (\frac{\pi \ap }{2} k_i\cdot k_j)$ terms in the string-theory KLT relations as speculated in the reference.

At the order of $\ap$, one may extract new representations\footnote{Note that alternative representations
with $8$ supercharges on both chiral halves can be extracted from the low-energy limit of one-loop string amplitudes in $K3$ orbifolds \cite{Gregori:1997hi, Berg:2016wux}.} for supersymmetrized matrix 
elements of $R^2$ from (\ref{R2R3.3}) by trading ${\cal A}_{{\rm YM}+F^3+F^4} $ for color-ordered 
amplitudes of ten-dimensional SYM and their dimensional reductions. These supersymmetrizations play a
key role in recent studies of divergences and duality anomalies of ${\cal N}=4$ 
supergravity \cite{Bern:2017tuc, Bern:2017rjw}.

%%%%%%%%%%%%%%%%%%%%%%%%%%%%%%%%%%%%%%%%%%%%%%%%
%%%%%%%%%%%%%%%%%%%%%%%%%%%%%%%%%%%%%%%%%%%%%%%%
%%%%%%%%%%%%%%%%%%%%%%%%%%%%%%%%%%%%%%%%%%%%%%%%
%%%%%%%%%%%%%%%%%%%%%%%%%%%%%%%%%%%%%%%%%%%%%%%%
%%%%%%%%%%%%%%%%%%%%%%%%%%%%%%%%%%%%%%%%%%%%%%%%

\section{Conclusions and outlook}

In this work, we have studied various representations for tree-level amplitudes of $D$-dimensional gauge 
theories with $\alpha' F^3+\alpha'^2 F^4$ deformations. Our results are independent on the choice of
gauge group and hold to the order of $\alpha'^2$, where the interplay of the $F^3$ and $F^4$-operators 
is known to result in the BCJ duality between color and
kinematics \cite{Broedel:2012rc}. While the BCJ duality has originally been explained by the realization
of the $\alpha' F^3+\alpha'^2 F^4$ operators from the open bosonic string, our work takes a different 
approach by identifying the seeds of the duality in the Berends--Giele currents of (\YMF).

We study the Berends--Giele currents of (\YMF) in the perturbiner formalism 
\cite{Rosly:1996vr, Rosly:1997ap, Selivanov:1997aq, Selivanov:1999as, Bardeen:1995gk}, 
where non-linear gauge transformations can be mapped to reparametrizations
of the scattering amplitudes \cite{Lee:2015upy}. We pinpoint a specific non-linear gauge transformation up to
the order of five on-shell legs which rearranges the naive Feynman-diagram output of the action such as to 
manifest the BCJ duality. Like this, we derive the BCJ relations among color-ordered amplitudes 
to the order of $\alpha'^2$ from purely kinematic considerations. Furthermore,
two kinds of explicit cubic-diagram parametrizations are given for (\YMF)-amplitudes
where the manifestly local numerators obey kinematic Jacobi relations.

Our construction is inspired by superspace kinematic factors of 
ten-dimensional SYM \cite{Lee:2015upy, Mafra:2015vca} 
whose properties were inferred from the conformal-field-theory 
description of the pure-spinor superstring \cite{Mafra:2010ir, Mafra:2011kj, Mafra:2014oia}. 
We identify extensions of these superspace-inspired structures to higher orders in $\alpha'$ and to operators $F^3, F^4$
that do not admit any supersymmetrization. It would be interesting to find a conformal-field-theory derivation
of the local multiparticle polarizations that drive our BCJ-duality-satisfying amplitude representations.
One possible starting point is to combine the worldsheet description of the bosonic string with the 
off-shell techniques of \cite{Fu:2018hpu}. Alternatively, it might be helpful to identify a vertex-operator
origin of the CHY formulae for $F^3$ amplitudes \cite{He:2016iqi}\footnote{Note that the CHY half-integrands
${\cal P}_n$  in \cite{He:2016iqi} with a puncture $z_j\in \mathbb C$ on the Riemann 
sphere for each external state $j=1,2,\ldots,n$
 and $z_{i,j}=z_i-z_j$ may be reproduced from the first order in $\ap$ of
 $$
 {\cal P}_n(z_j,k_j,e_j) = 
 \frac{ a_{12\ldots n-1}^\mu e_n^\mu }{z_{1,2}z_{2,3}\ldots z_{n-1,n}z_{n,1}}  \, \Big|_{(\ap)^1}+ {\rm perm}(2,3,\ldots,n{-}1) 
 \, .
 $$} along with generalizations
to higher orders in $\alpha'$.

A complementary approach to the $\alpha' F^3+\alpha'^2 F^4$ operators of the bosonic string
is suggested by the recent double-copy description of bosonic-string amplitudes \cite{Azevedo:2018dgo}:
After peeling off the worldsheet integrals that are common with superstring amplitudes, an 
all-order family of $\alpha'$-corrections of the bosonic string can be traced back to a massive 
gauge theory dubbed $(DF)^2+{\rm YM}$. The latter has been constructed in \cite{Johansson:2017srf} 
by imposing the BCJ duality on a collection of dimension-six interactions between gauge bosons and massive scalars,
and it should reproduce the (\YMF)-amplitudes in this work upon low-energy expansion.
It would be interesting
to study our results from the $(DF)^2+{\rm YM}$-perspective and to generalize them to arbitrary orders
in $\alpha'$ by integrating out its massive modes.

Convenient and Jacobi-satisfying representations of tree-level subdiagrams are helpful
for loop integrands of string- and field-theory amplitudes, see e.g.\ \cite{Mafra:2014gja, Mafra:2015mja, 
Berg:2016fui, He:2017spx}. Our results might guide the organization of tensor structures of loop amplitudes
in bosonic and heterotic string theories. This in turn could give input on loop integrands of half-maximal
supergravity and their interplay with evanescent matrix elements and anomalies \cite{Bern:2015xsa, 
Bern:2017puu, Bern:2017tuc, Bern:2017rjw}.

%%%%%%%%%%%%%%%%%%%%%%%%%%%%%%%%%%%%%%%%%%%%%%%%
%%%%%%%%%%%%%%%%%%%%%%%%%%%%%%%%%%%%%%%%%%%%%%%%
%%%%%%%%%%%%%%%%%%%%%%%%%%%%%%%%%%%%%%%%%%%%%%%%
%%%%%%%%%%%%%%%%%%%%%%%%%%%%%%%%%%%%%%%%%%%%%%%%
%%%%%%%%%%%%%%%%%%%%%%%%%%%%%%%%%%%%%%%%%%%%%%%%

\section*{Acknowledgements}

We are grateful to the organizers of the PSI Winter School 2018 where this work was initiated.
We would like to thank Freddy Cachazo, Sebastian Mizera, Michael David Morales Curi and
Barbara Skrzypek for stimulating discussions during the Winter School and various later stages
of this work. We are grateful to Johannes Br\"odel, Alfredo Guevara, Song He and Julio Parra-Martinez for
valuable discussions. OS thanks Carlos Mafra for collaboration on related topics and comments on a draft. 
This research was supported in part by Perimeter Institute for Theoretical Physics. 
Research at Perimeter Institute is supported by the Government of Canada through the 
Department of Innovation, Science and Economic Development Canada and by the 
Province of Ontario through the Ministry of Research, Innovation and Science.
The research is supported in part by the Swedish Research Council under 
grant 621-2014-5722 and the Knut and Alice Wallenberg Foundation under grant KAW 2013.0235.

%%%%%%%%%%%%%%%%%%%%%%%%%%%%%%%%%%%%%%%%%%%%%%%%
%%%%%%%%%%%%%%%%%%%%%%%%%%%%%%%%%%%%%%%%%%%%%%%%
%%%%%%%%%%%%%%%%%%%%%%%%%%%%%%%%%%%%%%%%%%%%%%%%
%%%%%%%%%%%%%%%%%%%%%%%%%%%%%%%%%%%%%%%%%%%%%%%%
%%%%%%%%%%%%%%%%%%%%%%%%%%%%%%%%%%%%%%%%%%%%%%%%

\appendix
\section{Properties of the cyclic building blocks}
\label{app:A}

In this appendix, we prove several properties of the building block $\mathfrak{M}_{A,B,C}$ 
defined in (\ref{pert3.12}) to first order in $\ap$. These proofs are based on transversality 
$k_P\cdot A_P=0$ and the truncation of (\ref{pert3.9}) to the first order in $\ap$,
\beq
k_P^\la F_P^{\la \mu} = \sum_{P=XY} \Big[  A_X^\la F_Y^{\la \mu} + 2\ap F^{\nu \la}_X F_Y^{\nu|\la}{}^{\mu} 
- (X\leftrightarrow Y) \Big] + {\cal O}(\ap^2)\, .
\label{eq:kf}
\eeq
Moreover, we will use the generalization of (\ref{eq:kf}) to higher mass dimension,
\begin{equation} \label{eq:knabf}
k^\mu_P F^{\la|\mu \nu}_P = \sum_{P=XY}\Big[ A^\mu_X F^{\la|\mu \nu}_Y + F^{\mu \nu}_X F^{\mu \lambda}_Y  - (X\leftrightarrow Y) \Big] +\mathcal{O}(\alpha') \, ,
\end{equation}
which follows from the definition $[\nabla^\mu,\nabla^\la]=-\mathds{F}^{\mu \la}$ and a corollary of the equations of motion, 
$[\nabla^\mu , [\nabla^\la, \mathds{F}^{\mu \nu}]] =  [\nabla^\la , [\nabla^\mu, \mathds{F}^{\mu \nu}]] 
 + [ [\nabla^\mu,\nabla^\la], \mathds{F}^{\mu \nu} ]$ $= -  [ \mathds{F}^{\mu \la}, \mathds{F}^{\mu \nu}] + {\cal O}(\ap)$.
By the Jacobi identity $[\nabla^\alpha,[\nabla^\mu,\nabla^\nu]]+{\rm cyc}(\mu,\nu,\alpha)=0$, the currents of 
higher mass dimension have the symmetries
\begin{equation} \label{eq:jac}
F^{\alpha|\mu \nu}_P +  F^{\mu|\nu \alpha}_P +  F^{\nu|\alpha \mu}_P = 0 \, .
\end{equation}
This can be inserted into (\ref{eq:knabf}) to find
\begin{equation} \label{new:knabf}
k^\mu_P F^{\mu| \nu \la}_P = \sum_{P=XY}\Big[A^\mu_X F^{\mu| \nu \la}_Y -2 F^{\mu \nu}_X F^{\mu \lambda}_Y - (X\leftrightarrow Y) \Big] +\mathcal{O}(\alpha') \, .
\end{equation}
By virtue of (\ref{eq:kf}), (\ref{eq:knabf}) and (\ref{new:knabf}), one can rewrite any contraction of
$F_P^{\mu\nu}$ and $F_P^{\mu |\nu \la}$ with the corresponding momentum $k_P$ in terms of deconcatenations.
We will always work to first order in $\ap$, but we will split the proofs into different orders for the 
convenience of the reader.

%%%%%%%%%%%%%%%%%%%%%%%%%%%%%%%%%%%%%%%%%%%%%%%%
%%%%%%%%%%%%%%%%%%%%%%%%%%%%%%%%%%%%%%%%%%%%%%%%
%%%%%%%%%%%%%%%%%%%%%%%%%%%%%%%%%%%%%%%%%%%%%%%%
%%%%%%%%%%%%%%%%%%%%%%%%%%%%%%%%%%%%%%%%%%%%%%%%
%%%%%%%%%%%%%%%%%%%%%%%%%%%%%%%%%%%%%%%%%%%%%%%%

\subsection{Appearance in the amplitudes}
\label{app:A.1}
The first property we wish to prove is 
\begin{equation} \label{eq:prop1zero}
\sum_{XY=12\ldots n-1} \Mfrak_{X,Y,n} = s_{12\ldots n-1} A^\mu_{12\ldots n-1} A^\mu_n=\sum_{XY=12 \ldots n-1} A^\mu_{[X,Y]} A^\mu_n \,,
\end{equation}
where, as a special case of (\ref{pert3.12}),
\begin{equation}
\begin{split}
\mathfrak{M}_{X,Y,n} &= \frac{1}{2} \left(A^\mu_X F^{\mu \nu}_Y A^\nu_n + A^\mu_Y F^{\mu \nu}_n A^\nu_X + A^\mu_n F^{\mu \nu}_X A^\nu_Y\right) -2\ap F^{\mu \nu}_X F^{\nu \lambda}_Y F^{\lambda \mu}_n\\
&+ \frac{\ap}{2} \Big(  F^{\mu | \nu \lambda}_X F^{\nu \lambda}_Y A^\mu_n +  F^{\mu | \nu \lambda}_Y F^{\nu \lambda}_n A^\mu_X + F^{\mu | \nu \lambda}_n F^{\nu \lambda}_X A^\mu_Y \\
 & \ \ \ \ \ \ \  -  F^{\mu| \nu \lambda}_X F^{\nu \lambda}_n A^\mu_Y -  F^{\mu | \nu \lambda}_Y F^{\nu \lambda}_X A^\mu_n -  F^{\mu | \nu \lambda}_n F^{\nu \lambda}_Y A^\mu_X \Big) \,
\end{split}\label{m1order}
\end{equation}
and
\begin{equation}
\begin{split}
A^\mu_{[X,Y]} &= \frac{1}{2} \left( A^\mu_Y (k_Y \cdot A_X) - A^\mu_X (k_X \cdot A_Y) + A^\nu_X F^{\nu \mu}_Y - A^\nu_Y F^{\nu \mu}_X \right)\\
&+ \alpha'(F^{\nu \lambda}_X  F^{\nu |\lambda \mu}_Y - F^{\nu \lambda}_Y  F^{\nu|\lambda \mu}_X) \,.
\end{split}\label{A1order}
\end{equation}
We first focus on the zeroth order of $\alpha'$. Notice how the last two terms on the first line of \eqref{A1order}, when contracted with $A_n^\mu$, exactly match the first and third terms on the first line of \eqref{m1order}. To see that the remaining terms are equal to each other, we notice that 
\begin{equation}
\begin{split}
A^\mu_Y F^{\mu \nu}_n A^\nu_X &= A^\mu_Y A^\nu_X (k^\mu_n A^\nu_n - k^\nu_n A^\mu_n) \\
&=-A^\mu_Y A^\nu_X (k^\mu_X A^\nu_n + k^\mu_Y A^\mu_n - k^\nu_X A^\mu_n - k^\nu_Y A^\mu_n) \\
&=( A_Y \cdot A_n)  (k_Y \cdot A_X)  -  (A_X \cdot A_n)(k_X \cdot A_Y) \, ,
\end{split}
\end{equation}
where we have used momentum conservation $ k^\mu_n = -k^\mu_X - k^\mu_Y$ and transversality. This matches with the missing terms at the $\ap^0$ order of \eqref{A1order}, upon contraction with $A_n^\mu$.

We now show that the same matching occurs between the terms of order $\ap$ in both expressions. Equating them leads to 
\begin{equation} \label{eq:blbl}
\begin{split}
 \sum_{XY=12 \ldots n-1} &\bigg\{ -2  \underbrace{F^{\mu \nu}_X F^{\nu \lambda}_Y F^{\lambda \mu}_n}_{\text{G}} + \frac{1}{2} \Big( \underbrace{ F^{\mu | \nu \lambda}_X F^{\nu \lambda}_Y A^\mu_n}_{\text{A}}  + \underbrace{ F^{\mu | \nu \lambda}_Y F^{\nu \lambda}_n A^\mu_X}_{\text{E}} \\
 & + \underbrace{ F^{\mu | \nu \lambda}_n F^{\nu \lambda}_X A^\mu_Y}_{\text{C}} - \underbrace{ F^{\mu | \nu \lambda}_X F^{\nu \lambda}_n A^\mu_Y}_{\text{F}} - \underbrace{  F^{\mu| \nu \lambda}_Y F^{\nu \lambda}_X A^\mu_n}_{\text{B}}  - \underbrace{ F^{\mu | \nu \lambda}_n F^{\nu \lambda}_Y A^\mu_X}_{\text{D}} \Big) \bigg\}\\
 & =  \sum_{XY=12 \ldots n-1}  (F^{\nu \lambda}_X  F^{\nu | \lambda \mu}_Y A^\mu_n - F^{\nu \lambda}_Y  F^{\nu | \lambda \mu}_X A^\mu_n) \,.
\end{split}
\end{equation} 
When using the Jacobi identity \eqref{eq:jac}, two of the terms on the left-hand side combine 
to cancel exactly the two on the right-hand side:
\begin{equation}
\begin{split}
\sum_{XY=12 \ldots n-1}\frac{1}{2}\left(\text{A} + \text{B}\right) =\sum_{XY=12 \ldots n-1} (-  F^{\nu |\lambda \mu}_X F^{\nu \lambda}_Y A^\mu_n +  F^{\nu |\lambda \mu}_Y F^{\nu \lambda}_X A^\mu_n)\,.
\end{split}
\end{equation} 
Now we have to make sure that the remaining terms C, D, E, F and G on the left-hand side 
of (\ref{eq:blbl}) cancel each other. The first two can be rewritten as
\begin{equation}
\begin{split}
\sum_{XY=12 \ldots n-1}\frac{1}{2} \left(\text{C} + \text{D}\right) &=  F^{\nu | \lambda \mu}_n \sum_{XY=12 \ldots n-1}  (A^\mu_X F^{\nu \lambda}_Y - A^\mu_Y F^{\nu \lambda}_X ) \\
&= -k^\nu_n F^{\lambda \mu}_n ( F^{\mu | \nu \lambda}_{12\ldots n-1} - k^\mu_{12\ldots n-1} F^{\nu \lambda}_{12\ldots n-1}) \\
&= -k^\nu_n F^{\lambda \mu}_n  F^{\mu | \nu \lambda}_{12\ldots n-1} \, ,
\end{split}
\end{equation}
using the form $ F^{\nu | \lambda \mu}_n = k^\nu_n F^{\lambda \mu}_n$ of the single-particle current
as well as momentum conservation $k^\mu_{12\ldots n-1} = - k^\mu_n$ and $k^\mu_n F^{\mu \nu}_n=0$ in passing
to the third line, cf.\ \eqref{eq:knabf}. 

The other two terms combine in a similar way,
\begin{align}
\sum_{XY=12 \ldots n-1}\frac{1}{2} \left(\text{E} + \text{F}\right) &= F^{\nu \lambda}_n \sum_{XY=12 \ldots n-1} ( A^\mu_X  F^{\nu | \mu \lambda}_Y - A^\mu_Y  F^{\nu | \mu \lambda}_X) \notag \\
&= F^{\nu \lambda}_n k^\mu_{12\ldots n-1}  F^{\nu | \mu \lambda}_{12\ldots n-1} - F_n^{\nu \lambda} \sum_{XY=12 \ldots n-1}(F^{\mu \lambda}_X F^{\mu \nu}_Y - F^{\mu \lambda}_Y F^{\mu \nu}_X)  \notag\\
&= - F^{\nu \lambda}_n k^\mu_{n}  F^{\nu | \mu \lambda}_{12\ldots n-1}
  - 2  \sum_{XY=12 \ldots n-1} F^{\nu \lambda}_n F^{\mu \lambda}_X F^{\mu \nu}_Y \\
  &=    \sum_{XY=12 \ldots n-1} \Big\{ {-}\frac{1}{2} \left(\text{C} + \text{D}\right)  +2 \text{G} \Big\}
  \, , \notag
\end{align}
where the second line follows from \eqref{eq:knabf}. In passing to the third line, we have again used momentum conservation 
and exploited antisymmetry $F_n^{\nu \lambda} = F_n^{[\nu \lambda]}$. Hence, the statement \eqref{eq:prop1zero} is 
proved to first order in $\alpha'$.

%%%%%%%%%%%%%%%%%%%%%%%%%%%%%%%%%%%%%%%%%%%%%%%%
%%%%%%%%%%%%%%%%%%%%%%%%%%%%%%%%%%%%%%%%%%%%%%%%

\subsection{Integration by parts}
\label{app:A.2}
The second property of $\mathfrak{M}_{A,B,C}$ we want to prove is the integration-by-parts identity
\begin{equation} \label{eq:integrationbyparts}
\sum_{A=XY} \mathfrak{M}_{X,Y,B} = \sum_{B=XY} \mathfrak{M}_{A,X,Y} \,,
\end{equation}
which translates into the following claim at the zeroth order in $\alpha'$,
\begin{equation} \label{eq:claim}
\begin{split}
\sum_{A=XY} &(A^\mu_X F^{\mu \nu}_Y A^\nu_B +A^\mu_Y F^{\mu \nu}_B A^\nu_X +A^\mu_B F^{\mu \nu}_X A^\nu_Y) \\
- \sum_{B=XY} &(A^\mu_A F^{\mu \nu}_X A^\nu_Y +A^\mu_X F^{\mu \nu}_Y A^\nu_A +A^\mu_Y F^{\mu \nu}_A A^\nu_X)={\cal O}(\ap) \, .
\end{split}
\end{equation}
We can rewrite the second term $\sim A_X A_Y$ in the first line using the antisymmetry of $F_B^{\mu \nu}$ and the
definition (\ref{pert3.6}) of $F_A^{\mu \nu}$:
% recursion for $F_A^{\mu \nu}$ in (\ref{pert3.6}):
\begin{align}
\sum_{A=XY} A^\mu_Y F^{\mu \nu}_B A^\nu_X &= \sum_{A=XY} \frac{1}{2} F^{\mu \nu}_B (A^\mu_Y A^\nu_X - A^\nu_Y A^\mu_X) \notag \\
%&= \frac{1}{2} F^{\mu \nu}_B F^{\mu \nu}_A - \frac{1}{2} F^{\mu \nu}_B k^\mu_A A^\nu_A + \frac{1}{2} F^{\mu \nu}_B k^\nu_A A^\mu_A \\
&= \frac{1}{2} F^{\mu \nu}_B F^{\mu \nu}_A - F^{\mu \nu}_B k^\mu_A A^\nu_A\, .
 \label{eq:bl1}
\end{align}
The analogous sum in the second line of (\ref{eq:claim}) has a similar term related by $A\leftrightarrow B$,
\begin{equation} \label{eq:bl2}
- \sum_{B=XY} A^\mu_Y F^{\mu \nu}_A A^\nu_X= - \frac{1}{2} F^{\mu \nu}_A F^{\mu \nu}_B + F^{\mu \nu}_A k^\mu_B A^\nu_B \, ,
\end{equation}
where $F^{\mu \nu}_A F^{\mu \nu}_B $ cancels against (\ref{eq:bl1}). For the remaining terms $F^{\mu \nu}_A k^\mu_B A^\nu_B - F^{\mu \nu}_B k^\mu_A A^\nu_A$, we apply momentum conservation $k^\mu_A + k^\mu_B=0$ and the 
relation \eqref{eq:kf} for $k_A^\mu F_A^{\mu \nu}$,
\begin{align}
& \sum_{A=XY} A^\mu_Y F^{\mu \nu}_B A^\nu_X - \sum_{B=XY} A^\mu_Y F^{\mu \nu}_A A^\nu_X=-F^{\mu \nu}_A k^\mu_A A^\nu_B + F^{\mu \nu}_B k^\mu_B A^\nu_A  \notag \\
&= - \sum_{A=XY} (A^\mu_X F^{\mu \nu}_Y A^\nu_B - A^\mu_Y F^{\mu \nu}_X A^\nu_B) + \sum_{B=XY} (A^\mu_X F^{\mu \nu}_Y A^\nu_A - A^\mu_Y F^{\mu \nu}_X A^\nu_A)  \label{skipped} \\
& \ -2\ap\sum_{A=XY} (F^{\mu \lambda}_X  F^{\mu | \lambda \nu}_Y - F^{\mu \lambda}_Y  F^{\mu | \lambda \nu}_X) A^\nu_B + 2\ap\sum_{B=XY} ( F^{\mu \lambda}_X  F^{\mu | \lambda \nu}_Y - F_Y^{\mu \lambda}  F_X^{\mu | \lambda \nu} ) A^\nu_A  \, .\notag
\end{align}
Inserting this into \eqref{eq:claim} and ignoring the ${\cal O}(\ap)$-term in the last line, we conclude that the property \eqref{eq:integrationbyparts} is indeed satisfied to zeroth order in $\alpha'$.

We now show that the property is still valid at the first order in $\alpha'$. In doing that, we have to combine
the last term of (\ref{skipped}) with the ${\cal O}(\ap)$-terms in the $\Mfrak_{A,B,C}$ of (\ref{eq:integrationbyparts}).
Hence, the leftover task is to prove that
\begin{equation} 
\label{leftov}
\begin{split}
0 &= \sum_{A=XY} \Big( 
 (F^{\mu | \lambda \nu}_X F^{\mu \lambda}_Y   -  F^{\mu | \lambda \nu}_Y F^{\mu \lambda}_X )A^\nu_B 
-2 F^{\mu \nu}_X F^{\nu \lambda}_Y F^{\lambda \mu}_B 
 \\
& \ \ + \frac{1}{2} \Big( \underbrace{ F^{\mu | \nu \lambda}_X F^{\nu \lambda}_Y A^\mu_B}_{\text{M}} + \underbrace{ F^{\mu | \nu \lambda}_Y F^{\nu \lambda}_B A^\mu_X}_{\text{I}}
 + \underbrace{ F^{\mu | \nu \lambda}_B F^{\nu \lambda}_X A^\mu_Y}_{\text{E}} \\
&\ \ \ \ \ \ - \underbrace{ F^{\mu | \nu \lambda}_X F^{\nu \lambda}_B A^\mu_Y}_{\text{J}} - \underbrace{ F^{\mu | \nu \lambda}_Y F^{\nu \lambda}_X A^\mu_B}_{\text{N}}  - \underbrace{ F^{\mu | \nu \lambda}_B F^{\nu \lambda}_Y A^\mu_X}_{\text{F}} \Big) \Big) \\
&- \sum_{B=XY} \Big(
 (F^{\mu | \lambda \nu}_X F^{\mu \lambda}_Y 
-  F^{\mu | \lambda \nu}_Y F^{\mu \lambda}_X ) A^\nu_A 
 -2 F^{\mu \nu}_A F^{\nu \lambda}_X F^{\lambda \mu}_Y 
\\
&\ \ +\frac{1}{2} \Big( \underbrace{ F^{\mu | \nu \lambda}_A F^{\nu \lambda}_X A^\mu_Y}_{\text{G}} + \underbrace{ F^{\mu | \nu \lambda}_X F^{\nu \lambda}_Y A^\mu_A}_{\text{C}} + \underbrace{ F^{\mu | \nu \lambda}_Y F^{\nu \lambda}_A A^\mu_X}_{\text{K}} \\
&\ \ \ \ \ \ - \underbrace{ F^{\mu | \nu \lambda}_A F^{\nu \lambda}_Y A^\mu_X}_{\text{H}} 
- \underbrace{ F^{\mu | \nu \lambda}_X F^{\nu \lambda}_A A^\mu_Y}_{\text{L}}
- \underbrace{ F^{\mu| \nu \lambda}_Y F^{\nu \lambda}_X A^\mu_A}_{\text{D}} \Big)  \Big)\, .
\end{split}
\end{equation}
Let us start by using the Jacobi identity \eqref{eq:jac} to rewrite the following terms:
\begin{align}
\sum_{A=XY}\frac{1}{2} \left(\text{M} + \text{N}\right) = \sum_{A=XY}  ( F^{\mu | \lambda \nu}_Y F^{\mu \lambda}_X   -  F^{\mu | \lambda \nu}_X F^{\mu \lambda}_Y ) A^\nu_B 
 \\
\sum_{B=XY}\frac{1}{2} \left(\text{C} + \text{D}\right) = \sum_{B=XY} ( F^{\mu | \lambda \nu}_Y F^{\mu \lambda}_X -  F^{\mu | \lambda \nu}_X F^{\mu \lambda}_Y )A^\nu_A 
\end{align}
The right-hand sides cancel the first two terms inside the sums in (\ref{leftov}) 
over $A=XY$ and $B=XY$. Then, using the recursive definition of $ F^{\mu | \nu \lambda}_P$
in \eqref{pert3.8}, we can write
\begin{equation}
\begin{split}
\sum_{A=XY}\frac{1}{2} \left(\text{E} + \text{F}\right) &=  F^{\nu | \lambda \mu}_B \sum_{A=XY} (A^\mu_X F^{\nu \lambda}_Y - A^\mu_Y F^{\nu \lambda}_X) \\
&= F^{\nu | \lambda \mu}_B ( k^\mu_A F^{\nu \lambda}_A -  F^{\mu | \nu \lambda}_A)\,.
\end{split}
\end{equation}
Since the contribution from $\text{G} + \text{H}$ takes the same form with $A\leftrightarrow B$,
the terms $ F^{\nu | \lambda \mu}_B F^{\mu | \nu \lambda}_A$ cancel from the combination
\begin{equation}
\sum_{A=XY}\frac{1}{2} \left(\text{E} + \text{F}\right) - \sum_{B=XY}\frac{1}{2} \left(\text{G} + \text{H}\right) = k^\mu_A F^{\nu \lambda}_A  F^{\nu | \lambda \mu}_B - k^\mu_B F^{\nu \lambda}_B  F^{\nu | \lambda \mu}_A \, .
\label{here1}
\end{equation}
Also, using Jacobi identity \eqref{eq:jac} as well as \eqref{eq:knabf}, we have
\begin{equation}
\begin{split}
\sum_{A=XY}\frac{1}{2} \left(\text{I} + \text{J}\right) &= F^{\nu \lambda}_B \sum_{A=XY} ( -A^\mu_X  F^{\nu | \lambda \mu}_Y + A^\mu_Y  F^{\nu | \lambda \mu}_X) \\
&= F^{\nu \lambda}_B \Big(k^\mu_A  F^{\nu | \mu \lambda}_A - \sum_{A=XY} ( F^{\mu \lambda}_X F^{\mu \nu}_Y - F^{\mu \lambda}_Y F^{\mu \nu}_X) \Big)  \label{here2}\\
&= k^\mu_A F^{\nu \lambda}_B  F^{\nu | \mu \lambda}_A -2 \sum_{A=XY} F^{\nu \lambda}_B F^{\mu \lambda}_X F^{\mu \nu}_Y
\end{split}
\end{equation}
and using the same manipulations
\begin{equation}
\sum_{B=XY}\frac{1}{2} \left(\text{K} + \text{L}\right) = k^\mu_B F^{\nu \lambda}_A  F^{\nu | \mu \lambda}_B - 2 \sum_{B=XY} F^{\nu \lambda}_A F^{\mu \lambda}_X F^{\mu \nu}_Y \, .
\label{here3}
\end{equation}
The terms of the form $k^\mu_{\cdot} F^{\nu \lambda}_{\cdot}  F^{\nu | \mu \lambda}_{\cdot}$ cancel between (\ref{here1}),
(\ref{here2}) and (\ref{here3}) by momentum conservation $k_A+k_B=0$. Finally, the contributions of the form $\sum_{A=XY} F^{\nu \lambda}_B F^{\mu \lambda}_X F^{\mu \nu}_Y$ cancel between (\ref{here2}), (\ref{here3}) and the leftover terms of (\ref{leftov}). This concludes our proof of the integration-by-parts identity (\ref{eq:integrationbyparts}) to the order of $\ap$.

%%%%%%%%%%%%%%%%%%%%%%%%%%%%%%%%%%%%%%%%%%%%%%%%
%%%%%%%%%%%%%%%%%%%%%%%%%%%%%%%%%%%%%%%%%%%%%%%%

\subsection{Gauge algebra}
\label{app:A.3}
Finally, we want to see how a non-linear gauge transformation (\ref{pert1.gau}) 
acts on $\mathfrak{M}_{X,Y,Z}$. To zeroth order in $\alpha'$, we get
\begin{equation} \label{here5}
\begin{split}
\delta_\Omega \mathfrak{M}_{X,Y,Z} &= \frac{1}{2} \delta_\Omega (A^\mu_X F^{\mu \nu}_Y A^\nu_Z + \mathrm{cyc}(X,Y,Z)) + {\cal O}(\ap)\\
&= \frac{1}{2} \Big( k^\mu_X \Omega_X F^{\mu \nu}_Y A^\nu_Z - \sum_{X=AB} (A^\mu_A \Omega_B - A^\mu_B \Omega_A) F^{\mu \nu}_Y A^\nu_Z \\
&- \sum_{Y=AB} (F^{\mu \nu}_A \Omega_B - F^{\mu \nu}_B \Omega_A ) A^\mu_X A^\nu_Z + A^\mu_X F^{\mu \nu}_Y k^\nu_Z \Omega_Z \\
&- \sum_{Z=AB} (A^\nu_A \Omega_B - A^\nu_B \Omega_A) A^\mu_X F^{\mu \nu}_Y + \mathrm{cyc}(X,Y,Z) \Big) + {\cal O}(\ap)\, .
\end{split}
\end{equation}
Let us look at the terms which are not inside a sum. We can start by grouping all the ones with 
the same $\Omega_{\cdot}$ coefficient, and use momentum conservation $k^\mu_X + k^\mu_Y + k^\mu_Z =0$:
\begin{align}
&k^\mu_X \Omega_X F^{\mu \nu}_Y A^\nu_Z + k^\nu_X \Omega_X F^{\mu \nu}_Z A^\mu_Y  + \mathrm{cyc}(X,Y,Z) \label{here4} \\
& \ \ = - \Omega_X \big( F^{\mu \nu}_Y A^\nu_Z (k^\mu_Y + k^\mu_Z)+ A^\mu_Y F^{\mu \nu}_Z (k^\nu_Y + k^\nu_Z) \big) + \mathrm{cyc}(X,Y,Z)  \, . \notag
\end{align}
We then rewrite these four terms via
\begin{align}
-\Omega_X k^\mu_Y F^{\mu \nu}_Y A^\nu_Z &= - \Omega_X \sum_{Y=AB}
 (A^\mu_A F^{\mu \nu}_B - A^\mu_B F^{\mu \nu}_A) A^\nu_Z + {\cal O}(\ap) \notag \\
%- \Omega_X A^\mu_Y F^{\mu \nu}_Z k^\nu_Z &= \Omega_X \sum_{Z=AB} (A^\mu_A F^{\mu \nu}_B - A^\mu_B F^{\mu \nu}_A) A^\nu_Y \\
-\Omega_X F^{\mu \nu}_Y k^\mu_Z A^\nu_Z &= -\frac{1}{2} \Omega_X F^{\mu \nu}_Y (k^\mu_Z A^\nu_Z - k^\nu_Z A^\mu_Z)
\label{eq:bb} \\
&= -\frac{1}{2} \Omega_X F^{\mu \nu}_Y F^{\mu \nu}_Z - \Omega_X F^{\mu \nu}_Y \sum_{Z=AB} A^\mu_A A^\nu_B \notag
%- \Omega_X F^{\mu \nu}_Z k^\nu_Y A^\mu_Y &= \frac{1}{2} \Omega_X F^{\mu \nu}_Z F^{\mu \nu}_Y + \Omega_X F^{\mu \nu}_Z \sum_{Y=AB} A^\mu_A A^\mu_B,
\end{align}
and the same identities with $(Y\leftrightarrow Z)$. In the first line, we rewrote $k^\mu_Y F^{\mu \nu}_Y$
using the zeroth order in $\ap$ of \eqref{eq:kf}. The last two lines of (\ref{eq:bb}) are based on the antisymmetry 
of $F_Y^{\mu \nu}$ and the definition \eqref{pert3.6} of $F_Z^{\mu \nu}$. Notice that $F^{\mu \nu}_Y F^{\mu \nu}_Z$
cancels in the antisymmetrization w.r.t.\ $Y\leftrightarrow Z$ in (\ref{here4}). Hence, all the leftover terms in (\ref{here5})
involve a sum over deconcatenations, either $\sum_{X=AB}$ or one of $(X\leftrightarrow Y,Z)$. We collect
all the expressions from the cyclic permutations in (\ref{here5}) where the sum is $\sum_{X=AB}$
\begin{equation} \label{here9}
\begin{split}
\delta_\Omega \mathfrak{M}_{X,Y,Z} &= \frac{1}{2} \sum_{X=AB} \Big( ( \Omega_A A^\mu_B - \Omega_B A^\mu_A) F^{\mu \nu}_Y A^\nu_Z + A^\mu_Z (\Omega_A F^{\mu \nu}_B - \Omega_B F^{\mu \nu}_A ) A^\nu_Y \\
&+ A^\mu_Y F^{\mu \nu}_Z ( \Omega_A A^\nu_B - \Omega_B A^\nu_A) + \Omega_Y \big[ (A^\mu_A F^{\mu \nu}_B - A^\mu_B F^{\mu \nu}_A) A^\nu_Z - F^{\mu \nu}_Z A^\mu_A A^\nu_B \big] \\
&- \Omega_Z \big[ (A^\mu_A F^{\mu \nu}_B - A^\mu_B F^{\mu \nu}_A)A^\nu_Y - F^{\mu \nu}_Y A^\mu_A A^\nu_B \big] \Big) + \mathrm{cyc}(X,Y,Z) + {\cal O}(\ap) \, .
\end{split}
\end{equation}
It turns out that the coefficient of each of the $\Omega$'s inside the above sum can be identified
as some $\mathfrak{M}_{P,Q,R}$ with various combinations of the three words:
\begin{equation}
\begin{split}
\delta_\Omega \mathfrak{M}_{X,Y,Z} &= \sum_{X=AB} \Big( \Omega_A \mathfrak{M}_{B,Y,Z} - \Omega_B \mathfrak{M}_{A,Y,Z} + \Omega_Y \mathfrak{M}_{A,B,Z} - \Omega_Z \mathfrak{M}_{Y,A,B} \Big)\\
& + \mathrm{cyc}(X,Y,Z) + {\cal O}(\ap) \, .
\end{split}
\end{equation}
The object inside the sum over $X=AB$ is totally antisymmetric in $A,B,Y,Z$ and 
can be identified as $\Omega_{A,B,C,D}$ as defined in (\ref{pert3.51}). Hence, the zeroth
order of the gauge transformation (\ref{here5}) can be written as
\begin{equation}\label{eq:gaugem}
\delta_\Omega \mathfrak{M}_{X,Y,Z}  = \sum_{X=AB} \Omega_{A,B,Y,Z} + \sum_{Y=AB} \Omega_{A,B,Z,X} + \sum_{Z=AB} \Omega_{A,B,X,Y}+ {\cal O}(\ap) \, .
\end{equation}
We now want to extend the proof of \eqref{eq:gaugem} to the first order in $\alpha'$. First of all, terms of ${\cal O}(\alpha')$ 
have been neglected when inserting \eqref{eq:kf} into the first term of \eqref{eq:bb}. Therefore, we carry forward the 
following terms in $\delta_{\Omega} \Mfrak_{X,Y,Z}$,
\begin{align}
&\Omega_X \sum_{Z=AB} (F^{\mu \lambda}_A  F^{\mu | \lambda \nu}_B - F^{\mu \lambda}_B  F^{\mu | \lambda \nu}_A) A^\nu_Y  \pm \mathrm{perm}(X,Y,Z) \notag \\
 =& -\Omega_Z \sum_{X=AB} (F^{\mu \lambda}_A  F^{\mu | \lambda \nu}_B - F^{\mu \lambda}_B F^{\mu | \lambda \nu}) A^\nu_Y \label{here7}  \\
&+\Omega_Y \sum_{X=AB} (F^{\mu \lambda}_A  F^{\mu | \lambda \nu}_B - F^{\mu \lambda}_B  F^{\mu | \lambda \nu}_A) A^\nu_Z + \mathrm{cyc}(X,Y,Z) \, , \notag
\end{align}
where we have spelt out all the terms of the same form $\sum_{X=AB}$ as in (\ref{here9}).
This needs to be combined with the gauge variation of the ${\cal O}(\alpha')$ terms in the
definition (\ref{pert3.12}) of $\Mfrak_{X,Y,Z}$:
\begin{align}
\text{L} &= -2 \delta_\Omega (F^{\mu \nu}_X F^{\nu \lambda}_Y F^{\lambda \mu}_Z) \notag
\\
&= 2 \sum_{X=AB} (F^{\mu \nu}_A \Omega_B - F^{\mu \nu}_B \Omega_A) F^{\nu \lambda}_Y F^{\lambda \mu}_Z + \mathrm{cyc}(X,Y,Z) \notag\\
\text{G} &= \frac{1}{2} \delta_\Omega ( F^{\mu | \nu \lambda}_X F^{\nu \lambda}_Y A^\mu_Z) \pm \mathrm{perm}(X,Y,Z)  \label{eq:aa}\\
&= \frac{1}{2} \Big( \sum_{X=AB} ( F^{\mu | \nu \lambda}_A \Omega_B -  F^{\mu | \nu \lambda}_B \Omega_A) F^{\nu \lambda}_Y A^\mu_Z  \notag \\
&\ \ \ \ - \sum_{Y=AB} (F^{\nu \lambda}_A \Omega_B - F^{\nu \lambda}_B \Omega_A)  F^{\mu | \nu \lambda}_X A^\mu_Z + \Omega_Z k^\mu_Z  F^{\mu | \nu \lambda}_X F^{\nu \lambda}_Y  \notag \\
&\ \ \ \ - \sum_{Z=AB} (A^\mu_A \Omega_B - A^\mu_B \Omega_A)  F^{\mu | \nu \lambda}_X F^{\nu \lambda}_Y \pm \mathrm{perm}(X,Y,Z) \Big)\, . \notag
\end{align}
The only term in (\ref{eq:aa}) which is not yet in the form of a deconcatenation sum will now be rewritten 
via momentum conservation $k^\mu_X + k^\mu_Y + k^\mu_Z=0$:
\begin{equation}
\Omega_Z k^\mu_Z  F^{\mu | \nu \lambda}_X F^{\nu \lambda}_Y = \underbrace{-\Omega_Z k^\mu_X  F^{\mu | \nu \lambda}_X F^{\nu \lambda}_Y}_{\text{C}} \underbrace{-\Omega_Z k^\mu_Y  F^{\mu | \nu \lambda}_X F^{\nu \lambda}_Y}_{\text{D}}.
\end{equation}
The first term calls for the relation \eqref{eq:knabf},
\begin{equation}
\text{C} = 
 -2 \Omega_Z \sum_{X=AB} (A^\mu_A  F^{\nu | \mu \lambda}_B - A^\mu_B  F^{\nu | \mu \lambda}_A
 +F^{\mu \lambda}_A F^{\mu \nu}_B - F^{\mu \lambda}_B F^{\mu \nu}_A) F^{\nu \lambda}_Y\, ,
\end{equation}
which we combined with the Jacobi identity (\ref{eq:jac}).
In the other term we use the definition (\ref{pert3.8}) of the currents $F_Y^{\mu|\nu\la}$,
\begin{equation}
\text{D} = -\Omega_Z  F^{\mu | \nu \lambda}_X \Big(  F^{\mu | \nu \lambda}_Y + \sum_{Y=AB} (A^\mu_A F^{\nu \lambda}_B - A^\mu_B F^{\nu \lambda}_A) \Big) \, .
\end{equation}
The first term $ F^{\mu | \nu \lambda}_X   F^{\mu | \nu \lambda}_Y$ cancels 
under the antisymmetrization w.r.t.\ $X,Y,Z$ of (\ref{eq:aa}),
\begin{equation}
\text{D} \pm \mathrm{perm}(X,Y,Z) = - \Omega_Y \sum_{X=AB} (A^\mu_A F^{\nu \lambda}_B - A^\mu_B F^{\nu \lambda}_A)  F^{\mu | \nu \lambda}_Z \pm \mathrm{perm}(X,Y,Z)\, ,
\end{equation}
such that all the terms in the quantity $\text{G}$ have been expressed 
via deconcatenation sums:
\begin{equation} \label{here12}
\begin{split}
\text{G} &= \frac{1}{2}  \sum_{X=AB} \Big(- ( F^{\mu | \nu \lambda}_A \Omega_B -  F^{\mu | \nu \lambda}_B \Omega_A ) F^{\nu \lambda} A^\mu_Z \\
&- (F^{\nu \lambda}_A \Omega_B - F^{\nu \lambda}_B \Omega_A )  F^{\mu | \nu \lambda}_Z A^\mu_Y - 2 \Omega_Z (A^\mu_A  F^{\nu | \mu \lambda}_B - A^\mu_B  F^{\nu | \mu \lambda}_A) F^{\nu \lambda}_Y \\
&- 2\Omega_Z (F^{\mu \lambda}_A F^{\mu \nu}_B - F^{\mu \lambda}_B F^{\mu \nu}_A) F^{\nu \lambda}_Y - \Omega_Y  (A^\mu_A F^{\nu \lambda}_B - A^\mu_B F^{\nu \lambda}_A)  F^{\mu | \nu \lambda}_Z \\
&-  (A^\mu_A \Omega_B - A^\mu_B \Omega_A)  F^{\mu | \nu \lambda}_Y F^{\nu \lambda}_Z \pm \mathrm{perm}(X,Y,Z) \Big) \, .
\end{split}
\end{equation}
Once we convert the permutation sum in (\ref{here12}) to a cyclic one,
\beq
f(X,Y,Z) + \mathrm{perm}(X,Y,Z) = f(X,Y,Z)- f(X,Z,Y) + \mathrm{cyc}(X,Y,Z)\, ,
\eeq
the result for $\text{G}$ is perfectly lined up with (\ref{here7}) and the expression for $\text{L}$ in (\ref{eq:aa}).
Hence, the overall ${\cal O}(\alpha')$ contribution to the gauge variation of $\Mfrak_{X,Y,Z}$ in (\ref{pert3.12})
reads
\begin{equation} \label{longeq}
\begin{split}
\delta_\Omega \mathfrak{M}_{X,Y,Z} \, \big|_{\ap^1}& = \frac{1}{2} \sum_{X=AB} \Big\{   2(F^{\mu \nu}_A \Omega_B - F^{\mu \nu}_B \Omega_A) F^{\nu \lambda}_Y F^{\lambda \mu}_Z \\
&- \frac{1}{2}( F^{\mu | \nu \lambda}_A \Omega_B -  F^{\mu | \nu \lambda}_B \Omega_A) F^{\nu \lambda}_Y A^\mu_Z + \frac{1}{2} (  F^{\mu | \nu \lambda}_A \Omega_B -  F^{\mu | \nu \lambda}_B \Omega_A) F^{\nu \lambda}_Z A^\mu_Y \\
&- \frac{1}{2} (F^{\nu \lambda}_A \Omega_B - F^{\nu \lambda}_B \Omega_A) F^{\mu | \nu \lambda}_Z A^\mu_Y + \frac{1}{2} (F^{\nu \lambda}_A \Omega_B - F^{\nu \lambda}_B \Omega_A)  F^{\mu | \nu \lambda}_Y A^\mu_Z\\
 &- \frac{1}{2} (A^\mu_A \Omega_B - A^\mu_B \Omega_A)  F^{\mu | \nu \lambda}_Y F^{\nu \lambda}_Z + \frac{1}{2} ( A^\mu_A \Omega_B - A^\mu_B \Omega_A) F^{\mu | \nu \lambda}_Z F^{\nu \lambda}_Y \\
 & + \Omega_Y \big[ (F^{\mu \lambda}_A  F^{\mu | \lambda \nu}_B - F^{\mu \lambda}_B  F^{\mu | \lambda \nu}_A) A^\nu_Z - \frac{1}{2} (A^\mu_A F^{\nu \lambda}_B - A^\mu_B F^{\nu \lambda}_A)  F^{\mu | \nu \lambda}_Z \\
 &\ \ \ \ \ \ + (A^\mu_A  F^{\nu | \mu \lambda}_B - A^\mu_B  F^{\nu | \mu \lambda}_A) F^{\nu \lambda}_Z + (F^{\mu \lambda}_A F^{\mu \nu}_B - F^{\mu \lambda}_B F^{\mu \nu}_A) F^{\nu \lambda}_Z \big] \\
 &+ \Omega_Z \big[ {-}(F^{\mu \lambda}_A  F^{\mu | \lambda \nu}_B - F^{\mu \lambda}_B  F^{\mu | \lambda \nu}_A) A^\nu_Y + \frac{1}{2} (A^\mu_A F^{\nu \lambda}_B - A^\mu_B F^{\nu \lambda}_A)  F^{\mu | \nu \lambda}_Y \\
 &\ \ \ \ \ \ - (A^\mu_A  F^{\nu | \mu \lambda}_B - A^\mu_B  F^{\nu | \mu \lambda}_A) F^{\nu \lambda}_Y - (F^{\mu \lambda}_A F^{\mu \nu}_B - F^{\mu \lambda}_B F^{\mu \nu}_A) F^{\nu \lambda}_Y \big]  \Big\} \\
 & + \mathrm{cyc}(X,Y,Z) \, .
\end{split}
\end{equation}
The coefficients of $\Omega_A$, $\Omega_B$, $\Omega_X$ and $\Omega_Y$ are identified
as the ${\cal O}(\alpha')$ terms of $\mathfrak{M}_{A,B,C}$. Hence, with the definition \eqref{pert3.51}
of $\Omega_{A,B,Y,Z}$, the expression in (\ref{longeq}) condenses to
\begin{equation}
\delta_\Omega \mathfrak{M}_{X,Y,Z}  \, \big|_{\ap^1} = \sum_{X=AB} \Omega_{A,B,Y,Z} \, \big|_{\ap^1} + \mathrm{cyc}(X,Y,Z)
\end{equation}
and confirms (\ref{eq:gaugem}) to also hold at the first order in $\ap$.

%%%%%%%%%%%%%%%%%%%%%%%%%%%%%%%%%%%%%%%%%%%%%%%%
%%%%%%%%%%%%%%%%%%%%%%%%%%%%%%%%%%%%%%%%%%%%%%%%
%%%%%%%%%%%%%%%%%%%%%%%%%%%%%%%%%%%%%%%%%%%%%%%%
%%%%%%%%%%%%%%%%%%%%%%%%%%%%%%%%%%%%%%%%%%%%%%%%
%%%%%%%%%%%%%%%%%%%%%%%%%%%%%%%%%%%%%%%%%%%%%%%%

\section{The explicit form of gauge scalars towards BCJ gauge}
\label{app:D}

%%%%%%%%%%%%%%%%%%%%%%%%%%%%%%%%%%%%%%%%%%%%%%%%
%%%%%%%%%%%%%%%%%%%%%%%%%%%%%%%%%%%%%%%%%%%%%%%%

\subsection{The local building block $h_{12345}$}
\label{app:D.1}

In this appendix, we spell out two representations of the local rank-five scalar $h_{12345}$ that arises in
the redefinition (\ref{pert4.19}) towards the multiparticle polarization $a^\mu_{12345}$.
The scalar $h_{12345}$ can be expressed in terms of the local building blocks $N_{X,Y,Z}$ defined in (\ref{pert3.12loc})
which are composed from multiparticle polarizations at rank $\leq 3$,
\begin{align}
h_{12345} &= \frac{1}{10} \big[N_{123,4,5}+N_{453,2,1} + N_{12,3,45}\big] 
+\frac{1}{60}   \big[ N_{1,2,3} \left(k_{123} \cdot a_{45}\right)- N_{3,4,5}\left(k_{345} \cdot a_{12} \right) \big]\notag \\
  & +\frac{1}{240} \big\{ \left(k_{1234} \cdot a_5\right)
   \big[2 N_{12,3,4}+N_{13,2,4}-N_{14,2,3}-N_{23,1,4}  +N_{24,1,3}+2 N_{34,1,2}\big]\notag \\
   &\ \ \ \ -\left(k_{1235} \cdot a_4\right)
   \big[2 N_{12,3,5}+N_{13,2,5}-N_{15,2,3}   - N_{23,1,5}+N_{25,1,3}+2  N_{35,1,2}\big] \notag \\
   &\ \ \ \ -\left(k_{2345} \cdot a_1 \right)
   \big[2 N_{54,3,2}+N_{53,4,2}-N_{52,4,3}-N_{43,5,2}+N_{42,5,3}+2   N_{32,5,4}\big]  \notag \\ 
   &\ \ \ \ + \left(k_{1345} \cdot a_2 \right)
   \big[2 N_{54,3,1}+N_{53,4,1}-N_{51,4,3}-N_{43,5,1}+N_{41,5,3}+2N_{31,5,4}\big]\big\} \notag \\
   & -\frac{1}{240}
   \left(k_{1245} \cdot a_3 \right)  \big[N_{1,4,5} \left(k_{145} {\cdot} a_2 \right)-N_{2,4,5} \left(k_{245} {\cdot} a_1\right)
    -N_{1,2,4} \left(k_{124} {\cdot} a_5 \right)+N_{1,2,5} \left(k_{125} {\cdot} a_4 \right)\big] \notag \\
 & +\frac{1}{40}  \left(k_{1245} \cdot a_3 \right) \big[ N_{12,4,5} - N_{45,1,2} \big] \, .
\end{align}
A more compact expression can be attained by additionally employing lower-rank scalars $h_{ijk}$ and $h_{ijkl}$,
\begin{align}
h_{12345} &=\frac{1}{10} \big[N_{123,4,5}+N_{453,2,1}+ N_{12,3,45} \big] 
+ \frac{1}{40} \left(k_{1245} \cdot a_3 \right)  \big[ N_{12,4,5}  + N_{45,2,1} \big]
\notag \\
    & +\frac{1}{10} \big[ h_{1234} \left(k_{1234} \cdot a_5\right)-h_{1235} \left(k_{1235} \cdot a_4\right)-h_{5432}\left(k_{2345}\cdot a_1\right)+h_{5431} \left(k_{1345}\cdot a_2\right)\big] \notag \\
   & +\frac{1}{40}
   \left(k_{1245} \cdot a_3 \right) \big[
   %{-}s_{45}\mathfrak{M}_{45,1,2}+s_{12} \mathfrak{M}_{12,4,5}
h_{452} \left(k_{245}\cdot
   a_1\right)  - h_{451}
   \left(k_{145} \cdot a_2\right) +h_{124} \left(k_{124} \cdot a_5 \right)-h_{125} \left(k_{125} \cdot a_4 \right)\big] \notag \\
    & +\frac{1}{10} \big[
   h_{123} \left(k_{123}\cdot a_{45}\right)-
   h_{453} \left(k_{345}\cdot a_{12} \right)\big] \, .
\end{align}

%%%%%%%%%%%%%%%%%%%%%%%%%%%%%%%%%%%%%%%%%%%%%%%%
%%%%%%%%%%%%%%%%%%%%%%%%%%%%%%%%%%%%%%%%%%%%%%%%

\subsection{An alternative expression for $H_{1234}$}
\label{app:D.0}

The gauge scalar $H_{1234}$ in (\ref{rk4H}) which relates Berends--Giele currents
in Lorenz and BCJ gauge via (\ref{pert4.mn}) admits the following alternatively representation
\begin{align}
s_{1234}H_{1234} &= \frac{1}{48}(k_{123}\cdot a_4)\Mfrak_{1,2,3}\left(\frac{3}{s_{123}}\left(\frac{1}{s_{12}}-\frac{1}{s_{23}}\right)+\frac{1}{s_{234}}\left(\frac{1}{s_{34}}-\frac{1}{s_{23}}\right) +\frac{2}{s_{12}s_{34}}\right) \notag \\
& + \frac{1}{48}(k_{234}\cdot a_1)\Mfrak_{2,3,4}\left(\frac{1}{s_{123}}\left(\frac{1}{s_{12}}-\frac{1}{s_{23}}\right)+\frac{3}{s_{234}}\left(\frac{1}{s_{34}}-\frac{1}{s_{23}}\right) +\frac{2}{s_{12}s_{34}}\right) \notag \\
& + \frac{1}{48}(k_{134}\cdot a_2)\Mfrak_{1,3,4}\left(\frac{1}{s_{123}}\left(\frac{1}{s_{23}}-\frac{1}{s_{12}}\right)+\frac{1}{s_{234}}\left(\frac{1}{s_{34}}-\frac{1}{s_{23}}\right) -\frac{2}{s_{12}s_{34}}\right) \notag \\
& + \frac{1}{48}(k_{124}\cdot a_3)\Mfrak_{1,2,4}\left(\frac{1}{s_{123}}\left(\frac{1}{s_{12}}-\frac{1}{s_{23}}\right)+\frac{1}{s_{234}}\left(\frac{1}{s_{23}}-\frac{1}{s_{34}}\right) -\frac{2}{s_{12}s_{34}}\right)  \\
& + \frac{1}{8}s_{12}\Mfrak_{12,3,4}\left(-\frac{1}{s_{123}s_{12}}+\frac{1}{s_{234}s_{34}} +\frac{2}{s_{12}s_{34}}\right) \notag\\
& + \frac{1}{8}s_{34}\Mfrak_{34,1,2}\left(-\frac{1}{s_{123}s_{12}}+\frac{1}{s_{234}s_{34}} -\frac{2}{s_{12}s_{34}}\right) \notag\\
& - \frac{1}{8}\Mfrak_{32,1,4}\left(\frac{1}{s_{123}}-\frac{1}{s_{234}}\right)- \frac{1}{8}s_{14}\Mfrak_{14,3,2}\left(\frac{1}{s_{123}s_{23}}-\frac{1}{s_{234}s_{23}}\right)  \, .
\notag
\end{align}

%%%%%%%%%%%%%%%%%%%%%%%%%%%%%%%%%%%%%%%%%%%%%%%%
%%%%%%%%%%%%%%%%%%%%%%%%%%%%%%%%%%%%%%%%%%%%%%%%

\subsection{The Berends--Giele version $H_{12345}$}
\label{app:D.2}

In this appendix, we spell out the rank-five generalization of the gauge scalars $H_P$ in (\ref{pert4.op})
that relate Berends--Giele currents in Lorenz and BCJ gauge via (\ref{pert4.qr}).
\begin{align}
s_{12345}H_{12345} &= \frac{1}{s_{1234}s_{234}} \bigg( \frac{h_{23415}}{s_{23}} - \frac{h_{34215}}{s_{34}} \bigg) + \frac{1}{s_{1234}s_{123}} \bigg( \frac{h_{23145}}{s_{23}} - \frac{h_{12345}}{s_{12}} \bigg) - \frac{ h_{12345} - h_{12435}}{s_{1234}s_{12}s_{34}} \notag \\
&+ \frac{N_{123,4,5}}{5 s_{12}s_{45}} \bigg( \frac{3}{2 s_{123}} + \frac{1}{s_{345}} \bigg) - \frac{1}{5 s_{123}s_{23}s_{45}} \bigg( \frac{3}{2} N_{231,4,5} + N_{541,3,2} \bigg)  \notag \\
&+ \big( k_{1234} \cdot a_5 \big) \frac{N_{34,1,2} - N_{12,3,4}}{4 s_{12}s_{34}} \bigg( \frac{1}{2 s_{1234}} + \frac{1}{5 s_{345}} \bigg) + \frac{1}{5 s_{123} s_{45}} \bigg( \frac{N_{23,1,45}}{s_{23}} - \frac{N_{12,3,45}}{s_{12}} \bigg)   \notag \\
&+\big( k_{1234} \cdot a_5 \big) h_{1234} \bigg( \frac{1}{5 s_{12}s_{45}} \bigg( \frac{3}{2 s_{123}} + \frac{1}{s_{345}} \bigg) + \frac{1}{2 s_{1234}} \bigg( \frac{1}{s_{123}s_{12}} - \frac{1}{s_{234}s_{34}} \bigg) \bigg) \notag \\
&+ \big( k_{1235} \cdot a_4 \big) \frac{h_{2135}}{5 s_{12}} \bigg( \frac{3}{2 s_{123}s_{45}} + \frac{1}{s_{345}} \bigg( \frac{1}{s_{45}} - \frac{1}{s_{34}} \bigg) \bigg)  \notag \\
&+ \big( k_{1234} \cdot a_5 \big) \frac{h_{3214}}{2 s_{23}} \bigg( \frac{3}{5 s_{123} s_{45}} + \frac{1}{s_{1234}} \bigg( \frac{1}{s_{123}} - \frac{1}{s_{234}} \bigg) \bigg) \notag \\
&+ \big( k_{123} \cdot a_{45} \big) \frac{h_{123}}{5 s_{45}} \bigg( \frac{3}{2 s_{123}} \bigg( \frac{1}{s_{12}} - \frac{1}{s_{23}} \bigg) + \frac{1}{s_{345}s_{12}} \bigg)   \notag \\
&+ \big( k_{145} \cdot a_{23} \big) \frac{h_{415}}{5 s_{123}s_{23}s_{45}} +  \big( k_{1235} \cdot a_4 \big) \frac{3 h_{2315}}{10 s_{123}s_{23}s_{45}}  + \big( k_{1245} \cdot a_3 \big)  \frac{h_{5412}}{5s_{123}s_{23}s_{45}}  \notag \\
&- \big( k_{1234} \cdot a_5 \big) \bigg[ \big( k_{123} \cdot a_4 \big) \frac{h_{123}}{4} \bigg( \frac{1}{s_{12} s_{34}} \bigg( \frac{1}{2 s_{1234}} + \frac{1}{5 s_{345}} \bigg) + \frac{1}{s_{1234} s_{123}} \bigg( \frac{1}{s_{12}} - \frac{1}{s_{23}} \bigg) \bigg) \notag \\
&\ \ \ \ + \big( k_{234} \cdot a_1 \big) \frac{h_{342}}{4} \bigg( \frac{1}{s_{12}s_{34}} \bigg( \frac{1}{2 s_{1234}} + \frac{1}{5 s_{345}} \bigg) + \frac{1}{s_{1234}s_{234}} \bigg( \frac{1}{s_{34}} - \frac{1}{s_{23}} \bigg) \bigg) \notag \\
&\ \ \ \ - \frac{\big( k_{124} \cdot a_3 \big) h_{124} + \big( k_{134} \cdot a_2 \big) h_{341} }{4 s_{12}s_{34}} \bigg( \frac{1}{2 s_{1234}} + \frac{1}{5 s_{345}} \bigg) \bigg]  \notag \\
&- \frac{\big( k_{1245} \cdot a_3 \big)}{20 s_{123} s_{12} s_{45}} \bigg( N_{12,4,5} - N_{45,1,2} +  \big( k_{124} \cdot a_5 \big) h_{124} - \big( k_{125} \cdot a_4 \big) h_{125} \notag \\
&\ \ \ \ - \big( k_{145} \cdot a_2 \big) h_{451} + \big( k_{245} \cdot a_1 \big) h_{452} \bigg) + \left( 12345 \rightarrow 54321 \right) 
\end{align} 
%

%%%%%%%%%%%%%%%%%%%%%%%%%%%%%%%%%%%%%%%%%%%%%%%%
%%%%%%%%%%%%%%%%%%%%%%%%%%%%%%%%%%%%%%%%%%%%%%%%
%%%%%%%%%%%%%%%%%%%%%%%%%%%%%%%%%%%%%%%%%%%%%%%%
%%%%%%%%%%%%%%%%%%%%%%%%%%%%%%%%%%%%%%%%%%%%%%%%
%%%%%%%%%%%%%%%%%%%%%%%%%%%%%%%%%%%%%%%%%%%%%%%%

\section{Deriving a BCJ representation for (\YMF) amplitudes}
\label{app:E}

This appendix is dedicated to the proof of (\ref{bcjgauge66}), an $n$-point amplitude representation for
(\YMF) with manifest BCJ duality. In comparison to (\ref{bcjgauge62}), the local master 
numerators are built from multiparticle polarizations of lower rank.
We start by deriving (\ref{bcjgauge66}) in the color-ordering $\sigma =1,2,\ldots,n$ from the amplitude
representation in (\ref{pert3.13}): By non-linear gauge invariance, one can transform the Berends--Giele
currents from Lorenz gauge to BCJ gauge, $\Mfrak_{1P,n-1Q,n} \rightarrow \sum_{\beta,\pi} \Phi(P|\beta)_1
 \Phi(Q|\pi)_{n-1} N_{1\beta,n-1\pi,n}$ and rewrite (\ref{pert3.13}) as
\begin{align}
&{\cal A}_{{\rm YM}+F^3+F^4}(1,2,3,\ldots,n{-}1,n) = \sum_{j=1}^{n-2} (-1)^{n-j} \Mfrak_{12\ldots j,\, n-1 n-2 \ldots  j+1,\, n} \notag \\
& \ \ \ = \sum_{j=1}^{n-2} (-1)^{n-j} \sum_{\beta \in S_{j-1}} \sum_{\pi \in S_{n-2-j}} \Phi( 23\ldots j| \beta )_1 \Phi( n{-}2\ldots j{+}1| \pi)_{n-1} N_{1\beta, \, n-1 \pi, \, n}  \label{bcjgauge67} \\
& \ \ \ = - \sum_{j=1}^{n-2} \sum_{\beta \in S_{j-1}} \sum_{\pi \in S_{n-2-j}}  \phi_{12\ldots j| 1\beta} \phi_{n-1,n-2\ldots j+1 | n-1 \pi} 
\Nfrak_{1,\beta|n| \tilde{\pi} ,n-1} \, , \notag
\end{align}
where $\beta$ and $\pi$ are understood to act on $2,3,\ldots,j$ and $n{-}2,n{-}3,\ldots,j{+}1$, respectively.
In passing to the last line, we have converted $N_{1\beta, \, n-1 \pi, \, n}=-N_{1\beta, \, n, \, n-1 \pi}=(-1)^{n-j-1} \Nfrak_{1,\beta|n| \tilde{\pi} ,n-1} $
via (\ref{bcjgauge65}), where $\tilde{\pi} = \pi(j{+}1),\ldots,\pi(n{-}2)$ is the reversal of  $\pi=\pi(n{-}2),\pi(n{-}3)\ldots \pi(j{+}1)$.

Now, it remains to check that the coefficients of the $\Nfrak_{\ldots}$ are identical in (\ref{bcjgauge66}) and (\ref{bcjgauge67}). 
The coefficients in (\ref{bcjgauge66}) can be rewritten using the Berends--Giele 
recursion (\ref{genpar23}) and (\ref{genpar24}) for doubly-partial amplitudes \cite{Mafra:2016ltu},
\begin{align}
&m(1,2,\ldots,n{-}1,n| 1,\rho(2,3,\ldots,j),n,\rho(j{+}1,\ldots,n{-}2),n{-}1) \notag \\
&= s_{12\ldots n-1} \phi_{12\ldots n-1 | \rho(j+1)\ldots \rho(n-2),n-1,1,\rho(2)\ldots \rho(j)} \notag  \\
&=  \sum_{XY=12\ldots n-1 \atop{ AB= \rho(j+1)\ldots \rho(n-2),n-1,1,\rho(2)\ldots \rho(j)}} (\phi_{X|A}\phi_{Y|B}-\phi_{Y|A}\phi_{X|B})
 \label{bcjgauge68} \\
&= - \phi_{123\ldots j|1\rho(2)\rho(3)\ldots \rho(j)} \phi_{j+1\ldots n-2,n-1|\rho(j+1) \ldots \rho(n-2) n-1}
\notag \\
&= - \phi_{123\ldots j|1\rho(2)\rho(3)\ldots \rho(j)} \phi_{n-1,n-2\ldots j+1| n-1  \rho(n-2) \ldots  \rho(j+1)}\, .
\notag
\end{align}
In the third step, we have used that any deconcatenation $12\ldots n{-}1=XY$ will have $1$ and
$n{-}1$ in different words $X$ and $Y$, such that $\rho(j{+}1)\ldots \rho(n{-}2),n{-}1,1,\rho(2)\ldots \rho(j)=AB$
must also be deconcatenated in a manner where $n{-}1$ and $1$ are separated. One would otherwise 
get a vanishing current $\phi_{P|Q}$ where $P$ is not a permutation of $Q$. The only admissible deconcatenation
in (\ref{bcjgauge68}) is $A=\rho(j{+}1)\ldots \rho(n{-}2),n{-}1$ and $B=1,\rho(2)\ldots \rho(j)$. After 
combining (\ref{bcjgauge68}) with (\ref{bcjgauge66}), the leftover task is to demonstrate the matching of the permutation sums
\begin{align}
&\sum_{\rho \in S_{n-3}}  \Nfrak_{1\rho(23\ldots j)|n| \rho(j+1\ldots n-2)n-1} \phi_{123\ldots j|1\rho(2)\rho(3)\ldots \rho(j)} \phi_{n-1,n-2\ldots j+1| n-1  \rho(n-2) \ldots  \rho(j+1)} \notag \\
&\ \ = \sum_{\beta \in S_{j-1}} \sum_{\pi \in S_{n-2-j}}  \phi_{12\ldots j| 1\beta} \phi_{n-1,n-2\ldots j+1 | n-1 \pi} 
\Nfrak_{1,\beta|n| \tilde{\pi} ,n-1} \, .
\label{bcjgauge69}
\end{align}
We exploit once more that $\phi_{P|Q}$ vanishes unless $P$ is a permutation of $Q$. Hence, the first 
line can only contribute via permutations $\rho \in S_{n-3}$ that do not mix the sets $2,3,\ldots,j$ and $j{+}1,\ldots,n{-}2$,
i.e.\ that factorize into $\beta \in S_{j-1}$ acting on $2,3,\ldots,j$ and $\pi \in S_{n-2-j}$ acting on $n{-}2,\ldots,j{+}1$
as seen in the second line. Finally, the relative flip between the permutation $\pi$ in the second current and $\tilde \pi$
in the $\Nfrak_{\ldots}$ in the second line of (\ref{bcjgauge69}) ties in with the analogous reversal 
of $\rho(j{+}1)$, $\rho(j{+}2), \ldots \rho(n{-}2)$ in the first line. 

So far, we have shown that (\ref{bcjgauge66}) and (\ref{bcjgauge67}) agree when $\sigma=1,2,\ldots,n$.
Given that the special footing of legs $1,n{-}1,n$ in the master numerators $\Nfrak_{1,\beta|n| \tilde{\pi} ,n-1}$ is inert 
under permutations of $2,3,\ldots,n{-}2$, one can literally repeat the above steps for $\sigma= 1,\tau(2,3,\ldots,n{-}2),n{-}1,n$ 
with $\tau \in S_{n-3}$. Like this, (\ref{bcjgauge66}) is demonstrated to hold for a BCJ 
basis of ${\cal A}_{{\rm YM}+F^3+F^4}(\sigma)$. For more general choices of $\sigma$, both sides
of (\ref{bcjgauge66}) obey the same BCJ relations, so the arguments of the proof extend to any $\sigma \in S_{n}$.

\bibliographystyle{JHEP}
%\bibliography{cites}{}

\providecommand{\href}[2]{#2}\begingroup\raggedright\endgroup

\end{document}